\documentclass[aip,reprint]{revtex4-2}

\usepackage[utf8]{inputenc}
\usepackage[T1]{fontenc}
\usepackage{mathptmx}       % Times font for math + text
\usepackage{graphicx}       % Figures
\usepackage{dcolumn}        % Decimal-aligned columns
\usepackage{bm}             % Bold math symbols
\usepackage{makecell}       % Table enhancements
\usepackage{xcolor}         % Colors

\usepackage{amsmath}
\usepackage{amssymb}

\usepackage{subcaption}     % For subfigures (better than subfig)

\usepackage{hyperref}
\hypersetup{colorlinks=true, linkcolor=blue, citecolor=blue, urlcolor=blue}

\usepackage[normalem]{ulem} % Underline / strikeout
\renewcommand{\emph}[1]{\textit{#1}} % Ensure emph = italics

\begin{document}

\preprint{AIP/123-QED}

\title[]{Vorticity Packing Effects on Long Time Turbulent Transport in Decaying Two-Dimensional Incompressible Navier-Stokes Fluids.}
% Force line breaks with \\
\author{ Snehanshu Maiti}
\email{snehanshu.maiti@gmail.com}
% \altaffiliation[Also at ]{Institute for Plasma Research, Bhat, Gandhinagar, Gujarat 382428, India.}%Lines break automatically or can be forced with \\
\affiliation{%
Institute for Plasma Research, Bhat, Gandhinagar, Gujarat 382428, India%\\This line break forced% with \\
}%
\author{ Shishir Biswas}%
% \email{Second.Author@institution.edu.}
\affiliation{ 
Department of Physics and Astronomy, University of Notre Dame, Notre Dame, IN, USA%\\This line break forced with \textbackslash\textbackslash
}%

\author{Rajaraman Ganesh}
% \homepage{http://www.Second.institution.edu/~Charlie.Author.}
\affiliation{%
Institute for Plasma Research, Bhat, Gandhinagar, Gujarat 382428, India%\\This line break forced% with \\
}%
\affiliation{%
Homi Bhabha National Institute, Training School Complex, Anushaktinagar, Mumbai 400094, India.%\\This line break forced% with \\
}%
%\altaffiliation[Also at ]{Homi Bhabha National Institute, Training School Complex, Anushaktinagar, Mumbai 400094, India.}%Lines break automatically or can be forced with \\

\date{\today}% It is always \today, today,

\begin{abstract}

 {Recent high-resolution, high-Reynolds-number simulations have shown that the initial total circulation, quantified by the vorticity packing fraction (VPF), strongly influences the late-time Eulerian statistical equilibria of decaying incompressible two-dimensional Navier--Stokes turbulence (Biswas et al., 2022, \textit{Physics of Fluids} \textbf{34}, 065101), revealing a transition from point-vortex–dominated to finite-size (patch-vortex) equilibria with increasing vortex packing, and emphasizing the role of of the classical exclusion principle
(i.e., incompressibility) and total circulation in determining the final statistical states. The present study examines how the associated Lagrangian tracer transport evolves with VPF across the early (linear--nonlinear turbulence onset), intermediate (turbulence development), and late (coherent dipole evolution) stages, and how it correlates with the corresponding Eulerian states. Turbulence, triggered by the Kelvin--Helmholtz instability and sustained by inverse energy cascades, forms large-scale coherent vortices that govern long-time transport. Tracer dynamics, analyzed via mean-square displacement and position--velocity probability distribution functions (PDFs), reveal that increasing VPF accelerates turbulence onset, drives a transition from sub- to super-diffusive transport with decreasing anisotropy in the intermediate stage, and determines late-time behavior dominated by either orbital coherent vortex trapping (sub-diffusive) or linear translational dipole motion (super-diffusive). These distinct long-time transport characteristics, evolving from sub- to super-diffusive behavior with increasing vorticity packing, demonstrate a strong correspondence between the transition from point-vortex– to finite-size–vortex–dominated Eulerian equilibria and the underlying Lagrangian transport in decaying incompressible 2D Navier–Stokes turbulence.}

\end{abstract}

\maketitle

%\begin{quotation}
%The ``lead paragraph'' is encapsulated with the \LaTeX\ 
%\verb+quotation+ environment and is formatted as a %single paragraph before the first section heading. 
%(The \verb+quotation+ environment reverts to its usual meaning after the first sectioning command.) 
%Note that numbered references are allowed in the lead paragraph.
%
%The lead paragraph will only be found in an article being prepared for the journal \textit{Chaos}.
%`\end{quotation}

\section{Introduction}

A recent study, ``Long time fate of two-dimensional incompressible high Reynolds number Navier–Stokes turbulence: A quantitative comparison between theory and simulation,'' by \textit{Biswas \textit{et al.}}, \textit{Physics of Fluids} \textbf{34}, 065101 (2022), employed high-resolution, high-Reynolds-number direct numerical simulations (DNS) to systematically examine the role of initial total circulation, quantified via the vortex packing fraction, in determining the late-time relaxed states of decaying two-dimensional (2D) incompressible Navier-Stokes turbulence \cite{BG2022}. The DNS results were compared with statistical mechanical models of point-vortex theory \cite{Montgomery1992} and finite-sized patch vortex theory, based on the Kuz’min–Miller–Roberts–Sommeria (KMRS) model \cite{Kuzmin1982, Miller1990, Robert1991}. In point vortex theory, vortices are treated as discrete points, incompressibility is neglected, and only the total energy and zero total circulation are conserved, yielding the classical sinh–Poisson relationship between vorticity and stream function. In contrast, patch vortex theory treats vortices as finite-sized patches and enforces incompressibility by conserving regions of zero and nonzero vorticity, predicting a generalized relaxed state described by KMRS theory. The study showed that for loosely packed vortex configurations (low initial vorticity packing fraction), the final DNS state agrees quantitatively with the point-vortex theory, whereas for tightly packed vortex configurations (high initial vorticity packing fraction), the final DNS state agrees quantitatively with the KMRS predictions based on finite-sized “patch vortices,” which account for incompressibility, while deviating substantially from the point-vortex (sinh–Poisson) results that neglect incompressibility. As the packing fraction increases, the late-time DNS states shift from agreement with the point-vortex predictions toward the KMRS predictions, reflecting the transition from a point-vortex limit to a finite-size vortex–dominated regime and highlighting the critical role of initial total circulation and the classical exclusion principle (i.e., incompressibility) in governing the late-time statistical mechanics of decaying incompressible 2D Navier–Stokes turbulence.

Building upon the findings of Biswas et al.\cite{BG2022}, the present study investigates the associated Lagrangian passive tracer particle transport dynamics in decaying 2D incompressible  Navier–Stokes turbulence, examining how the initial vorticity packing fraction and circulation influence transport across different stages of fluid evolution—early (from linear to nonlinear onset of turbulence), intermediate (turbulence development), and late (evolution of the largest coherent structures)—and exploring the possible correlation between late-time Eulerian statistical equilibrium structures and the underlying Lagrangian transport. 

In fluid turbulence, the Eulerian framework describes the evolution of fluid quantities—such as velocity, vorticity, and stream function—at fixed spatial locations, whereas the Lagrangian framework tracks individual fluid parcels along their trajectories, providing direct insight into scale-dependent particle transport, mixing and dispersion. 
Transport in such systems can be quantified by evolving passive tracer particles within the Eulerian velocity fields and analyzing statistical measures such as mean square displacement (MSD) or absolute dispersion, time-dependent diffusion coefficients, and the probability distribution functions (PDFs) of particle positions and velocities, along with their correlations.

To this end, a dedicated passive tracer-particle solver (TP) has been developed, tested, and integrated with the existing in-house GPU-based 2D hydrodynamic fluid solver, GHD2D \cite{Rupakthesis, BG2022}, designed for studying 2D neutral fluids. The particle solver has been benchmarked against standard results of passive tracer-particle transport in a 2D kinematic chaotic system consisting of a regular lattice of non-stationary kinematic eddies \citep{FL2022}, ensuring its reliability. Using this combined GHD2D-TP tool, which enables the simultaneous evolution of tracer particles and the underlying fluid flow, we investigate the transport properties of two-dimensional incompressible Navier–Stokes turbulent flows initiated by Kelvin–Helmholtz instabilities, allowed to decay freely, and their associated Lagrangian transport over long times. Particle transport is examined across early (short), intermediate, and late (long) timescales. The nature of the resulting turbulence and it's long time dynamics is modulated by varying the initial conditions, particularly the vorticity packing fraction at the initial time. To characterize the resulting transport, we compute the time evolution of mean-square displacement (MSD) or absolute dispersion and the time-dependent diffusion coefficients of tracer particles, along with the time evolution of the probability distribution functions (PDFs) of their positions and velocities, and their correlations, to assess how turbulent transport depends on the initial vorticity distribution.

In Section II, we describe the governing equations of the problems studied in this article, the numerical solvers used and their benchmarking results, as well as the simulations employed in this study. Section III presents our results, highlighting several distinct regimes of fluid and particle motion that emerge during the evolution of turbulence, and how these regimes are influenced by the initial vorticity packing fraction.
In particular, we report the time evolution of mean- square displacement (MSD, or absolute dispersion) and the time-dependent diffusion coefficients of passive tracer particle transport, along with the time evolution of probability distribution functions (PDFs) of particle positions and velocities in both spatial directions, and their correlations. These statistical measures offer valuable insights into the nature of the underlying turbulent structures and transport dynamics. Finally, in Section IV, we summarize our findings, present the main conclusions of the study, and report whether a strong correlation exists between late-time Eulerian statistical equilibrium structures and the underlying Lagrangian transport in 2D decaying turbulence.

\section{Equations and Numerical Solvers}

In this section, we outline the governing equations for decaying two-dimensional (2D) incompressible Navier–Stokes (NS) turbulence and the associated passive tracer particle transport, describe the numerical solvers used to solve these equations and their benchmarking, and detail the simulation setup, parameters, and subsequent analysis of fluid fields and particle trajectories employed in this study. \\

\hspace{-0.3cm}\textbf{Eulerian fluid dynamics} \\

\textbf{Governing equations} \\

The dynamics of decaying 2D incompressible  hydrodynamic turbulence in the absence of external forcing is governed by the dimensionless Navier–Stokes equation and the solenoidal (divergence-free) condition on the velocity field:

\begin{equation}
    \frac{\partial \vec{u}}{\partial t} + (\vec{u} \cdot \vec{\nabla}) \vec{u}
    = - {\vec{\nabla} p} + \frac{1}{R_E} \nabla^2 \vec{u};  \quad \quad  \nabla . \vec u = 0
    \label{eq:navier_stokes}
\end{equation}

\hspace{-0.33cm}where $\vec u$ is the fluid velocity field and $p$ is the fluid pressure field. All quantities in Eq.~\eqref{eq:navier_stokes} are non dimensional. The Reynolds number, $R_E$, is a dimensionless parameter that quantifies the ratio of inertial to viscous forces, and is defined as $R_E =\frac{u_0 L_0}{\nu}$ where $u_0$ and $L_0$ are characteristic velocity and length scales, respectively, and $\nu$ is the kinematic viscosity. Larger values of $R_E$ indicate a greater dominance of inertial effects over viscous effects, corresponding to more turbulent flow conditions.

To simplify analysis, the velocity field $\vec{u}(x,y,t)$ can be reformulated in terms of scalar quantities: the vorticity field $\omega(x,y,t)$, representing local fluid rotation, and the stream function field $\psi(x,y,t)$, which remains constant along streamlines. This formulation yields the scalar vorticity equation:

     \begin{equation}
        \frac{\partial \omega}{\partial t}  = [\psi, \omega]  +  \frac{1}{R_E} {\nabla}^2 \omega
        \label{eq:navier_stokes_scalar}
        \end{equation}

The relationships among  velocity, vorticity, and stream function are:

  \begin{equation}
     \omega = {\hat z} \cdot (\nabla \times {\vec u});  \hspace{0.5cm} \omega= - {\nabla}^2 \psi; \hspace{0.5cm} \vec u= \nabla \times \psi{\hat z}
  \end{equation}

  \hspace{-0.38cm} where ${\vec u}$ is the in-plane 2D velocity vector and ${\hat z}$ is the out-of-the-board unit vector. 
Here, $[\psi, \omega]$ in Eq.~\eqref{eq:navier_stokes_scalar} denotes the Poisson bracket, defined as:
 
        \begin{equation}
        [\psi, \omega]= \partial _x \psi \partial _y \omega - \partial _y \psi \partial _x \omega \hspace{0.1cm} 
        \label{eq:poisson_bracket}
        \end{equation}

 The dynamics of 2D incompressible NS turbulence are typically initiated by an instability, leading to the evolution of the vorticity and stream function fields into a turbulent state through nonlinear interactions. This turbulence undergoes an inverse energy cascade, ultimately forming large-scale coherent structures. To reproduce this behavior, we solve the Navier-Stokes equations (Eqs.~\eqref{eq:navier_stokes}, ~\eqref{eq:navier_stokes_scalar}) and describe the corresponding numerical methods and setup in the following section. \\

%\vspace{1cm}
\newpage

\textbf{Numerical solver and simulations setup} \\

We have used an existing GPU-based  two-dimensional incompressible hydrodynamic solver, GHD2D \cite{Rupakthesis, BG2022, Biswas2024}, developed in-house at the Institute for Plasma Research, to solve Eqs.~\eqref{eq:navier_stokes}--\eqref{eq:poisson_bracket} and study decaying two-dimensional incompressible fluid turbulence in a doubly periodic square domain with Cartesian coordinates. This solver is capable of performing state-of-the-art simulations at high grid resolutions (up to 2048²) \cite{BG2022} and at very high Reynolds numbers, both of which have been employed in our study. The solver employs the pseudospectral method, which represents the solution using global basis functions, such as Fourier modes in the case of periodic domains. This approach leverages the high accuracy of spectral techniques for computing spatial derivatives, making it particularly well-suited for problems with periodic boundary conditions and smooth solution fields. While linear terms and derivatives are efficiently handled in spectral space, nonlinear terms are computed in physical space after performing an inverse Fast Fourier Transform (FFT). This hybrid treatment allows for computational efficiency while retaining spectral accuracy. To suppress aliasing errors that arise during nonlinear product evaluation, the standard 2/3 dealiasing rule\cite{Patterson1971} is applied, whereby the highest one-third of the spectral modes are set to zero. The time integration is carried out in spectral space using a second-order Adams–Bashforth method\cite{Adams1855, Bashforth1883}, ensuring stable and accurate evolution of the solution in time. The solver uses CUDA based FFT library [cuFFT library]\cite{nvidia_cufft} to perform Fourier transforms.

We generate various turbulent flow regimes using GHD2D at a high grid resolution of 2048 × 2048 by systematically varying the initial vorticity packing fraction, achieved through altering the number of initial vorticity strips in the simulation domain.
 The vorticity distribution is initialized using parallel, alternating strips of positive and negative vorticity, simulating oppositely directed jets. The number of strips is varied from 2 to 20. The computational domain is a square of size \( L = 2\pi \), and each strip has a width \( d = \pi/16 \). The region between the strips is non-circulating (zero vorticity). The vorticity packing fraction (VPF), defined as the ratio of the area occupied by circulating regions to the total domain area, is given by:

    \begin{equation}
        VPF = \frac{n \cdot d \cdot L}{L^2}
    \end{equation}

  \hspace{-0.33cm}  where \( n \) is the number of strips. The packing fractions corresponding to n = 2, 4, 8, 16, and 20 strips are 6.25\%, 12.5\%, 25\%, 50\%, and 62.5\%, respectively.

    The circulation is defined as \( C = \int \omega \, dx \, dy \). For the positive vorticity regions, the total circulation is:

    \begin{equation}
        C_+ = n_+ \cdot d \cdot L \cdot \omega_+
    \end{equation}

 \hspace{-0.33cm}   Since the number of positive and negative strips is equal, we have \( C_+ = C_- \), and thus the net circulation is \( C = C_+ + C_- = 0 \).

The fluid system is perturbed by adding a small-amplitude perturbation to the vorticity field:

   \begin{equation}
        \delta \omega = \sum_{m=1}^{64} \alpha \cos(mx + \phi_m)
   \end{equation}

\hspace{-0.33cm}where $\alpha$ denotes the perturbation amplitude (assumed to be small), \( m \) is the mode number, and \( \phi_m \) represents the phase associated with each mode. In our simulations, the perturbation amplitude $\alpha$ is set to 
0.01. The phase \( \phi_m \) is randomly drawn from a uniform distribution between \( -\pi \) and \( \pi \). The fluid system, along with the tracer particles, is integrated using a time step of $\Delta t = 10^{-3}$ which is sufficient to resolve the dynamics at a resolution of $2048^2$, and evolved until $T = 3000$, at which point a steady state is attained and examined over an extended period.
  \\

\textbf{Fluid field analysis} \\

The Eulerian evolution of the fluid field can be visualized through the time evolution of the vorticity and stream function fields (see Fig.~\ref{fig:VS}). The underlying fluid structure of 2D hydrodynamic  turbulence is composed of rotating vortices embedded within regions of background fluid deformation. The Okubo–Weiss parameter, $Q(x,y,t)$, quantifies the relative importance of rotation and deformation in such turbulent flows.  The Okubo--Weiss parameter $Q(x, y, t)$ quantifies the relative importance of rotation and deformation in two-dimensional turbulence. It is defined as
\begin{equation}
    Q = S^2 - \omega^2,
\end{equation}
where $S^2 = S_1^2 + S_2^2$, 
\begin{equation}
    S_1(x, y, t) = \partial_x u_x - \partial_y u_y, \quad 
    S_2(x, y, t) = \partial_x u_y + \partial_y u_x.
\end{equation}
This parameter distinguishes two distinct flow regimes: 
(1) \textit{elliptic domain} ($Q < 0$), where rotation dominates deformation ($\omega^2 > S^2$); and 
(2) \textit{hyperbolic domain} ($Q > 0$), where deformation dominates rotation ($\omega^2 < S^2$) (see Fig.~\ref{fig:OW}).

The late-time Eulerian statistical equilibria, derived from entropy–maximization principles, relate vorticity $(\omega)$ and stream function $(\psi)$ as follows.
For the point-vortex model (sinh–Poisson formulation):
\begin{equation}
\omega_{\mathrm{PV}} = \alpha \sinh(-\beta \psi),
\end{equation}
and for the finite-size (patch) vortex or KMRS model:
\begin{equation}
\omega_{\mathrm{KMRS}} =
\frac{A e^{-B\psi} - C e^{B\psi}}{1 + [A e^{-B\psi} + C e^{B\psi}]}.
\end{equation}

It has been demonstrated that\cite{BG2022}, while at low vortex packing fractions (VPF) the late-time Eulerian statistical equilibria agree well with the point-vortex (sinh–Poisson) theory, at high VPF they conform more closely to the finite-size or patch-vortex (KMRS) theory. \\

\newpage
\textbf{ Benchmarking numerical simulations and study of Kelvin-Helmholtz instability} \\

To validate the accuracy of our numerical simulations using GHD2D, we compute the Kelvin–Helmholtz instability (KHI) growth rates for an oppositely directed broken-jet type initial configuration from our GHD2D fluid simulations and compare them with the well-established analytical predictions \cite{D1961}. The Kelvin-Helmholtz instability (KHI) is a hydrodynamic instability that occurs when there is a velocity difference across the interface between two fluids, or a velocity shear within a single fluid, leading to the formation of vortex roll-up patterns and turbulence. The viscous growth rate of the Kelvin-Helmholtz instability for a step shear or broken jet profile in hydrodynamics, in the incompressible limit, was analytically derived by Drazin \cite{D1961}. The formula for the growth rate, as obtained by Drazin \cite{D1961}, is given as follows:

  \begin{equation}
\gamma = \frac{k_x U_0}{3} \left[ \sqrt{3} - 2\frac{k_x}{R_E} - 2 \sqrt{\left( \frac{k_x}{R_E} \right)^2 + 2\sqrt{3} \frac{k_x}{R_E} } \right] \label{eq:KHI}
\end{equation}

\begin{figure*}
    \centering
    \begin{subfigure}[t]{0.49\textwidth}  % align top
        \centering
        \includegraphics[width=\textwidth]{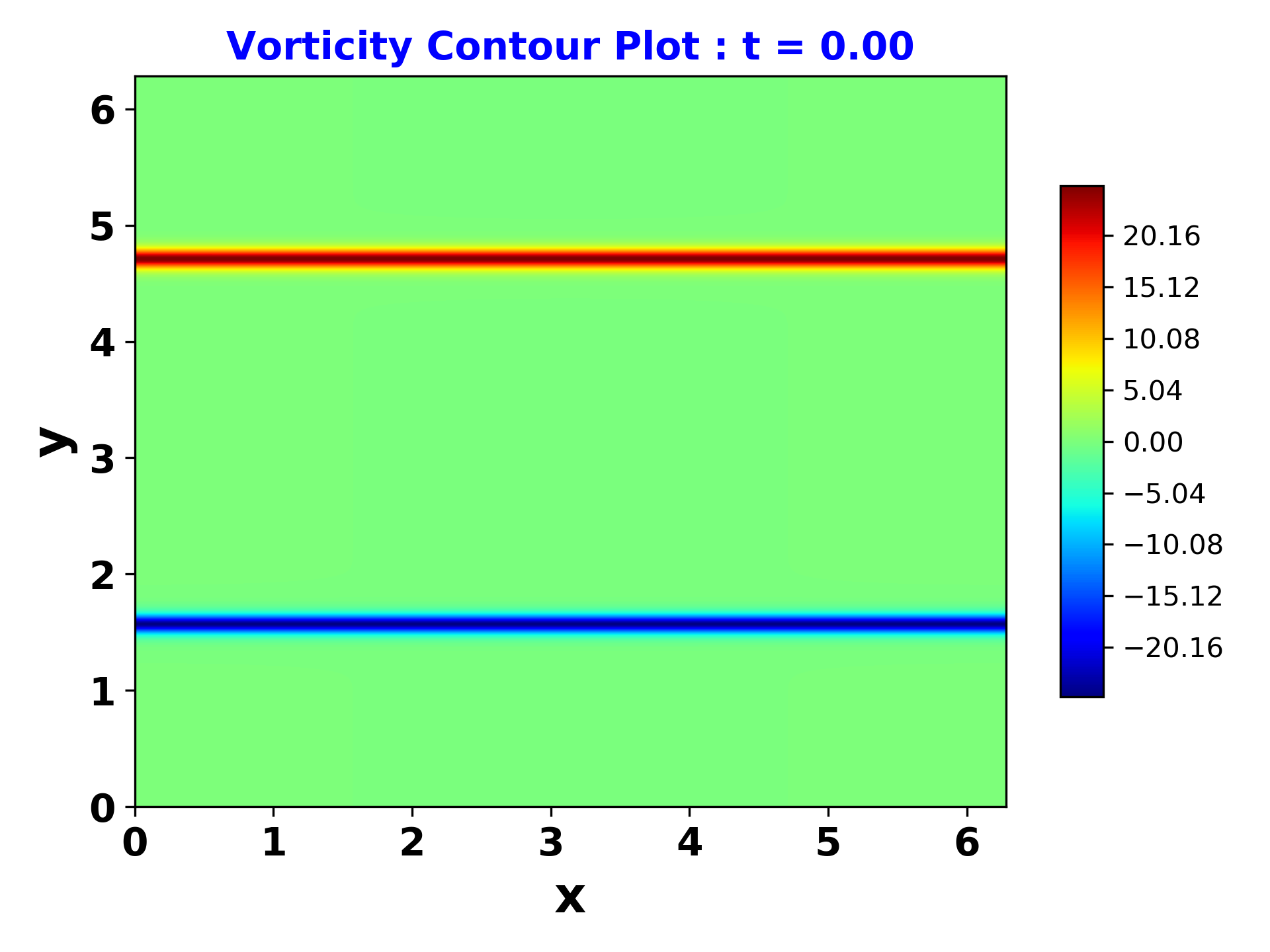}
        \caption{}
        \label{brokenjet_ic}
    \end{subfigure}
    \hfill
    \begin{subfigure}[t]{0.49\textwidth}  % align top
        \centering
       \raisebox{0.34cm}{\includegraphics[width=\textwidth]{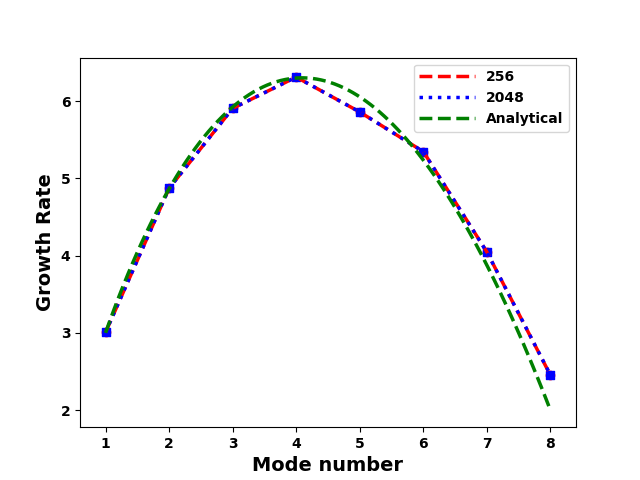}}
        \caption{}
        \label{growth_analytical_fitting}
    \end{subfigure}
    \caption{(a) Initial condition: Two finite strips of fluid with alternating vorticity, flowing in opposite directions, forming discontinuous shear layers (commonly referred to as “oppositely directed broken jets”) to study Kelvin–Helmholtz instability. The Reynolds number of the flow is $R_E = 122$, and the peak vorticity is $\omega_0 = \pm 25$. (b) The growth rates of the Kelvin–Helmholtz instability in the oppositely directed broken jets problem, computed using the GHD2D solver for various mode numbers, are in excellent agreement with the analytical results of Drazin \textit{et al.} in the two-strip configuration. Simulations were performed at grid resolutions of $256^2$ and $2048^2$, and the results are found to be invariant with respect to resolution.l
    \label{brokenjet_benchmark}}
\end{figure*}

\begin{figure*}
     \centering
     \begin{subfigure}[b]{0.49\textwidth}
         \centering
         \includegraphics[width=\textwidth]{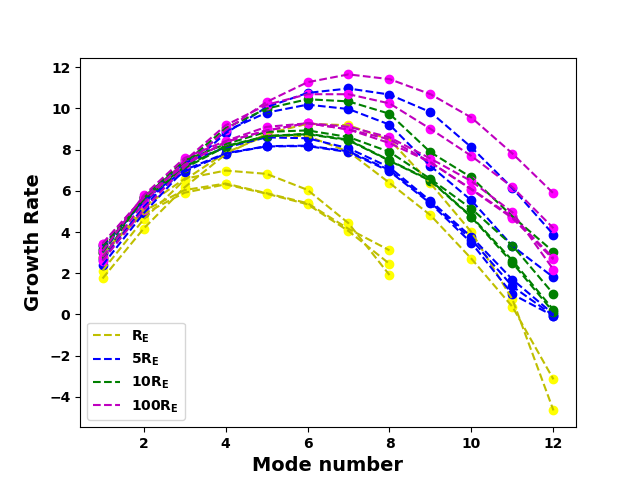}
        \caption{}
         \label{growth_tapered}
    \end{subfigure}
    \hfill
     \begin{subfigure}[b]{0.49\textwidth}
         \centering
         \includegraphics[width=\textwidth]{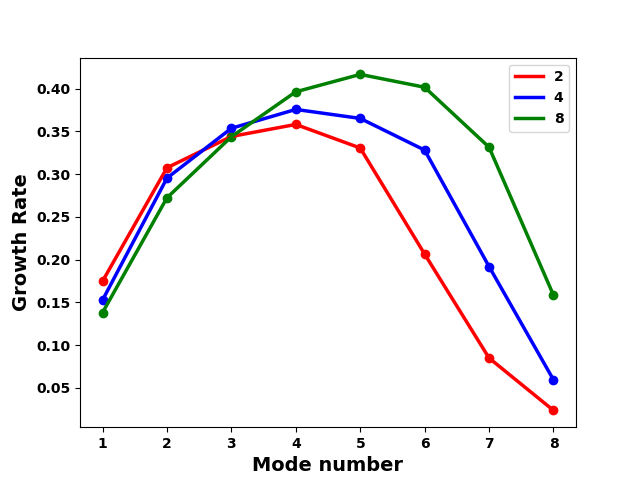}
        \caption{}
        \label{growth_sharp}
    \end{subfigure}
    \caption{Variation of the Kelvin–Helmholtz instability growth rate with mode number for different combinations of Reynolds number ($R_E = 122$ and its multiples) and initial vorticity strip numbers (2, 4, 8, 16, 20), using a \textbf{tapered profile} with vorticity $\omega_0 = \pm 25$. Each color represents a distinct Reynolds number, and within each color group, different lines correspond to increasing strip numbers (2 to 20). For a given $R_E$, the growth rate increases with strip number. Each bell-shaped curve highlights the effect of both Reynolds number and strip count on the peak growth rate and the dominant mode. (b) Variation of the Kelvin–Helmholtz instability growth rate with mode number for a \textbf{sharp profile} of the initial vortex strips at a $R_E$ = 228576 and  $\omega_0=\pm 1$.
    \label{growth_sharp_and_tapered}
 }
\end{figure*}

\hspace{-0.33cm}where $k_x$ is the dimensionless wave-number, $U_0$ is the dimensionless shear velocity and $R_E$ denotes the dimensionless Reynolds number. %The Reynolds number $R_E$ in Eq.~\eqref{eq:KHI}  is defined as $R_E = U_0d/ \eta$, where $d$ is the shearing length scale and $\eta$ is the shear viscosity. 
We compute the KHI growth rates for various mode numbers using our numerical simulations with the GHD2D solver. In this setup, the initial condition consists of two oppositely directed jets (broken jets) with  peak initial vorticities $\omega_0=\pm 25$ (see Fig. \ref{brokenjet_ic}) and a tapered distribution profile  to trigger the instability. In the tapered profile, the vorticity reaches its maximum at the center of the strips and decreases smoothly and symmetrically outward, resulting in a bell-shaped distribution instead of abrupt variations (see Fig. \ref{sharp_vs_tapered_initial}). For a $R_E=122$, the growth rate as a function of mode number from our simulations shows good agreement with the analytical predictions given by Eq.~\eqref{eq:KHI}. This comparison is illustrated in Fig. \ref{growth_analytical_fitting}, where results from both low (grid size 256) and high (grid size 2048) resolution simulations are presented. These benchmarking results are consistent with those reported by the original developers of GHD2D \cite{Rupakthesis, BG2022}. We carry out additional investigations of the Kelvin–Helmholtz (KH) instability growth rate by varying the Reynolds number ($R_E$ = 5, 10, 100 $R_E$) and the number of initial strips (2, 4, 8, 16, 20). Figure \ref{growth_tapered} shows results for a tapered profile, where the KHI growth rate increases with $R_E$. For a given $R_E$, the growth rates are similar for configurations with fewer strips (2, 4, 8) but increase significantly for higher strip numbers (16 and 20).

To demonstrate the robustness of our findings with respect to the initial vorticity profile, we further examine the effect of strip shape by comparing a tapered profile with a sharp-edged profile, using parameters employed in our long-time 2D incompressible Navier--Stokes simulations (${R_E} = 228576$, $\omega_0 = \pm 1$) (see Fig. \ref{sharp_vs_tapered_initial}). In the sharp- edge profile, the vorticity exhibits abrupt transitions between regions of high and low values, leading to steep gradients and a discontinuous-like structure. 
Figure \ref{sharp_vs_tapered_initial} shows the vorticity distributions for both tapered and sharp-edged profiles, with identical positions, widths, and circulations, and a peak vorticity value of $\omega_0 = 1$, enabling a direct comparison of their impact on the slower evolution of turbulence.
The resulting growth rates obtained from both profiles at high $R_E$ = 228576 are similar, although the  growth rate for sharp-edged broken jet initial conditions is slightly higher than that for tapered profile(see Fig. \ref{growth_sharp_vs_tapered}).
The sharp-edge case is characterized by the introduction of strong shear layers at the jet boundaries, which generate a broader spectrum of unstable modes, particularly at smaller spatial scales (higher wavenumbers), due to the presence of discontinuous vorticity jumps.  The high-wavenumber modes typically exhibit rapid growth, leading to a faster onset of the KH instability as compared to the other configuration.  Consequently, the transition to turbulence is more rapid and efficient in the sharp-edge case. Based on this outcome, we choose to proceed with the sharp profile in subsequent studies. Additionally, we investigate the influence of the number of initial vortex strips, or equivalently the initial vorticity packing fraction (VPF), on the Kelvin–Helmholtz (KH) instability growth rates across different mode numbers. For the sharp-edged profile at $R_E = 228576$, the KH growth rate across modes increases with the number of vortex strips or VPF (see Fig. \ref{growth_sharp}). \\

\begin{figure*}
     \centering
     \begin{subfigure}[b]{0.49\textwidth}
         \centering
         \includegraphics[width=\textwidth]{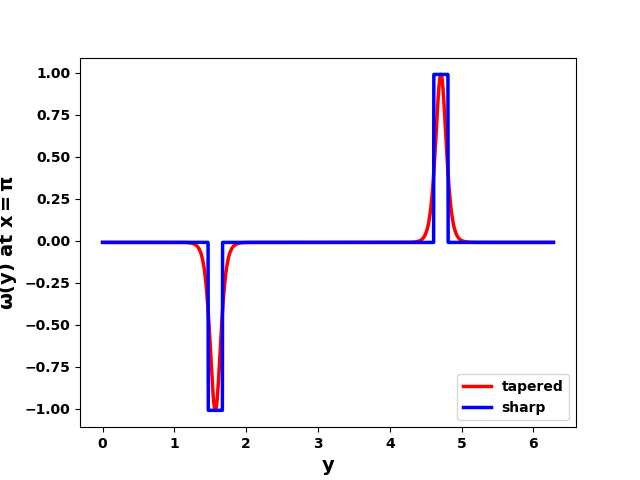}
        \caption{}
         \label{sharp_vs_tapered_initial}
    \end{subfigure}
    \hfill
     \begin{subfigure}[b]{0.49\textwidth}
         \centering
          \includegraphics[width=\textwidth]{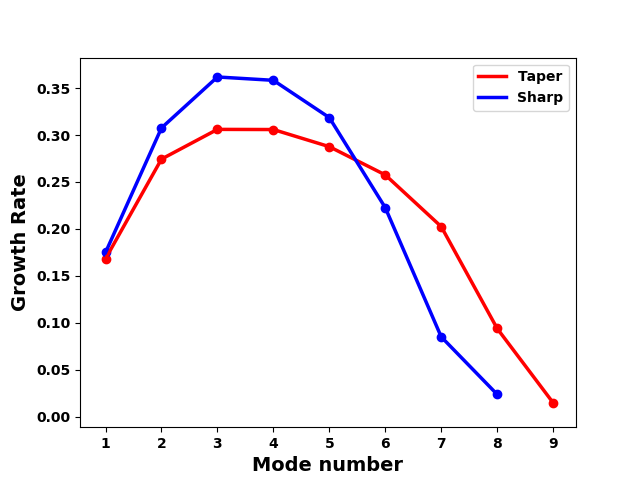}
        \caption{}
       \label{growth_sharp_vs_tapered}
    \end{subfigure}
    \caption{(a) Comparison of vorticity profiles for the two-strip configuration in the oppositely directed broken-jet problem, implemented using sharp and tapered initial conditions. (b) Comparison of the growth rates of the Kelvin–Helmholtz instability for sharp and tapered initial conditions in a fluid with $R_E = 228{,}576$ and $\omega_0 = \pm 1$, showing overall similar behavior.
    \label{growth_sharp_and_tapered_full}
 }
\end{figure*}

\iffalse
\textcolor{red}{}
%\newpage
\begin{figure*}
%     \centering
\begin{subfigure}[b]{0.33\textwidth}
         \centering
 %        \includegraphics[width=\textwidth]{1a.png}
 %        \caption{}
 %        \label{growth_analytical_fitting}
     \end{subfigure}
     \begin{subfigure}[b]{0.33\textwidth}
         \centering
         \includegraphics[width=\textwidth]{2A.png}
         \caption{}
         \label{sharp_vs_tapered_initial}
     \end{subfigure}
     \hfill
     \begin{subfigure}[b]{0.33\textwidth}
 %        \centering
         \includegraphics[width=\textwidth]{2B.png}
         \caption{}
         \label{growth_sharp_vs_tapered}
     \end{subfigure}
     \caption{Figure (a) shows the benchmarking of our GHD2D solver against the analytical results of Drazin et al., demonstrating close agreement in the growth rates of the Kelvin–Helmholtz instability across varying mode numbers for a 2 strip oppositely directed broken jet problem having a tapered vorticity distribution profile. The $R_E$ of the fluid is 122, $\omega_0=\pm 25$  and simulations has been performed at resolutions 256 and 2048. Figure (b) compares the vorticity profiles of the 2 strip oppositely directed broken jet problem implemented using sharp and tapered initial conditions. Figure (c) compares the growth rates of KH instability implemented using sharp and tapered initial conditions for a fluid with $R_E$ = 228576 and $\omega_0=\pm 1$ which shows comparatively similar behavior.}
\end{figure*}

\fi
%\textcolor{red}{}

\hspace{-0.3cm}\textbf{Lagrangian passive tracer particle dynamics}     \\   

\textbf{ Governing equations}     \\   

The Lagrangian description involves tracking passive tracer particles that move with the flow but do not influence the fluid dynamics. The motion of such particles is governed by the following equations:

    \begin{equation}
        \frac{dx}{dt}=  u(x,y,t) = \frac{\partial{\psi}}{\partial{y}}; \quad \quad 
        \frac{dy}{dt}=  v(x,y,t) = - \frac{\partial{\psi}}{\partial{x}}
         \label{eq:adv_eqn}
        \end{equation} 

\hspace{-0.33cm}Here, \( x(t) \) and \( y(t) \) represent the instantaneous positions of the tracer particle as functions of time, \( u(x, y, t) \) and \( v(x, y, t) \) are the instantaneous scalar fluid velocity field components in the \( x \)- and \( y \)-directions, respectively, and \( \psi(x, y, t) \) is the instantaneous stream-function field of the fluid, from which the incompressible velocity field is derived. \\

\textbf{Numerical solver and simulations setup}     \\  

\begin{figure*}

     \begin{subfigure}{0.49\textwidth}
       \centering\includegraphics[width=\textwidth]{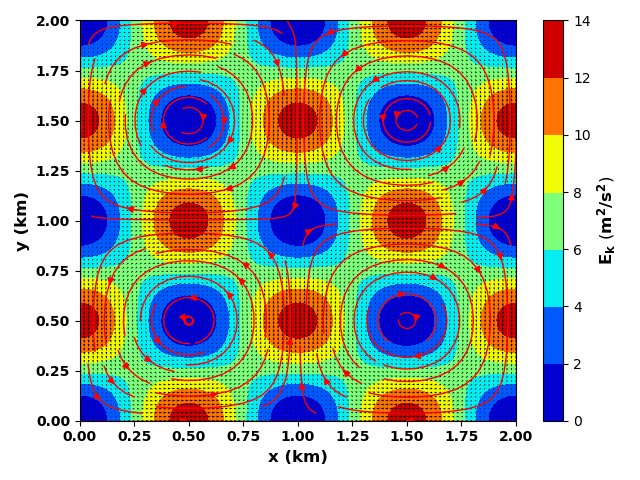}
        \caption{}
        \label{fig:kinv}
    \end{subfigure}
    \hfill
    \begin{subfigure}{0.49\textwidth}
         \centering \includegraphics[width=\textwidth]{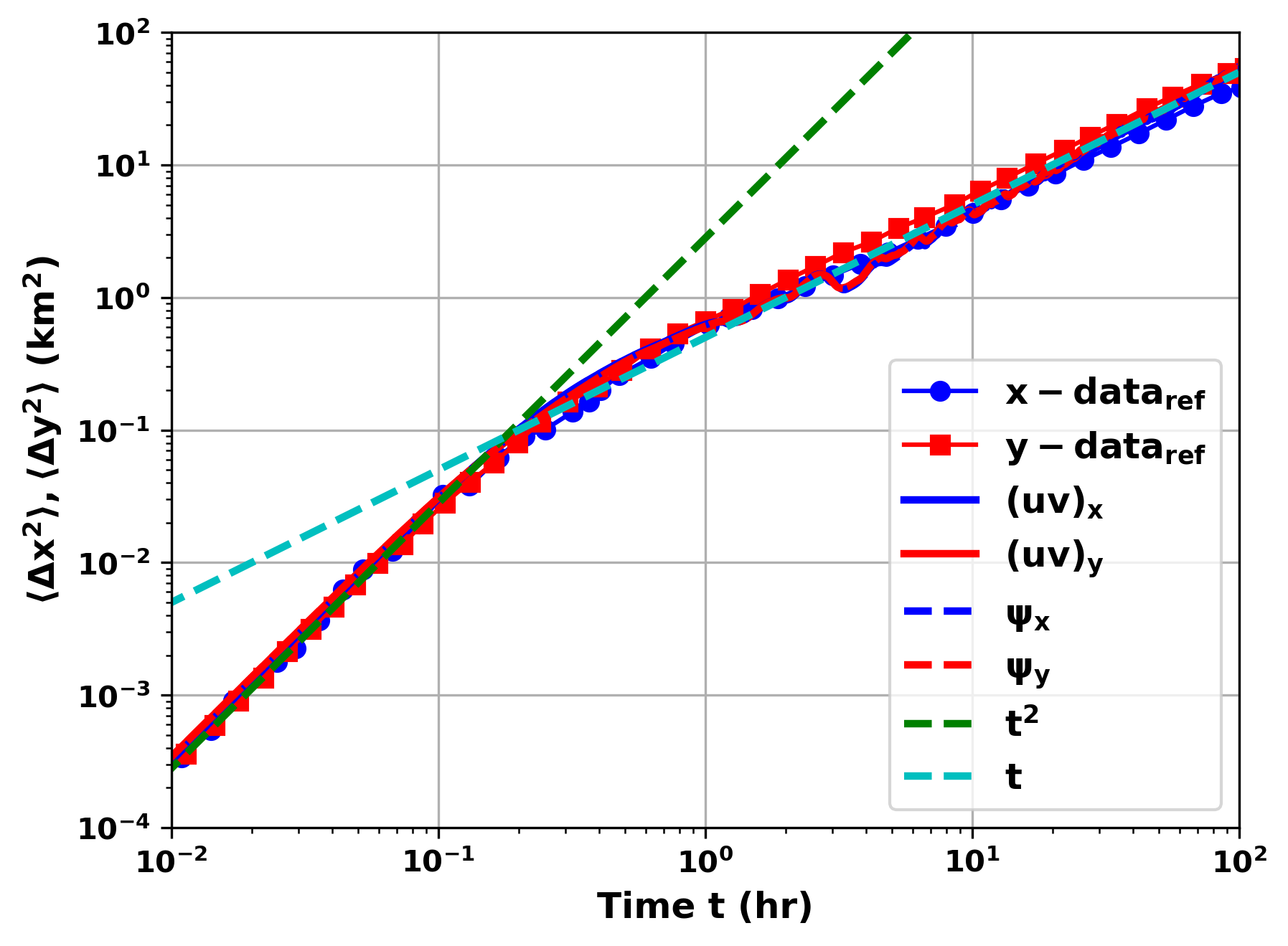}
        \caption{}
         \label{fig:kinp}
     \end{subfigure}
        \caption{ (a) Initial conditions of the 2D kinematic flow described by the velocity functions in Eqs.~\eqref{eq:u}--\eqref{eq:v}. The flow exhibits a lattice of non-steady square cells, as shown by the contour map of the total kinetic energy $E_k$. Streamlines (red lines) indicate the instantaneous flow directions, while quiver arrows represent the velocity vectors at discrete points, illustrating both the magnitude and orientation of the flow.  
        (b) Transport results (single-particle mean-square displacement (MSD) or absolute dispersion) from our tracer particle solver, showing 1,000 advected particles in the kinematic chaotic flow (see Fig. \ref{fig:kinv}) compared with the results of Folgia et al. \cite{FL2022}, which has been reproduced with the corresponding author’s permission. This figure shows results obtained from both velocity ($\mathrm{(uv)_x}, \mathrm{(uv)_y}$) and stream function ($\mathrm{(\psi)_x}, \mathrm{(\psi)_y}$) interpolation  and illustrates a similar behavior to Folgia et al. \cite{FL2022}($\mathrm{(x\text{-}data)_{ref}}, \mathrm{(y\text{-}data)_{ref}}$) , with an initial ballistic regime ($\propto \mathrm{t}^2$) transitioning into a normal diffusive regime ($\propto \mathrm{t}$), and an eddy turnover time of approximately 1~hr.}
           \label{fig:LagBM}
\end{figure*}

To solve the above passive tracer particle transport equations, Eq.~\eqref{eq:adv_eqn},  and to address the objectives of the present study, we have developed a custom particle-tracking solver to compute passive tracer particle trajectories from a given velocity or stream-function fluid field. The particle transport equations are integrated using the fourth-order Runge--Kutta (RK4) algorithm, which has a local truncation error of $\mathcal{O}(h^5)$ and a global accuracy of $\mathcal{O}(h^4)$\cite{Press2007}, and is chosen for its high accuracy and numerical stability. Particle trajectories are obtained using both velocity field and stream function field interpolations, implemented via a linear interpolation scheme\cite{Press2007}, with both methods yielding similar results for particle trajectories (see Fig.~\ref{fig:kinp}).

This particle solver has been benchmarked against tracer transport in a kinematic two-dimensional chaotic flow, consisting of a regular lattice of non-stationary kinematic eddies\cite{FL2022} (see Fig.~\ref{fig:kinv}), as described later. It has been further integrated with the two-dimensional GPU-based hydrodynamics solver (GHD2D), enabling the simultaneous evolution of tracer particles and the underlying fluid flow. The resulting coupled framework, GHD2D-TP (Tracer Particle), allows for detailed analysis of Lagrangian transport processes in 2D turbulence problems. Simulations are performed over long durations (up to $T$ = 3000 in simulation units) to ensure the system reaches a statistically steady state. The time step for both fluid and particle solvers is fixed at $\Delta t = 10^{-3}$. A total of 1000 tracer particles are employed to analyze dispersion characteristics. \\

\textbf{Particle trajectory analysis} \\

Once the particle trajectories are obtained, turbulent transport is quantified through the time evolution of the ensemble-averaged mean-square displacement (MSD) of tracer particles. The mean-square displacements (MSD) in the $x$ and $y$ directions, 
$\langle \Delta x^2 \rangle$, $\langle \Delta y^2 \rangle$, 
and the total (two-dimensional) MSD, $\langle \Delta r^2 \rangle$, 
are defined as:
\begin{equation}
\langle \Delta x^2(t) \rangle = \langle [x(t) - x_0]^2 \rangle, \quad
\langle \Delta y^2(t) \rangle = \langle [y(t) - y_0]^2 \rangle, \quad
\end{equation}
\begin{equation}
\langle \Delta r^2(t) \rangle = \langle \Delta x^2(t) + \Delta y^2(t) \rangle,
\label{eq:msd}
\end{equation}
where $x_0$ and $y_0$  denote the initial particle positions, and $\langle \cdot \rangle$
indicates an ensemble average over all tracer particles.  The MSD typically follows a power-law scaling of the form
\begin{equation}
\langle \Delta r^2(t) \rangle \propto t^{\alpha},
\end{equation}
where the exponent $\alpha$ characterizes the transport regime: $\alpha < 1$ denotes subdiffusive transport (slower than normal diffusion), $\alpha = 1$ corresponds to normal diffusion, and $\alpha > 1$ indicates superdiffusive or enhanced transport.

The corresponding directional diffusion coefficients $D_x$  and $D_y$, and the total diffusion coefficient D, are then obtained from the long-time growth rate of the MSD as:

\begin{equation}
D_x = \frac{\langle [x(t) - x_0]^2 \rangle}{2t}, \quad
D_y = \frac{\langle [y(t) - y_0]^2 \rangle}{2t}, \quad
D = D_x + D_y .
\label{eq:diff_coeff}
\end{equation} 

At asymptotically long times, the effective diffusion coefficient is defined as
\begin{equation}
D_{\infty} = \lim_{t \to \infty} \frac{ \langle [x(t) - x_0]^2 \rangle + \langle [y(t) - y_0]^2 \rangle }{2t}.
\end{equation}

In addition to the diffusion characteristics given by Eq.~\eqref{eq:diff_coeff}, we also analyze the time evolution of the probability distribution functions of tracer positions, $n(x)$ and $n(y)$, and tracer velocities, $n(v_x)$ and $n(v_y)$, in each direction to gain further insight into transport and mixing in turbulent flows. \\

\textbf{Numerical solver benchmarking} \\

We benchmark our tracer particle solver by simulating tracer particle transport in a kinematic two-dimensional chaotic flow, consisting of a regular lattice of non-stationary kinematic eddies (see Fig.~\ref{fig:kinv}), as studied by Folgia et al.~\cite{FL2022}. The stream-function $\psi(x,y,t)$ and velocity $u(x,y,t)$, $v(x,y,t)$ representation of the flow is given as\citep{FL2022}:

\begin{equation}
\psi= \frac{\alpha} { k} {sin({k[x-\epsilon sin(\omega t)])} {sin(k[y-\epsilon sin(\omega t+\phi)])}}
\end{equation}
\begin{equation}
u = {\alpha}  {sin({k[x-\epsilon sin(\omega t)])} {cos(k[y-\epsilon sin(\omega t+\phi)])}}
\label{eq:u}
\end{equation}
\begin{equation}
v = -{\alpha}  {cos({k[x-\epsilon sin(\omega t)])} {sin(k[y-\epsilon sin(\omega t+\phi)])}}
 \label{eq:v}
\end{equation}

The model setup for the reference problem \citep{FL2022} consists of a square simulation domain of length 
\( L = 2~\mathrm{km} \), with a spatial resolution of \(\Delta x = 0.02~\mathrm{km}\). 
The corresponding wavenumber is \( k = 2\pi/L = \pi~\mathrm{km}^{-1} \), yielding an eddy size of 
\( L/2 = 1~\mathrm{km} \). The maximum flow speed is set to \(\alpha = 3.6~\mathrm{km\,hr^{-1}}\). 
The oscillation parameters of the stream function—\(\varepsilon\), \(\omega\), and \(\phi\)—are chosen 
to ensure a fully chaotic Lagrangian flow regime, with 
\(\omega \approx \alpha / (2\pi k^{-1}) = 1.8~\mathrm{hr^{-1}}\), 
\(\varepsilon/L = 0.3\), and \(\phi = 0\) for the simplest case. 
A total of 1000 passive tracer particles are used in our simulation, compared to 10,000 particle pairs in the 
reference study. The kinematic flow and the associated tracer particles are evolved simultaneously over a total 
simulation time of 100 hours, with a time step of \(\Delta t = 10^{-3}~\mathrm{hr}\) (i.e., 3.6 seconds). 
Particle trajectories are computed using multiple methods: directly from the analytical velocity and 
stream-function fields, as well as via interpolation from both fields, and the resulting transport statistics are compared. The transport characteristics obtained from all methods are found to be in close agreement (see Fig.~\ref{fig:kinp}). In all cases, the dynamics 
exhibit ballistic behavior at short times, transitioning to normal diffusion at later times, with an eddy turnover time of approximately one hour. The overall transport behavior of our simulations is consistent with the findings of Folgia et al. \citep{FL2022} (see Fig.~\ref{fig:kinp}).

\section{Turbulent transport}

In this section, we discuss results obtained from numerical experiments exploring long-time turbulent transport by varying the initial number of vortex strips: 2, 4, 8, 16, and 20, corresponding to initial vorticity packing fractions (VPF) of 6.25\%, 12.5\%, 25\%, 50\%, and 62.5\%, focusing on three regimes: linear growth, developing turbulence, and the steady-state dipole-motion phase. To characterize particle transport across these regimes, we analyze the mean square displacement (MSD) or absolute diffusion and time-dependent diffusion coefficients to quantify dispersion and mixing. We also examine the evolution of particle position and velocity probability distribution functions (PDF), along with their correlations, to capture key aspects of the underlying dynamics. We report a comparative analysis of turbulence evolution, transport behavior, and associated distribution functions across varying initial vorticity packing fractions, highlighting both common trends and case-specific features. \\

\textbf{A. Initial condition}\\

\begin{figure*}
    \centering
    \begin{subfigure}[b]{0.245\textwidth}
        \centering
         \includegraphics[width=5.5cm]{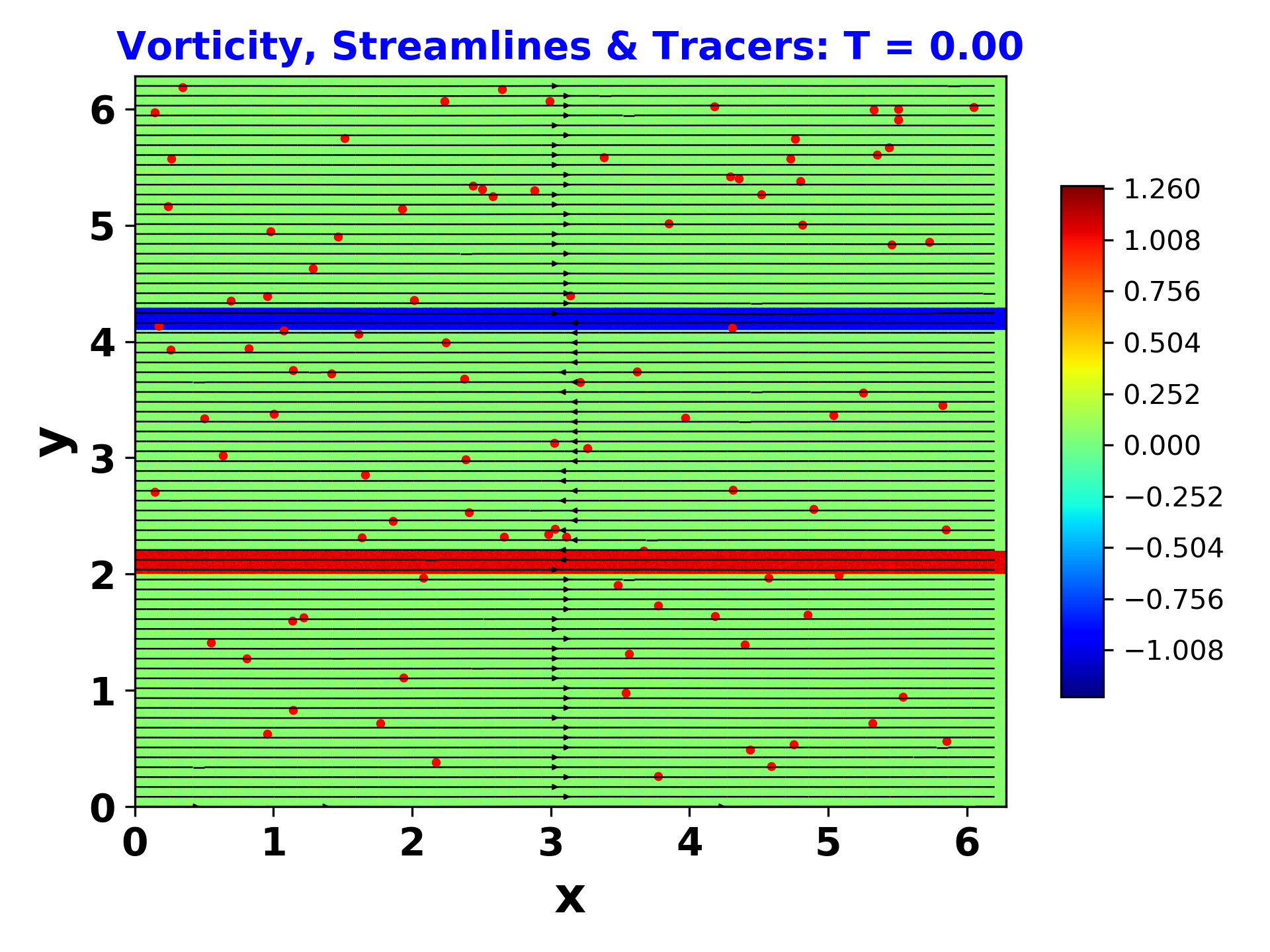}
        \caption{\textbf{2 strips}}
 %       \label{aa}
    \end{subfigure}
    \centering
    \begin{subfigure}[b]{0.245\textwidth}
        \centering
         \includegraphics[width=5.5cm]{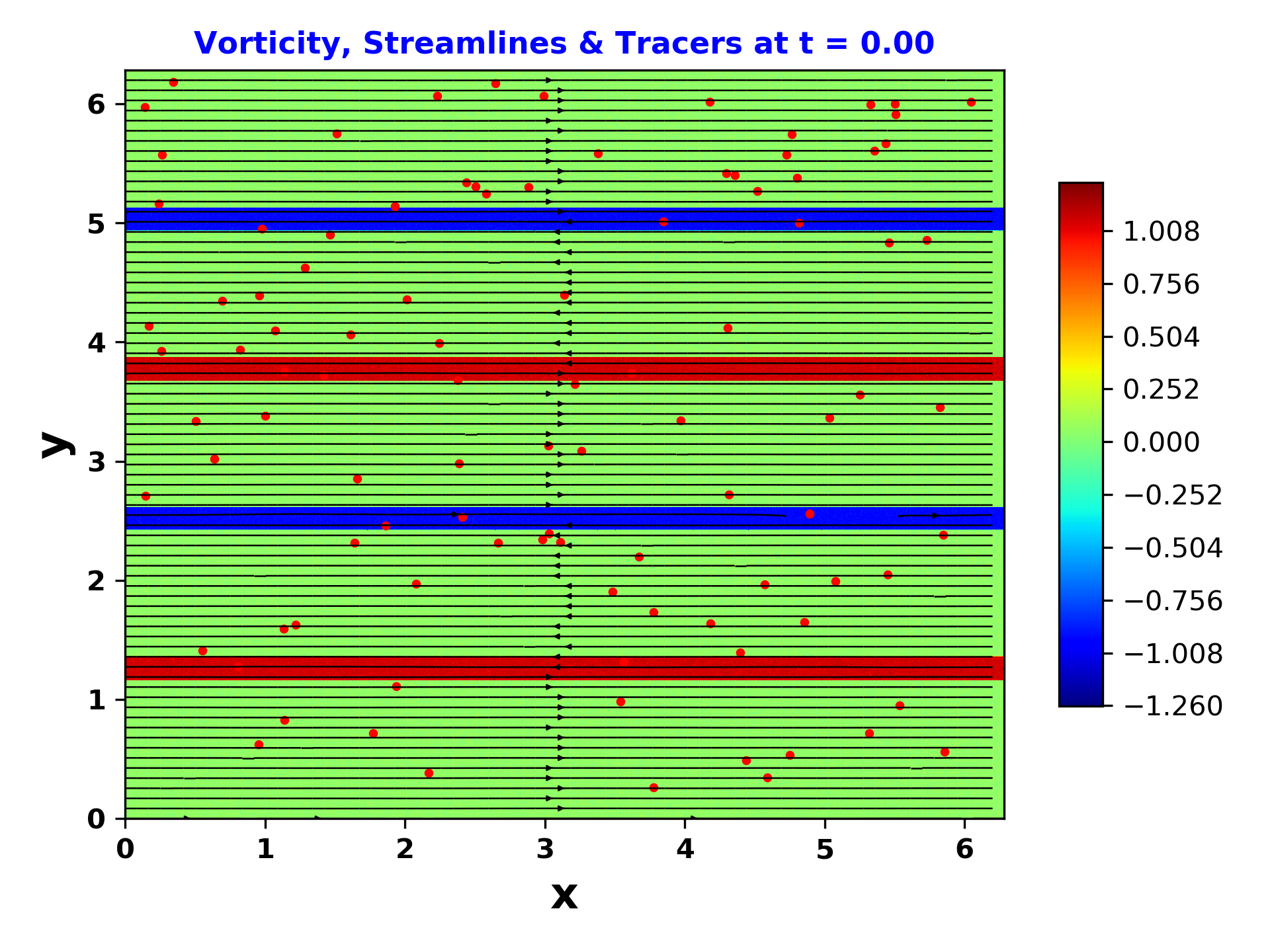}
        \caption{\textbf{4 strips}}
  %      \label{aa}
    \end{subfigure}
    \centering
    \begin{subfigure}[b]{0.245\textwidth}
        \centering
         \includegraphics[width=5.5cm]{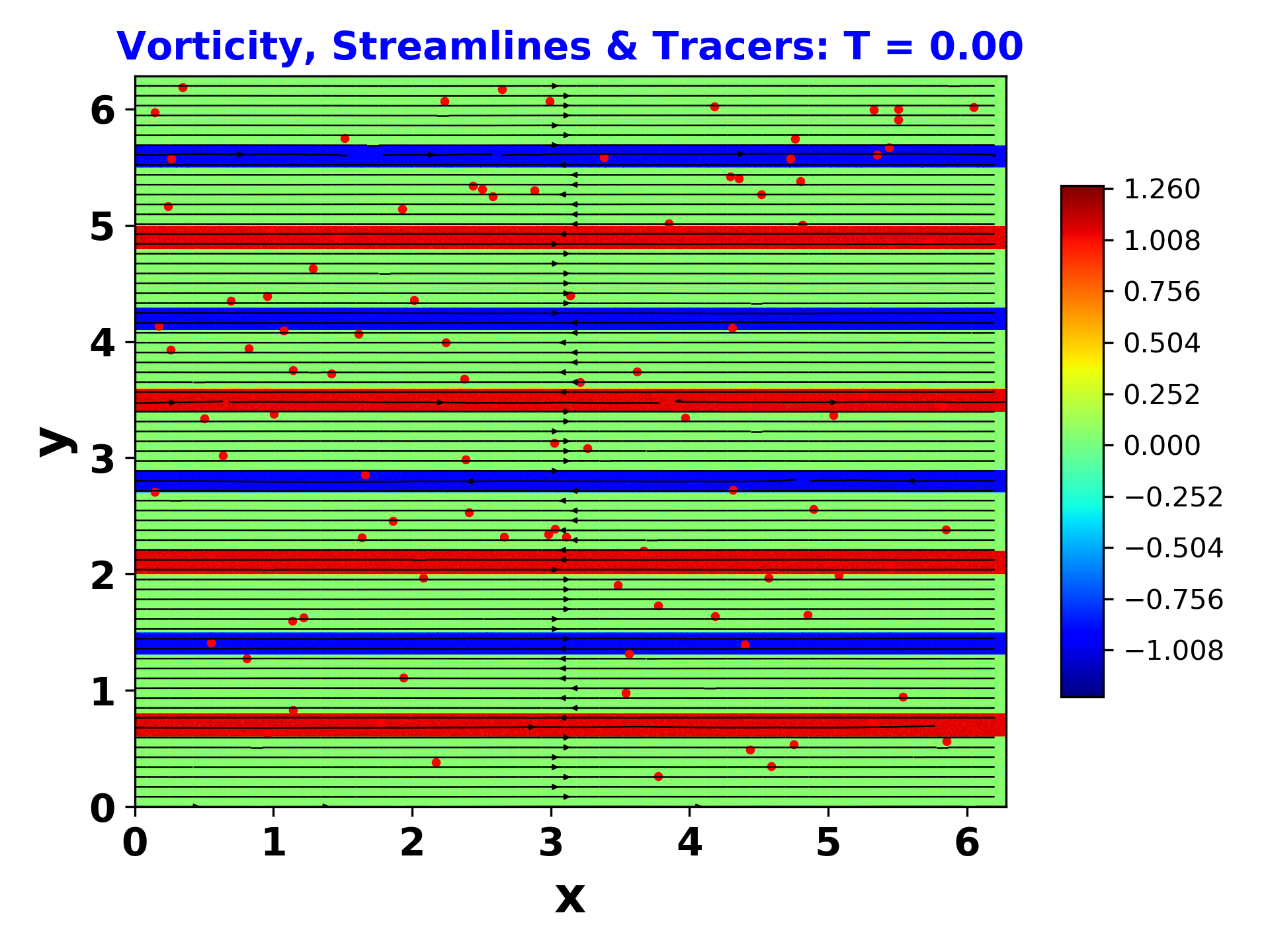}
        \caption{\textbf{8 strips}}
 %       \label{aa}
    \end{subfigure}
    \centering
    \begin{subfigure}[b]{0.245\textwidth}
        \centering
         \includegraphics[width=5.5cm]{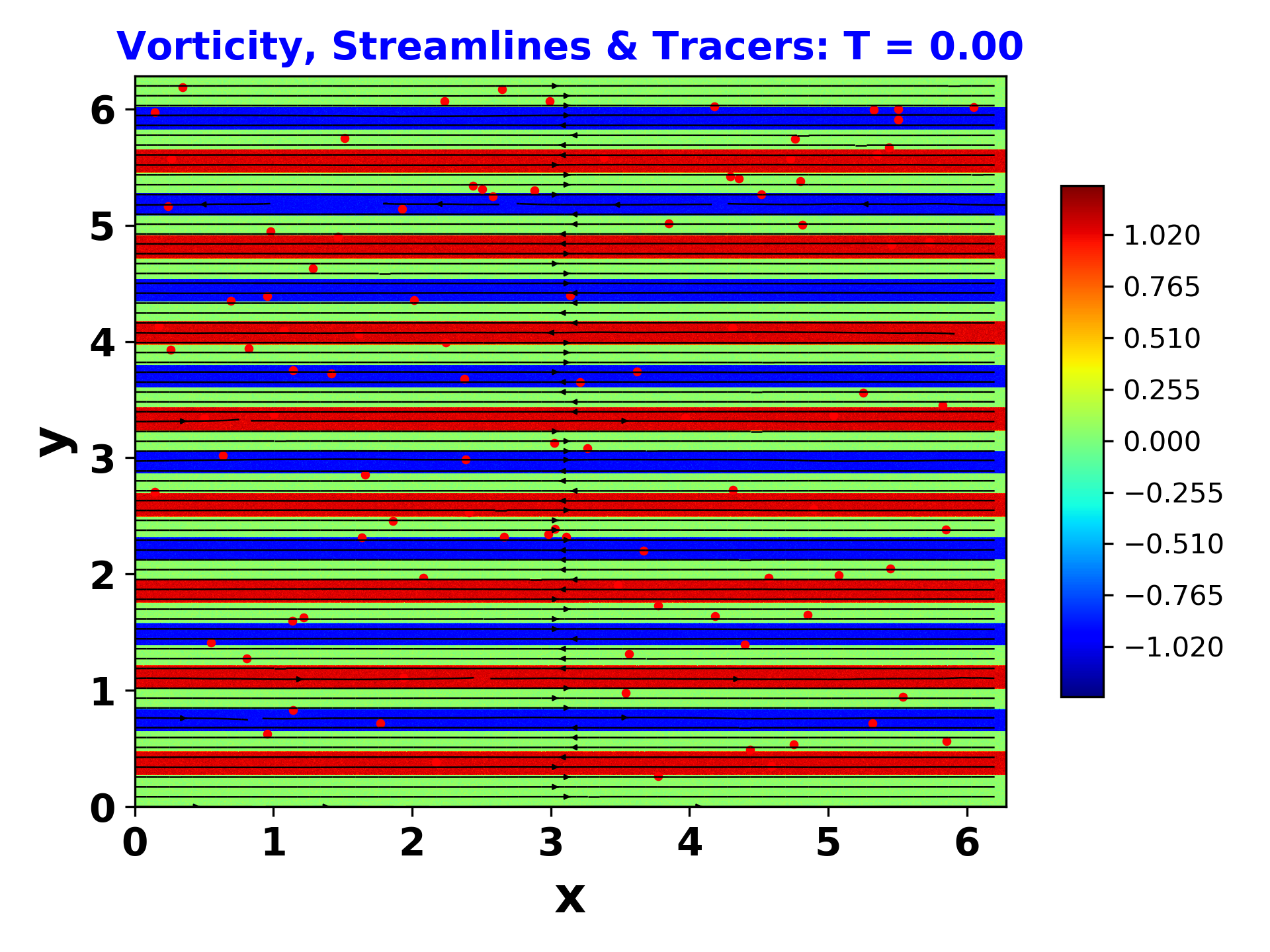}
        \caption{\textbf{16 strips}}
  %      \label{aa}
    \end{subfigure}
   \caption{ Initial vorticity, stream function and associated tracer particles distribution at $T$ = 0 for various cases of turbulence in our numerical simulations. (a) 2 strip configuration with lowest vorticity packing fraction (VPF 6.25\%) (b) 4 strip configuration with low VPF (12.5\%)   (c) 8 strip configuration with moderate VPF (25\%)  (d) 16 strip configuration with high VPF (50\%). The vorticity value for blue, red and green regions are -1, 1 and 0 respectively.}
\label{fig:IC}
\end{figure*}

\begin{figure*}
\centering
  \includegraphics[width=\textwidth]{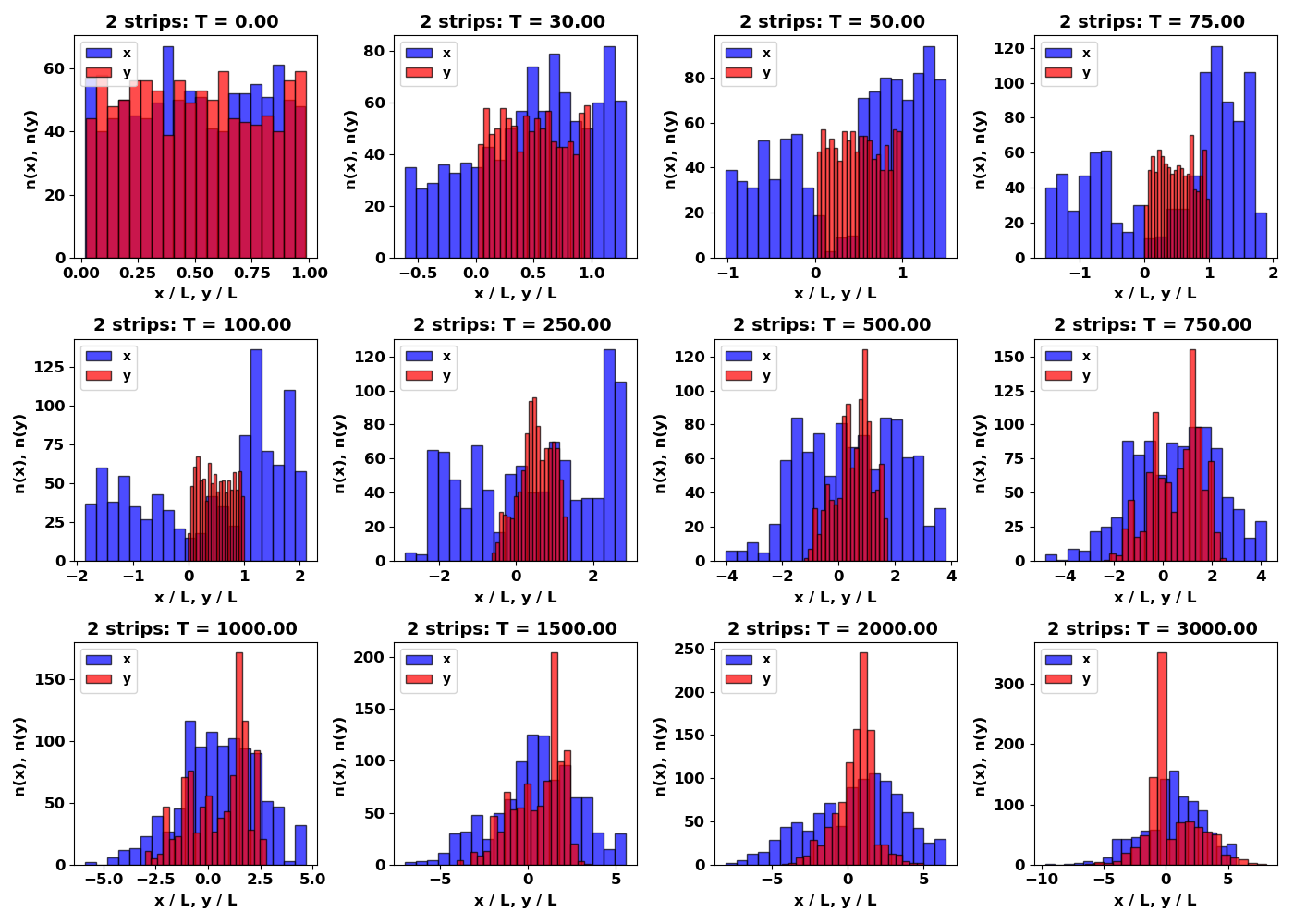}
%  \vspace{1em}
  \caption{Temporal evolution of the position PDFs of tracer particles along the $x$ and $y$ directions for the two-strip initial condition, corresponding to the lowest VPF (6.25\%). The position distributions $n(x)$ and $n(y)$ start from a flat uniform distribution at $T = 0$ in both directions; $n(y)$ attains Gaussianity at a relatively long time ($T \approx 200$), while $n(x)$ remains bimodal under dominant shear until an even longer time ($T \approx 400$) before becoming near-Gaussian. The weak $y$-directional KHI results in weak fluid deformation, and the late onset of weak turbulence leads to a much slower evolution toward Gaussianity in both directions, with the distributions remaining unequal and  non-overlapping, reflecting associated sub-diffusive and anisotropic particle transport (see intermediate times, Fig.~\ref{fig:tr_all_xy}, 2-strip case); at long times ($T \gtrsim 500$), the distributions are near-Gaussian but unequal, with minimal overlap, reflecting sustained anisotropic transport (see late times, Fig.~\ref{fig:tr_all_xy}, 2-strip case).}
  \label{fig:pdf_2}
\end{figure*}

\begin{figure*}
\centering
  \includegraphics[width=\textwidth]{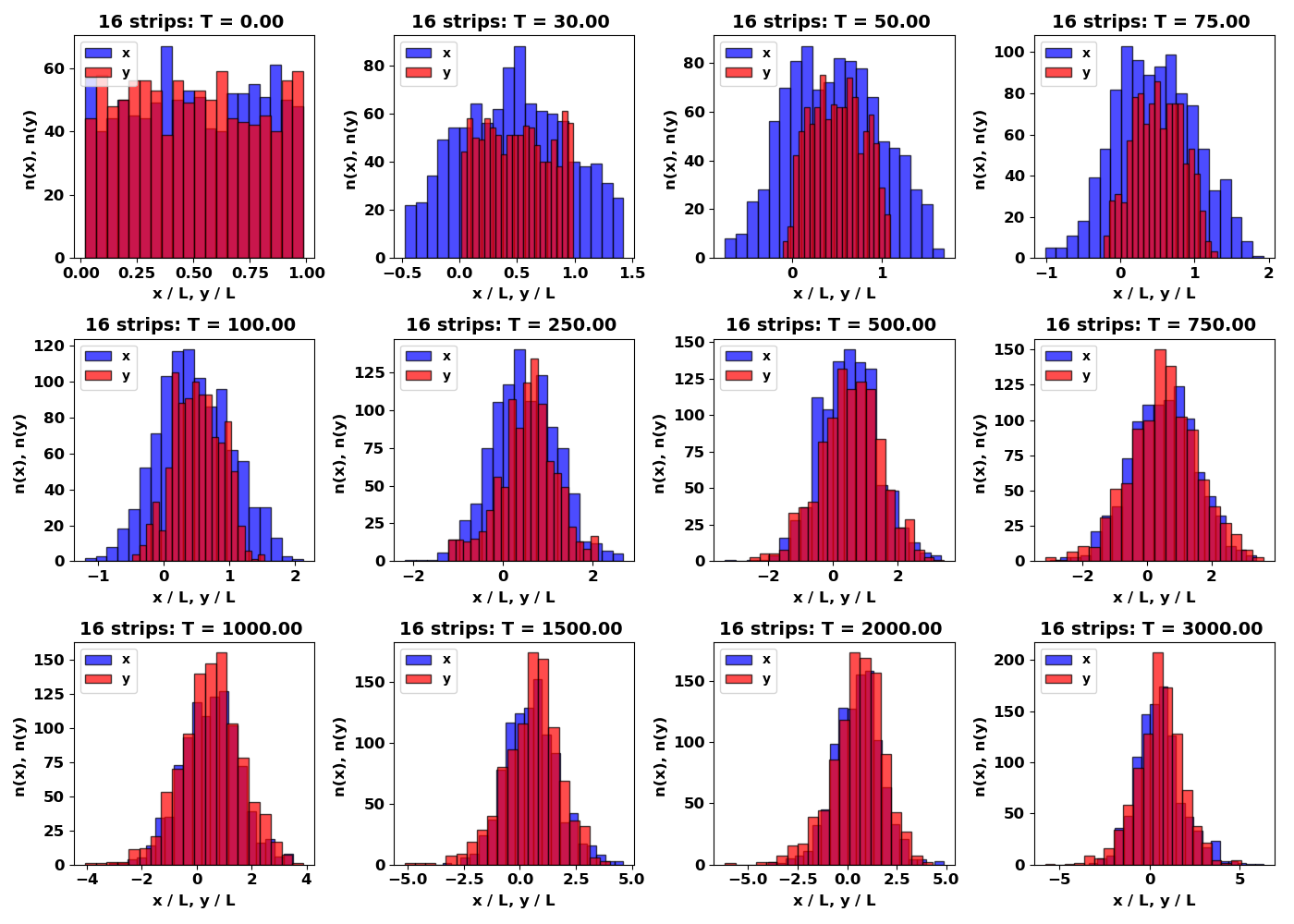}
 % \vspace{1em}
  \caption{Temporal evolution of the position PDFs of tracer particles along the $x$ and $y$ directions for the sixteen-strip initial condition, corresponding to a very high VPF (50\%). The position distributions $n(x)$ and $n(y)$ start from a flat uniform distribution at $T = 0$ in both directions and rapidly become Gaussian ($T \approx 40$ for $n(y)$, $T \approx 50$ for $n(x)$) in both directions. The strong $y$-directional KHI results in strong fluid deformation and the quick onset of strong turbulence, leading to the fastest evolution toward Gaussianity in both directions. By $T \approx 300$, they achieve symmetric overlap, reflecting associated super-diffusive and isotropic total particle transport (see intermediate times, Fig.~\ref{fig:tr_all_xy}, 16-strip case); and at long times ($T > 750$), during dominant dipole motions, the distributions remain Gaussian, symmetric, overlapping, and nearly equal, reflecting sustained isotropic transport (see late times, Fig.~\ref{fig:tr_all_xy}, 16-strip case).}
  \label{fig:pdf_16}
\end{figure*}

\begin{figure*}
\centering
  \includegraphics[width=\textwidth]{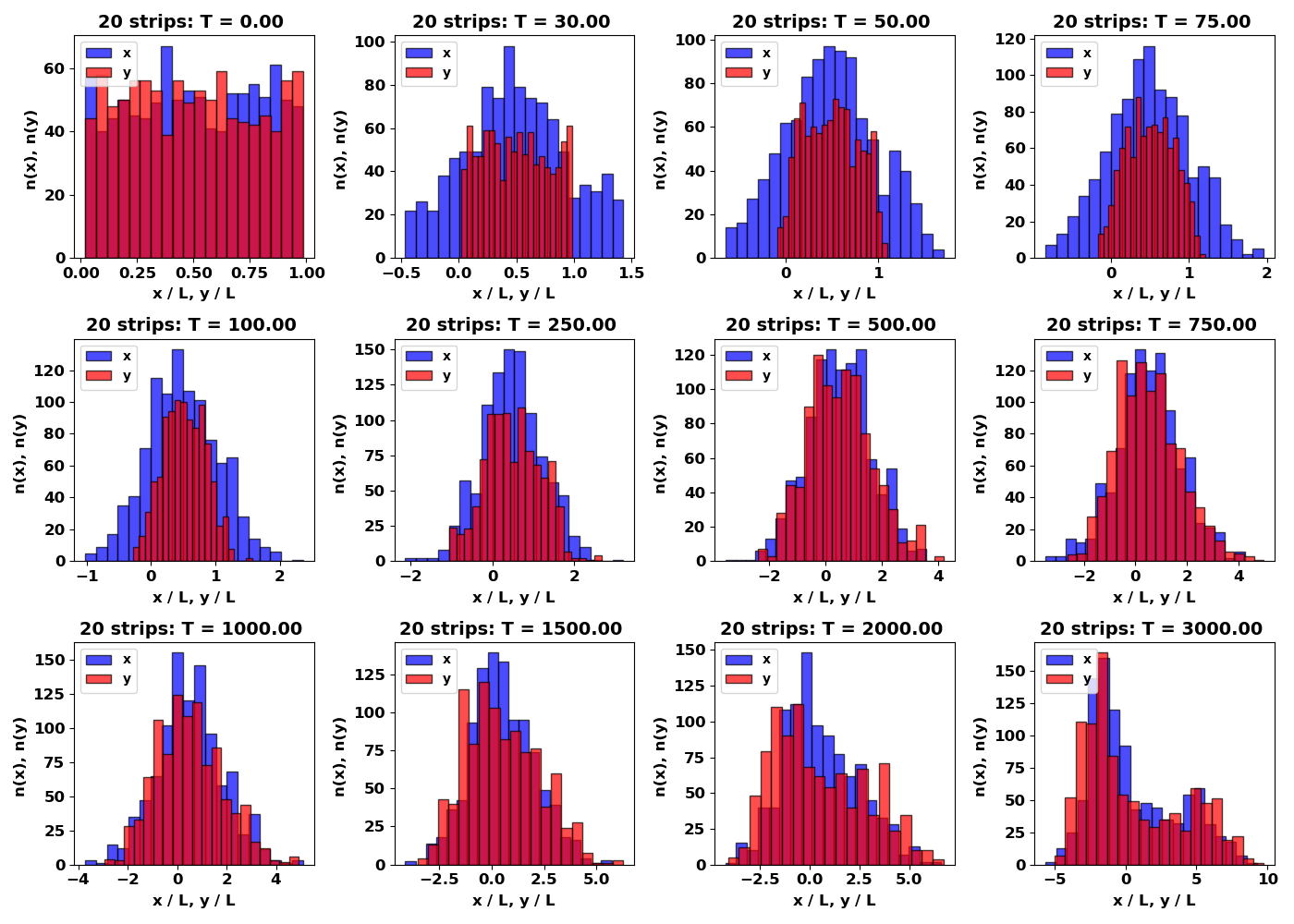}
 % \vspace{1em}
  \caption{Temporal evolution of the position PDFs of tracer particles along the $x$ and $y$ directions for the twenty-strip initial condition, corresponding to a very high VPF (62.5\%). The position distributions $n(x)$ and $n(y)$ start from a flat uniform distribution at $T = 0$ in both directions and rapidly become Gaussian ($T \approx 40$ for $n(y)$, $T \approx 50$ for $n(x)$) in both directions. The strong $y$-directional KHI results in strong fluid deformation and the quick onset of strong turbulence, leading to the fastest evolution toward Gaussianity in both directions. By $T \approx 250$, they achieve symmetric overlap, reflecting associated super-diffusive and isotropic total particle transport (see intermediate times, Fig.~\ref{fig:tr_all_xy}, 20-strip case); and at long times ($T > 2000$), during dominant dipole motions, the distributions develop heavy tails, signaling a departure from normality due to the on-set of super-diffusive behavior, while still maintaining overlap consistent with isotropic transport (see late times, Fig.~\ref{fig:tr_all_xy}, 20-strip case).}
  \label{fig:pdf_20}
\end{figure*}

We begin by describing the {initialization} of the 2D fluid and tracer particle system. At time {$T$ = 0}, the fluid vorticity is arranged in alternating clockwise and anticlockwise circulation strips aligned along the $x$-direction. The number of strips determines the initial vorticity distribution and packing fraction and influences the development of turbulence and coherent structures. The two-strip configuration corresponds to the lowest VPF of $6.25\%$, the four-strip configuration represents a modest increase to $12.5\%$, the eight-strip configuration a further increase to $25\%$, while the sixteen- and twenty-strip configurations correspond to the highest VPF of $50\%$ and $62.5\%$, respectively. Simultaneously, tracer particles are uniformly and randomly distributed across the 2D domain (see Fig. \ref{fig:IC}), resulting in flat and uniform position distribution functions, $n(x)$ and $n(y)$, in both the $x$ and $y$ directions (see $T = 0$, Figs.~\ref{fig:pdf_2}, \ref{fig:pdf_16}, and \ref{fig:pdf_20}).  All tracer particles start with the initial velocity of the fluid. \\

\textbf{B. Linear Regime}\\

\begin{figure*}
    \centering
    \begin{subfigure}[b]{0.49\textwidth}
        \centering
        \includegraphics[width=\textwidth]{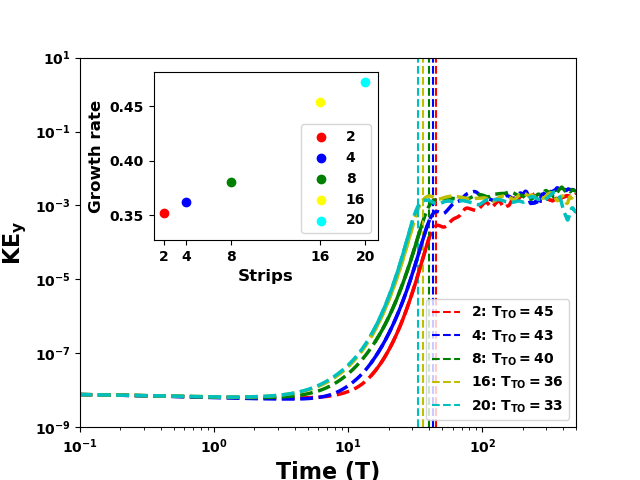} 
        \caption{}
        \label{fig:KE_y}  % <-- after \caption
    \end{subfigure}
    \hfill
    \begin{subfigure}[b]{0.49\textwidth}
        \centering
        \includegraphics[width=\textwidth]{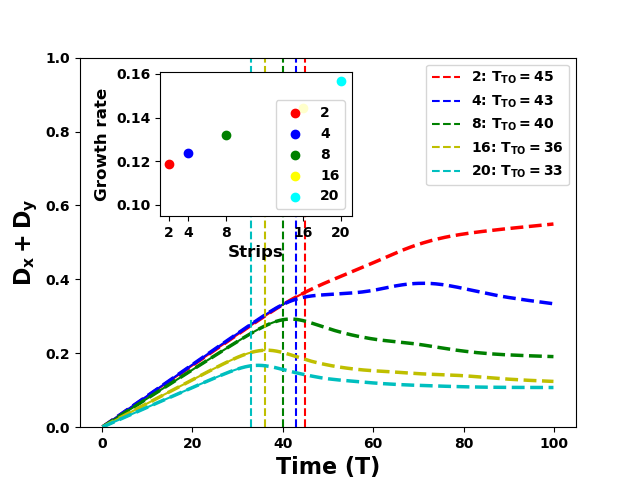}
        \caption{}
        \label{fig:Tr_tot_lin}  % <-- after \caption
    \end{subfigure}
    \caption{(a) Temporal evolution of the $y$-direction kinetic energy ($KE_y$) of the turbulent flow initialized with different initial circulations (corresponding to varying numbers of vorticity strips).
(b) Corresponding evolution of passive tracer transport. The insets in both panels show the growth rates for the respective cases of initial strip numbers. Vertical lines in both figures denote the turbulence onset time, $T_{TO}$, marking the transition of the flow from the linear to the nonlinear regime and the accompanying change in tracer transport behavior. The $T_{TO}$ values obtained from the fluid and tracer dynamics are found to coincide for each initial vorticity configuration.}
    \label{fig:KE_transport}
\end{figure*}

\begin{figure*}
\centering
  \includegraphics[width=\textwidth]{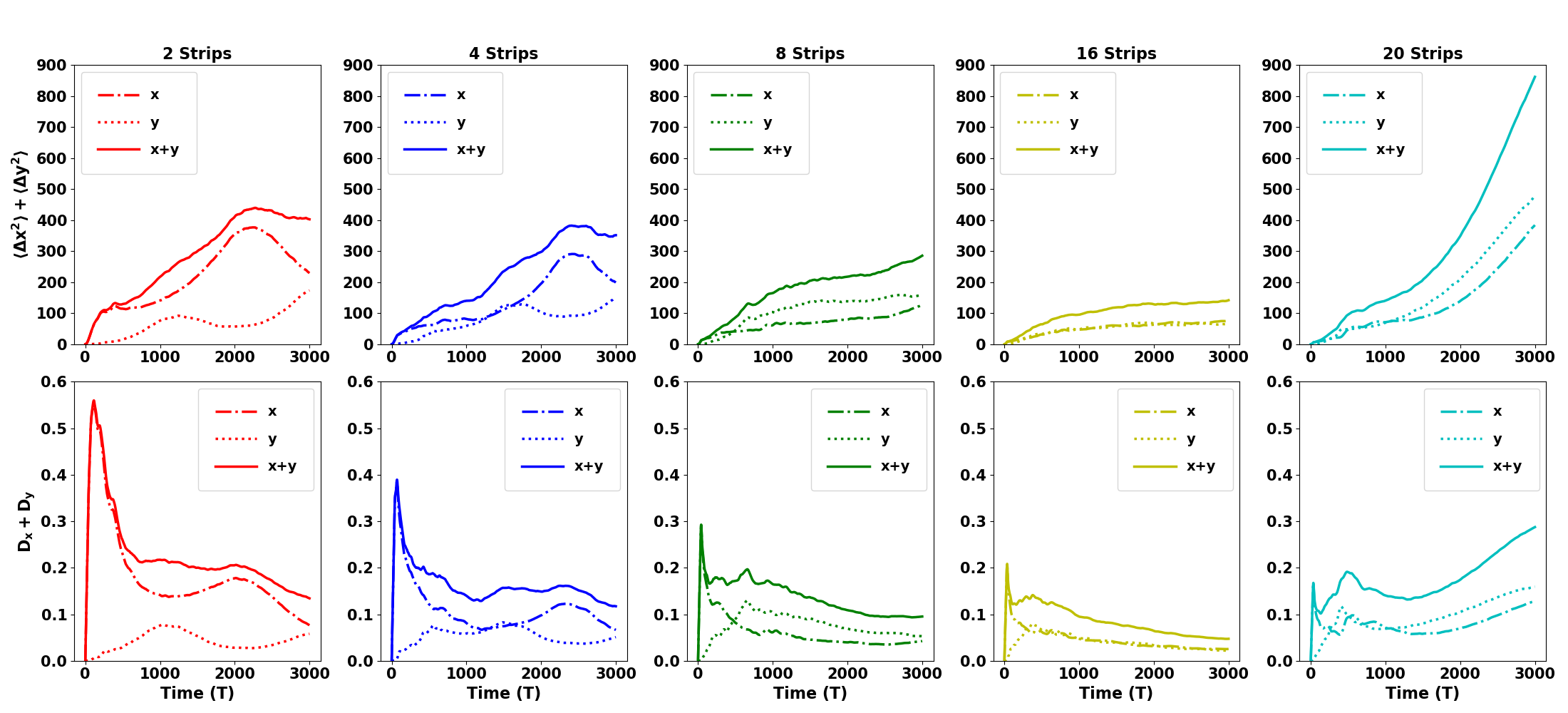}
%  \vspace{1em}
  \caption{ Transport of passive tracer particles in turbulent flows initiated with different initial vortex packing fractions, corresponding to 2, 4, 8, 16, and 20 vorticity strips (6.25\%, 12.5\%, 25\%, 50\%, and 62.5\%, respectively). The top-row panels show the evolution of the single-particle mean-square displacement (MSD) for increasing vortex packing fractions, while the bottom-row panels depict the corresponding diffusion coefficients ($D_x$ + $D_y$). The dot–dashed line denotes transport along the $x$ direction, the dashed line represents transport along the $y$ direction, and the solid line indicates the total transport. These transport results show that increasing VPF reduces the initial ballistic transport, drives the total particle transport from subdiffusive to superdiffusive after the onset of turbulence and from an anisotropic to an isotropic regime, and similarly leads to subdiffusive to superdiffusive behavior at long times during dipole-dominated motion. Although the overall transport generally decreases with increasing VPF, in the extreme case of the highest VPF (62.5\%), the total transport in the late stages becomes the largest among all cases due to anomalous superdiffusive behavior.}
  \label{fig:tr_all_xy}
\end{figure*}

\begin{figure*}
%     \centering
     \begin{subfigure}[b]{0.49\textwidth}
         \centering
         \includegraphics[width=\textwidth]{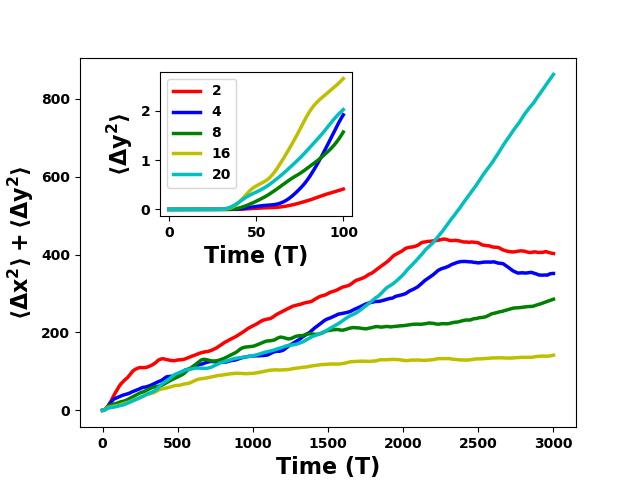}
         \caption{}
         \label{fig:Tottra}
     \end{subfigure}
     \hfill
     \begin{subfigure}[b]{0.49\textwidth}
 %        \centering
         \includegraphics[width=\textwidth]{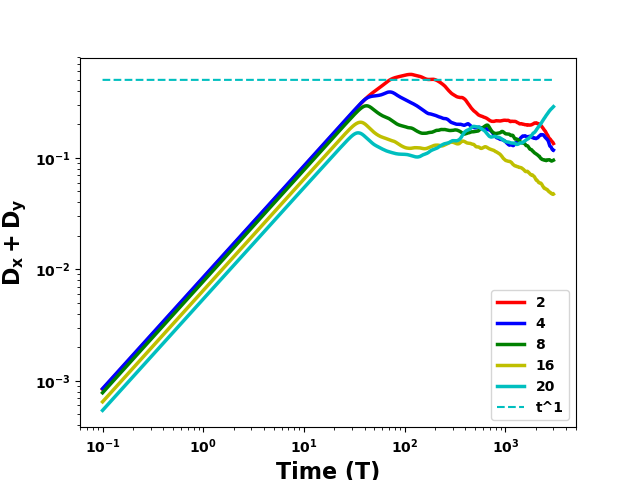}
         \caption{}
  \label{fig:Tottrb}
     \end{subfigure}
     \caption{ (a) Total transport of tracer particles—represented by the single-particle mean-square displacement (MSD) or absolute dispersion evolution—during turbulence development, shown in linear–linear axes. Different colors indicate transport in flows initialized with circulations of varying VPF, ranging from low (2 strips) to high (20 strips). While the overall particle transport gradually decreases with increasing vortex packing fraction, in the extreme case of the highest VPF (62.5\%), the total transport in the late stages becomes the largest among all cases due to anomalous superdiffusive behavior. The inset highlights the linear regime of $y$-direction transport for comparison with the initial linear growth rate. (b) Corresponding total diffusion coefficients, $(D_x + D_y)$, for the respective cases. Increasing VPF reduces the initial ballistic transport, leads to a transition from subdiffusive to superdiffusive transport after the onset of turbulence, and similarly results in subdiffusive to superdiffusive behavior at long times during dipole-dominated motion.}
     \label{fig:TotTransport}
\end{figure*}

As the fluid and tracer particle system evolves from $T$ = 0, shear-driven coherent flows in the $x$-direction induce a left–right drift of tracer particles. This initial stage is characterized by linear fluid motion  and ballistic tracer particle transport, with both motions primarily governed by the imposed shear flow. This ballistic motion leads to the most rapid and significant growth in $x$-direction transport—and in overall total tracer particle displacement—observed throughout the entire evolution of the flow (see Fig.~\ref{fig:tr_all_xy}, bottom row, and Fig.~\ref{fig:Tottrb}). During this phase, total transport grows quadratically with time, dominated by tracer motion in the $x$-direction (see Fig. \ref{fig:Tr_tot_lin}), reflecting the influence of shear flow before vortex interactions and turbulent mixing set in. 
The strength and duration of the shear-driven anisotropic ballistic motion and the associated linear regime progressively decrease as the initial vorticity packing fraction (VPF) increases (see Fig.~\ref{fig:tr_all_xy} and Fig.~\ref{fig:Tottrb}). For the lowest VPFs (6.25\% and 12.5\%), the linear regime persists up to $T \approx 45$ and $T \approx 43$, respectively, while the strong shearing motion continues up to approximately $T \approx 80$ for VPF 6.25\% and $T \approx 60$ for VPF 12.5\%. With increasing VPF, both the linear regime and the shear-dominated motion shorten further, lasting $T \approx 40$ for VPF 25\% and $T \approx 30$ for VPFs 50\% and 62.5\%, until the growing Kelvin–Helmholtz instability leads to the formation of vortex rolls and mergers that reduce the shear flow. The transition of the fluid from the linear to the nonlinear regime, marking the onset of turbulence, coincides precisely with the corresponding tracer particle transport, which also shifts from the linear to the nonlinear regime at the same time,  $T_{TO}$, as seen in Fig.~\ref{fig:KE_transport}.

The position PDF of advected particles, $n(x)$ and $n(y)$, initially uniform along both $x$ and $y$, evolves under the influence of the flow. At early times, the $x$-directed shear induces ballistic motion, causing the $x$-position PDF, $n(x)$, to broaden rapidly and extend beyond the box length, while the $y$-direction PDF, $n(y)$, remains flat and confined within the box due to limited transverse motion, reflecting the shear-driven anisotropic nature of early-stage transport (see $T = 30$–50, Fig.~\ref{fig:pdf_2} for the VPF (6.25\%) case).

\begin{figure*}
\centering
  \includegraphics[width=\textwidth]{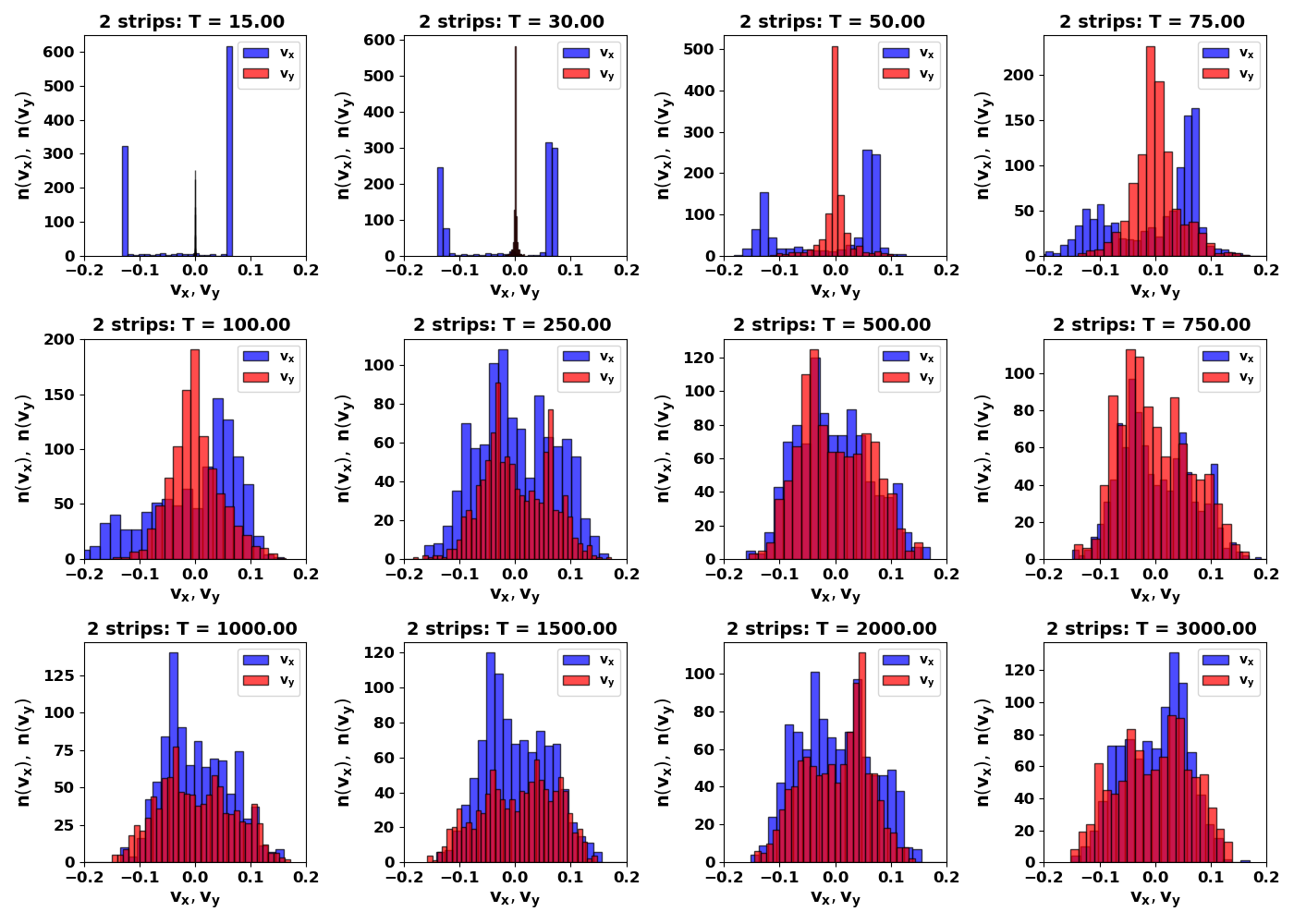}
%  \vspace{1em}
  \caption{Temporal evolution of the velocity distribution functions of tracer particles along the $x$ and $y$ directions for the two-strip initial condition, corresponding to the lowest VPF (6.25\%). Initially, the velocity PDFs are shear-dominated: $n(v_x)$ is bimodal, while $n(v_y)$ is spike-like as the fluid begins to evolve. As turbulence develops, $n(v_y)$ becomes approximately Gaussian around $T \approx 75$--$100$, whereas $n(v_x)$ remains shear-dominated and bimodal until $T \lesssim 100$, gradually broadening thereafter to attain Gaussianity. The delayed Gaussianization, along with the remaining differences and non-overlap between $n(v_x)$ and $n(v_y)$, reflects sub-diffusive, anisotropic transport. By $T \approx 500$, both PDFs become bimodal again due to the motion of counter-rotating dipoles, representing the characteristic rotational velocities of the flow.}
  \label{fig:vdf_2}
\end{figure*}

\begin{figure*}
\centering
  \includegraphics[width=\textwidth]{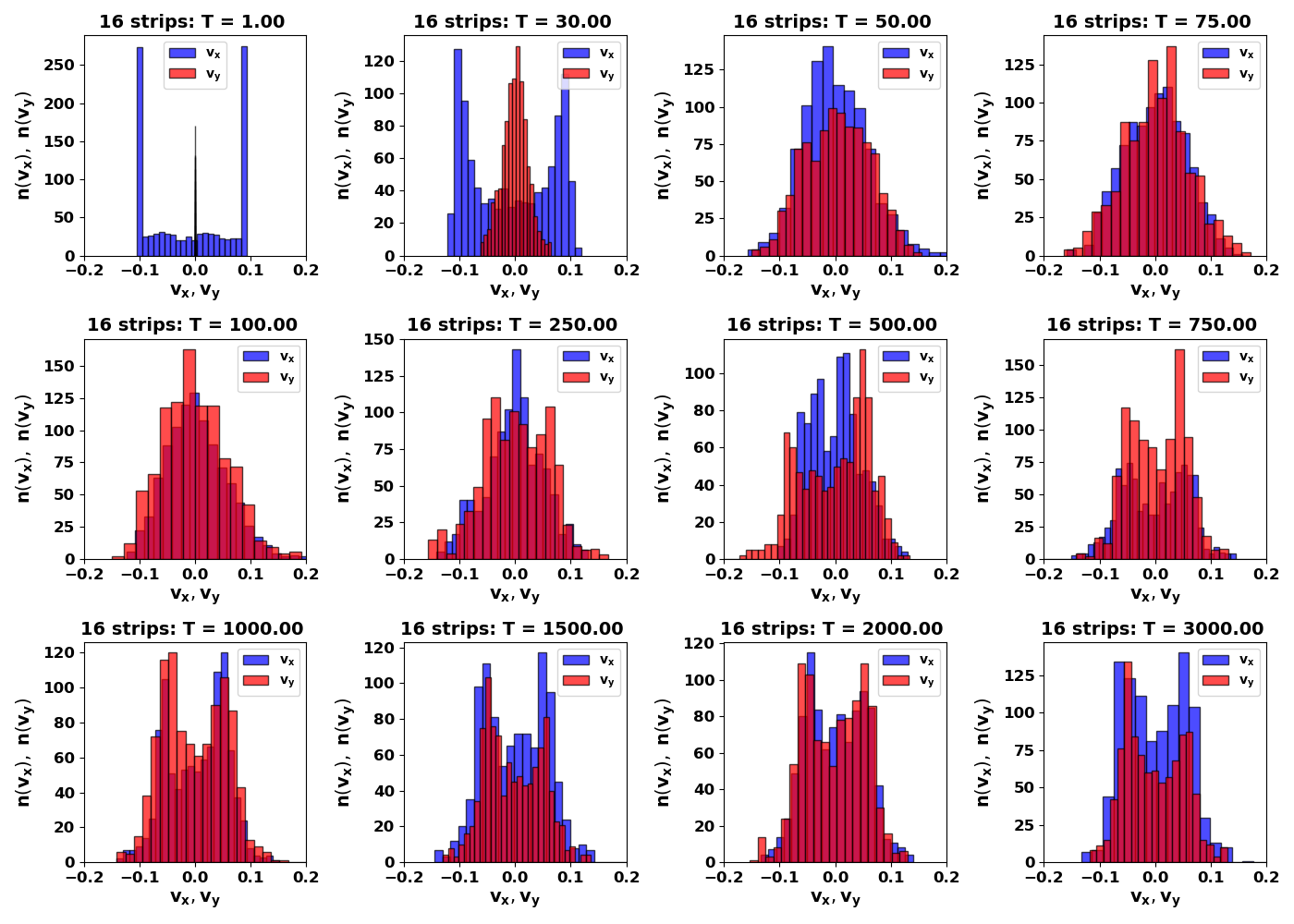}
%  \vspace{1em}
  \caption{Temporal evolution of the velocity distribution functions of tracer particles along the $x$ and $y$ directions for the sixteen-strip initial condition, corresponding to a very high VPF (50\%). The velocity PDFs are initially shear-dominated, with $n(v_x)$ bimodal and $n(v_y)$ spike-like as the fluid begins to evolve. As turbulence develops, both $n(v_x)$ and $n(v_y)$ rapidly Gaussianize by $T \approx 10$--$30$, become equal and overlapping by $T \lesssim 30$, and reflect associated super-diffusive, isotropic transport, before later becoming bimodal again by $T \approx 750$ due to the motion of counter-rotating dipoles that represent their characteristic rotational velocities.
}
  \label{fig:vdf_16}
\end{figure*}

The velocity PDF initially exhibits a bimodal $n(v_x)$ distribution, with peaks near the mean shear velocities of the counter-directed jets, while $n(v_y)$ remains sharply peaked around zero, reflecting the onset of small transverse fluctuations  that are only beginning to emerge (see Fig. \ref{fig:vdf_2} and Fig. \ref{fig:vdf_16}). \\

\textbf{C. Turbulence Onset and Development} \\

\begin{figure*}
    \centering
    \begin{subfigure}[b]{0.33\textwidth}
        \centering
         \includegraphics[width=\textwidth]{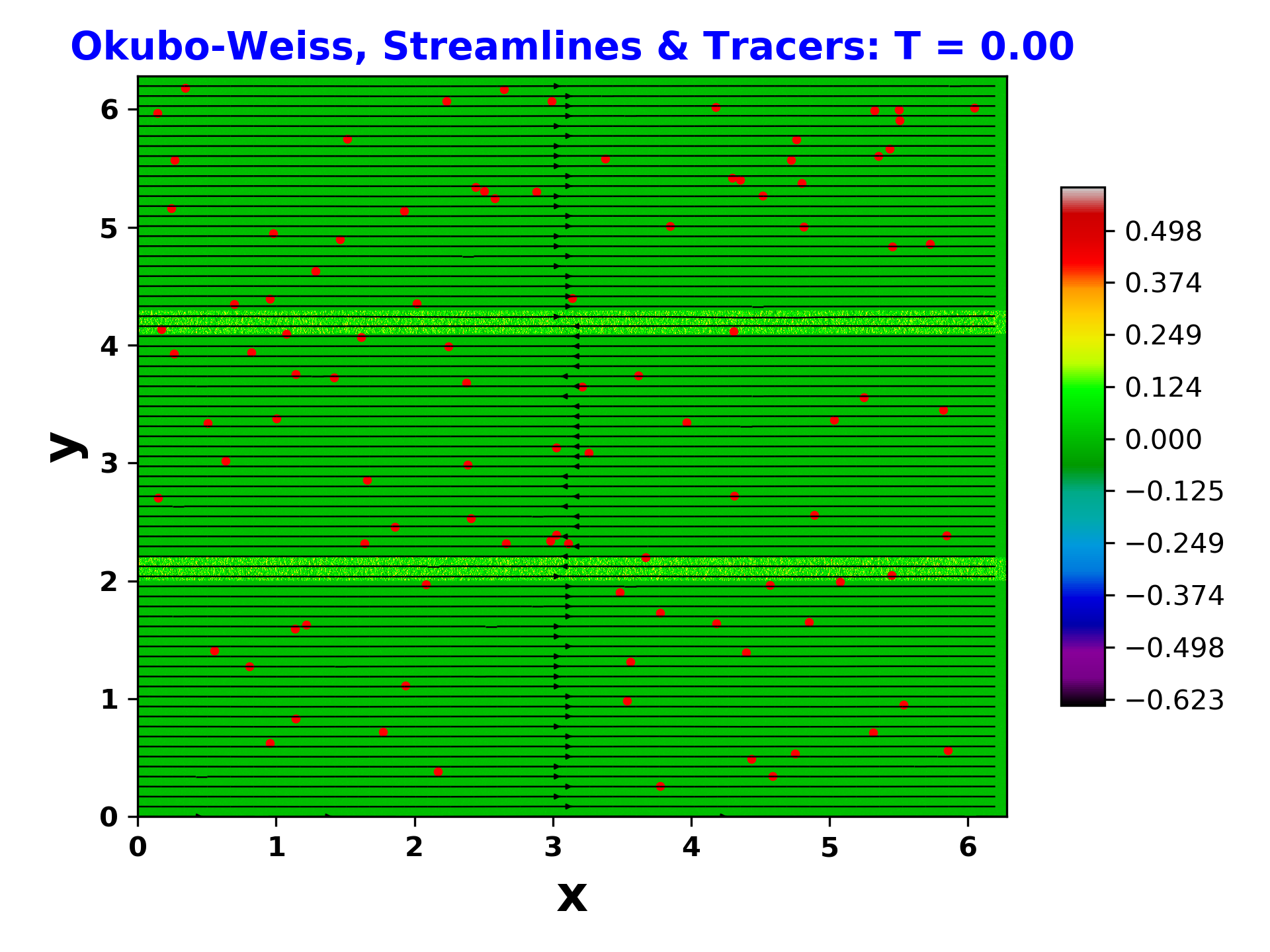}
        \caption{\textbf{T = 0}}
        \label{Okubua}
    \end{subfigure}
    \hfill
    \begin{subfigure}[b]{0.33\textwidth}
        \centering
         \includegraphics[width=\textwidth]{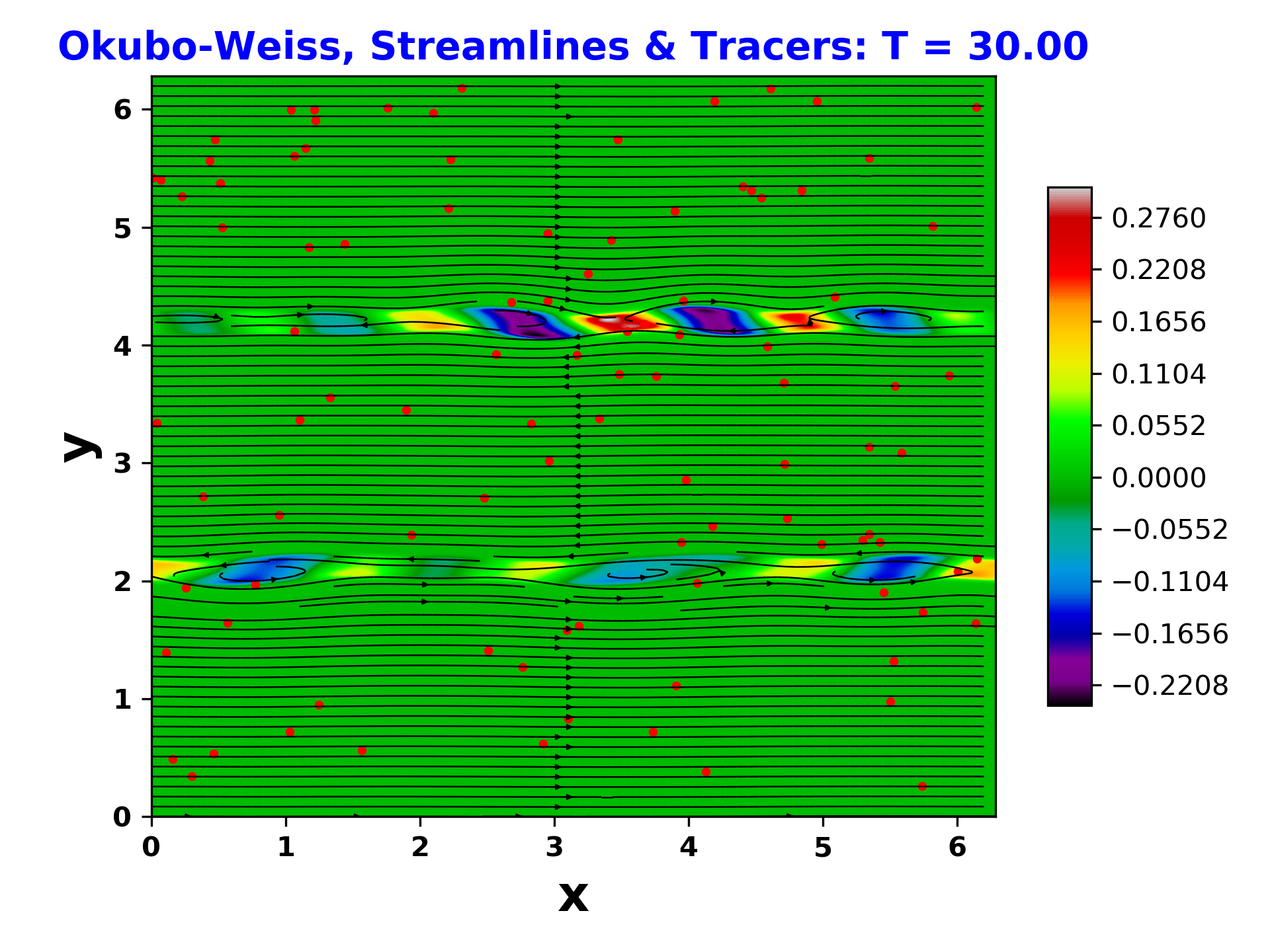}
        \caption{\textbf{T = 30}}
        \label{fig:Okubub}
    \end{subfigure}
    \hfill
    \begin{subfigure}[b]{0.33\textwidth}
        \centering
         \includegraphics[width=\textwidth]{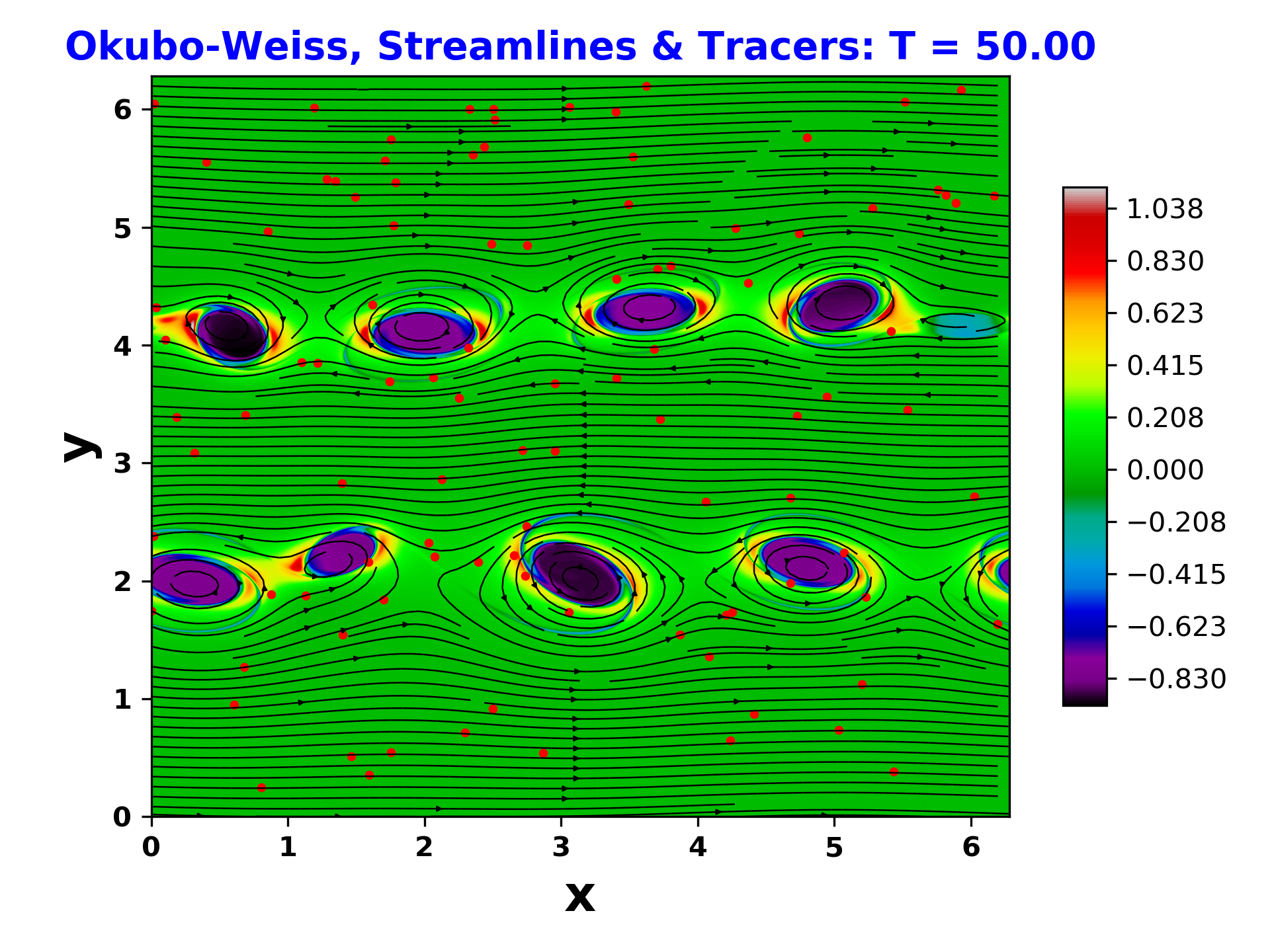}
        \caption{\textbf{T = 50}}
        \label{fig:Okubuc}
    \end{subfigure}
    \hfill
    \begin{subfigure}[b]{0.33\textwidth}
        \centering
         \includegraphics[width=\textwidth]{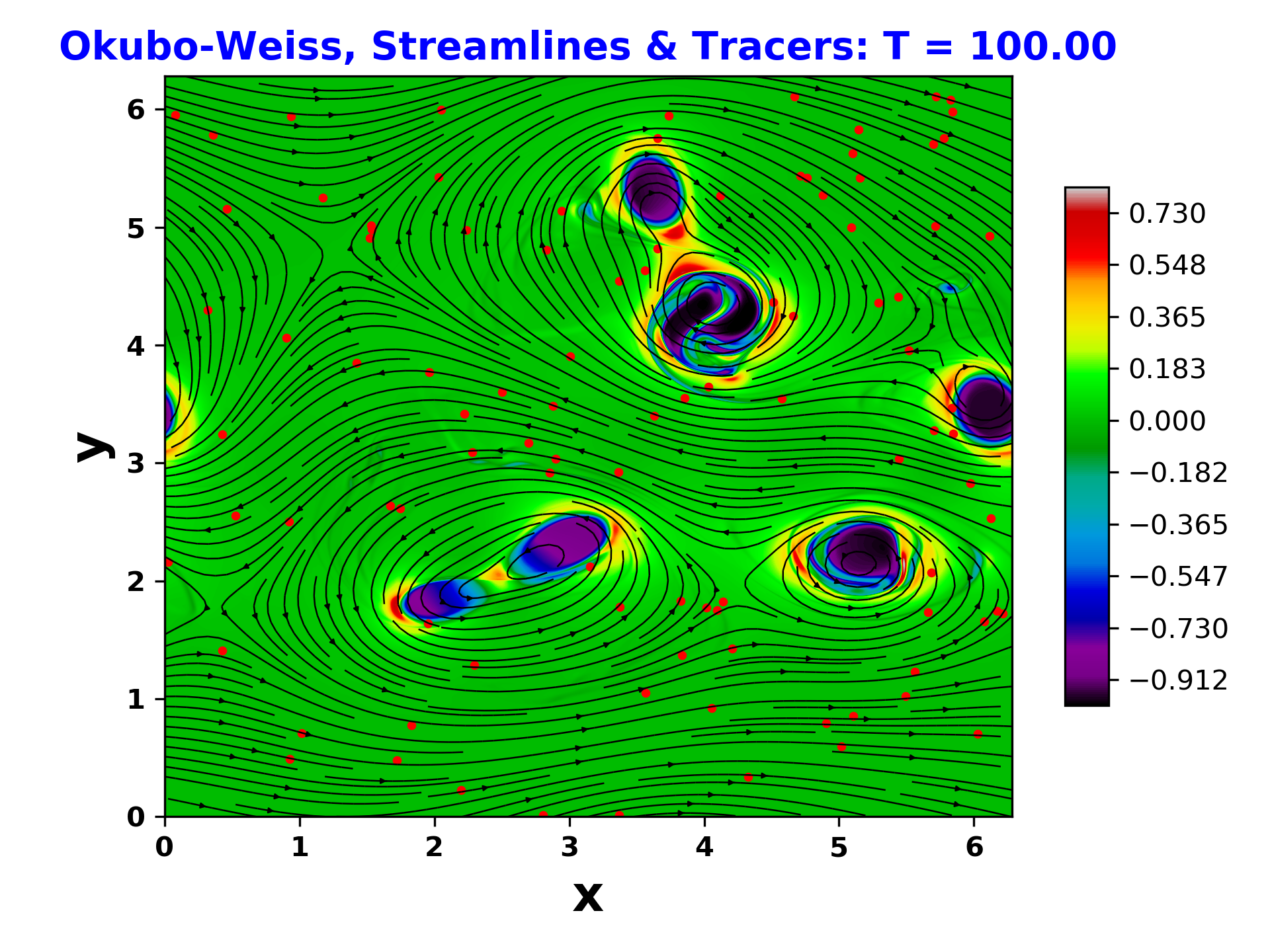}
        \caption{\textbf{T =100}}
        \label{fig:Okubud}
    \end{subfigure}
    \hfill
    \begin{subfigure}[b]{0.33\textwidth}
        \centering
         \includegraphics[width=\textwidth]{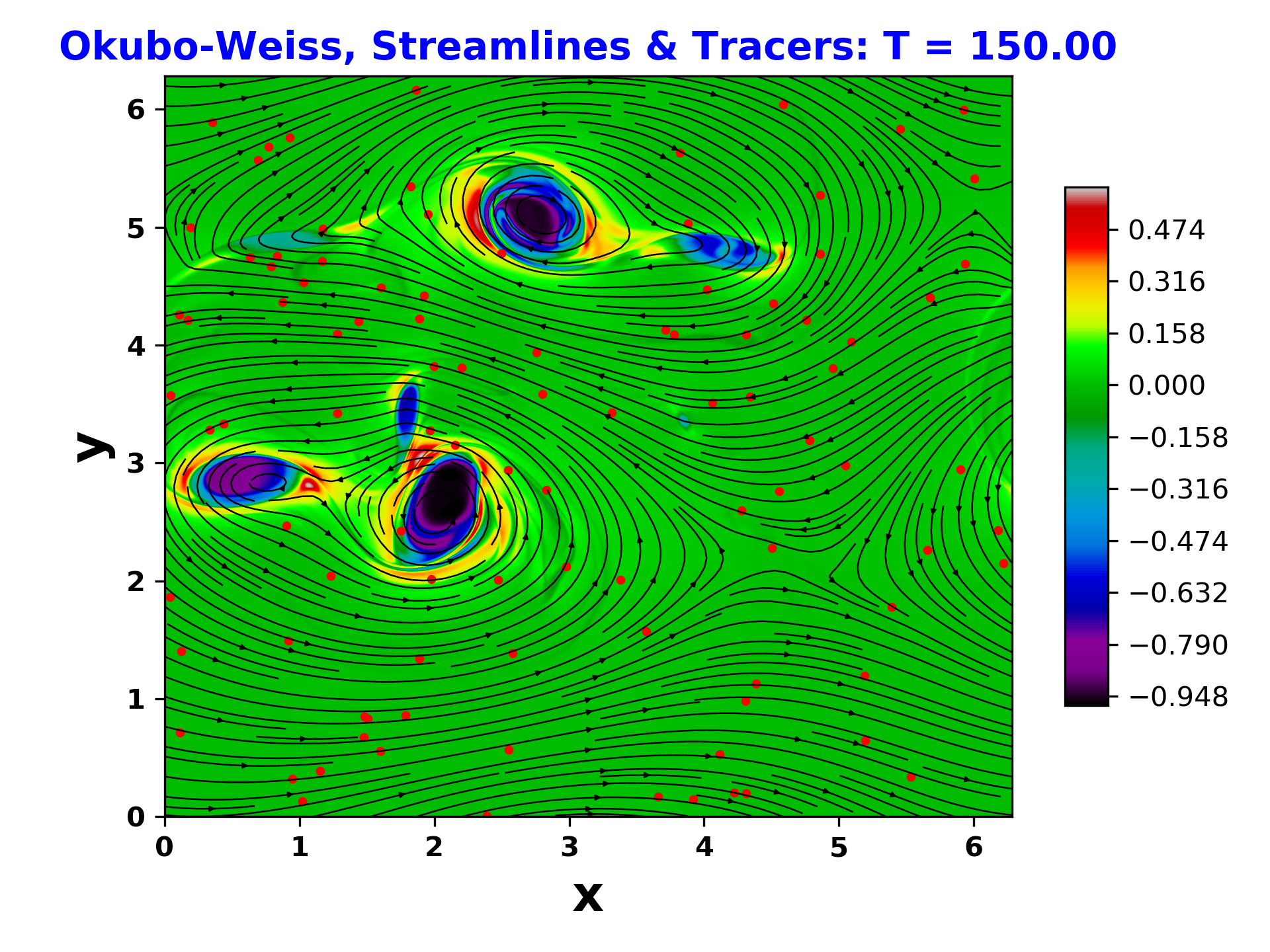}
        \caption{\textbf{T = 150}}
        \label{fig:Okubue}
    \end{subfigure}
    \hfill
    \begin{subfigure}[b]{0.33\textwidth}
        \centering
         \includegraphics[width=\textwidth]{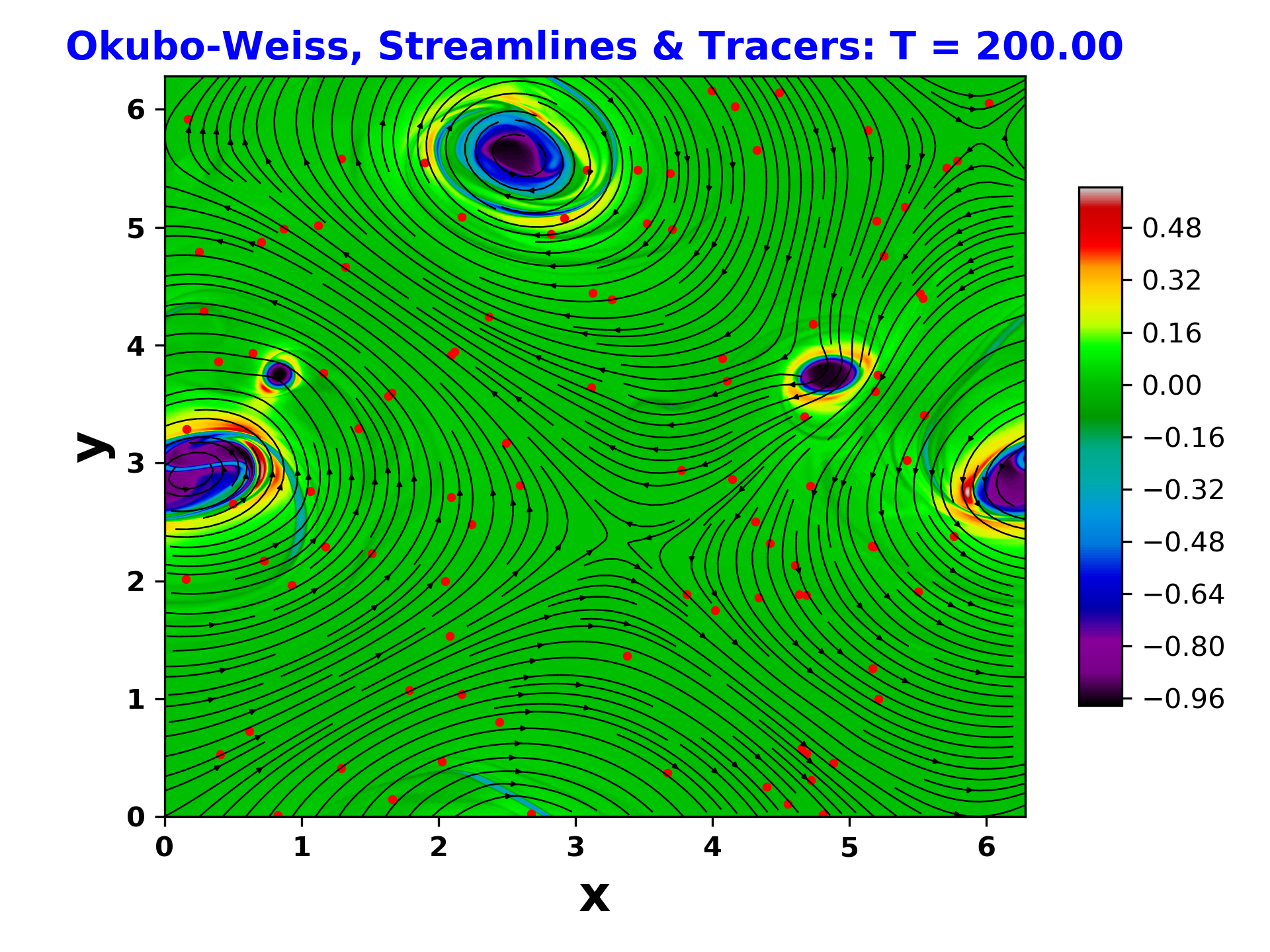}
        \caption{\textbf{T = 200}}
        \label{fig:Okubuf}
    \end{subfigure}
    \hfill
    \begin{subfigure}[b]{0.33\textwidth}
        \centering        \includegraphics[width=\textwidth]{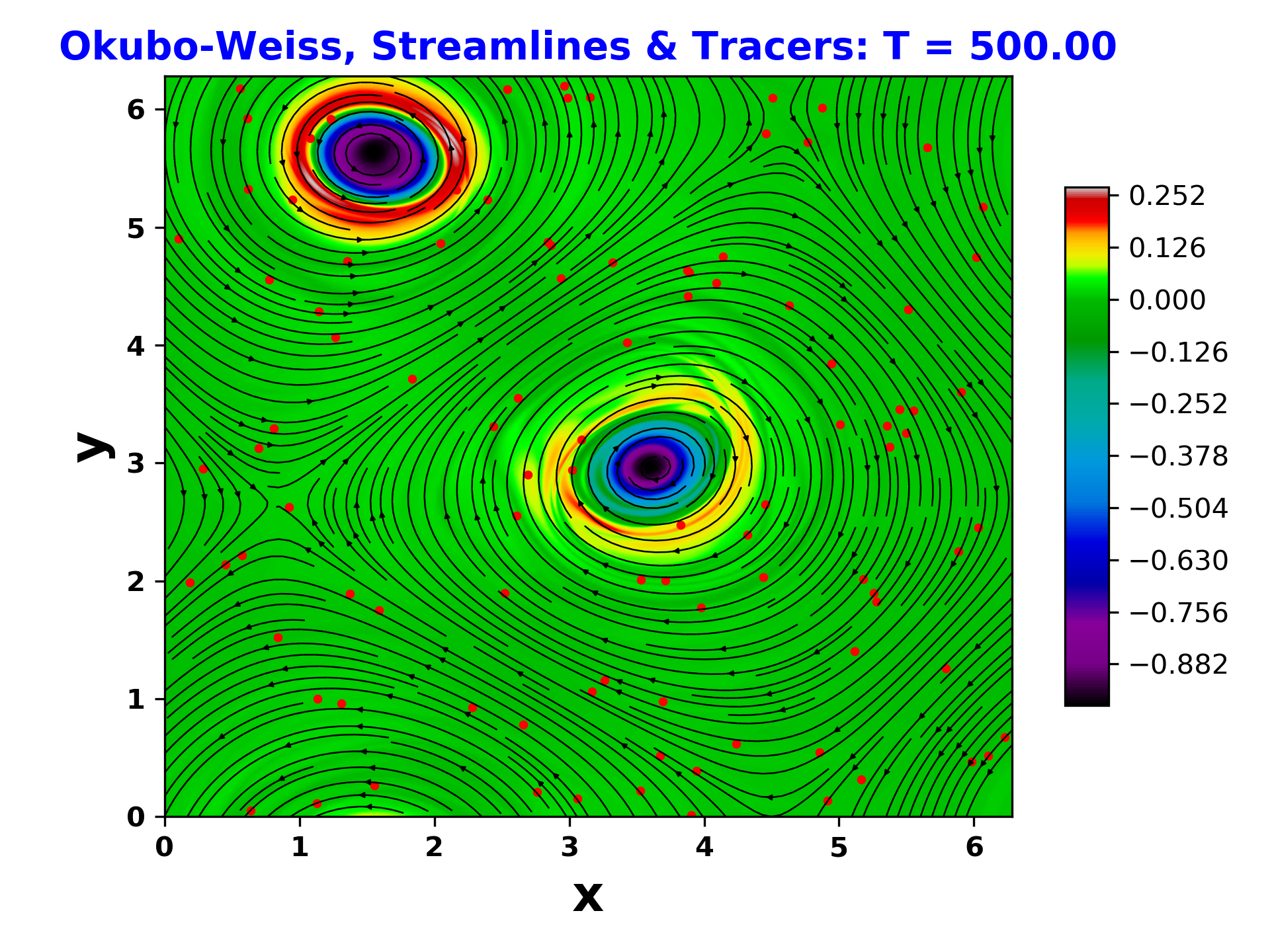}
        \caption{\textbf{T = 500}}
        \label{fig:Okubug}
    \end{subfigure}
    \hfill
    \begin{subfigure}[b]{0.33\textwidth}
        \centering
         \includegraphics[width=\textwidth]{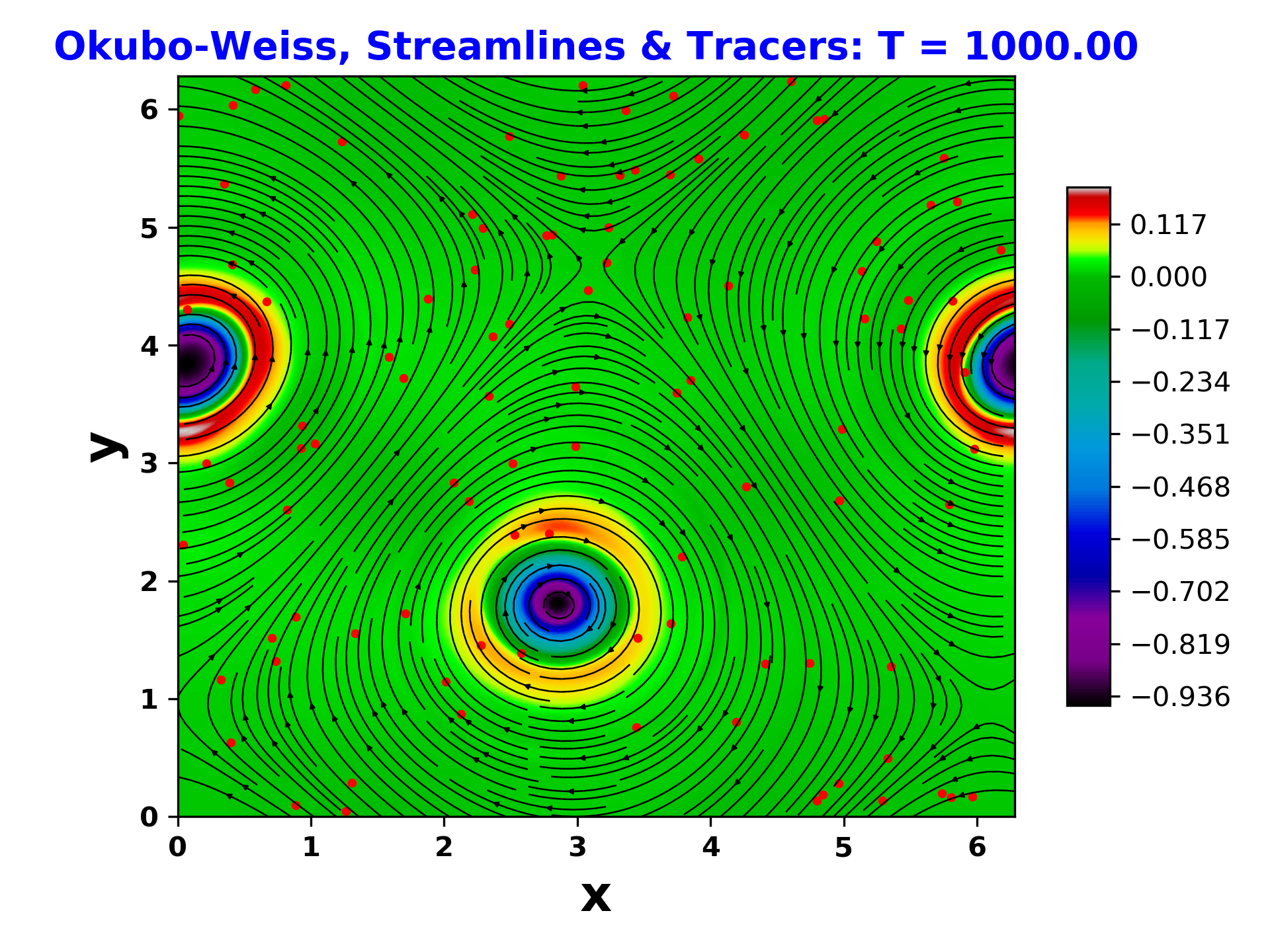}
        \caption{\textbf{T = 1000}}
        \label{fig:Okubuh}
    \end{subfigure}
    \hfill
    \begin{subfigure}[b]{0.33\textwidth}
        \centering
         \includegraphics[width=\textwidth]{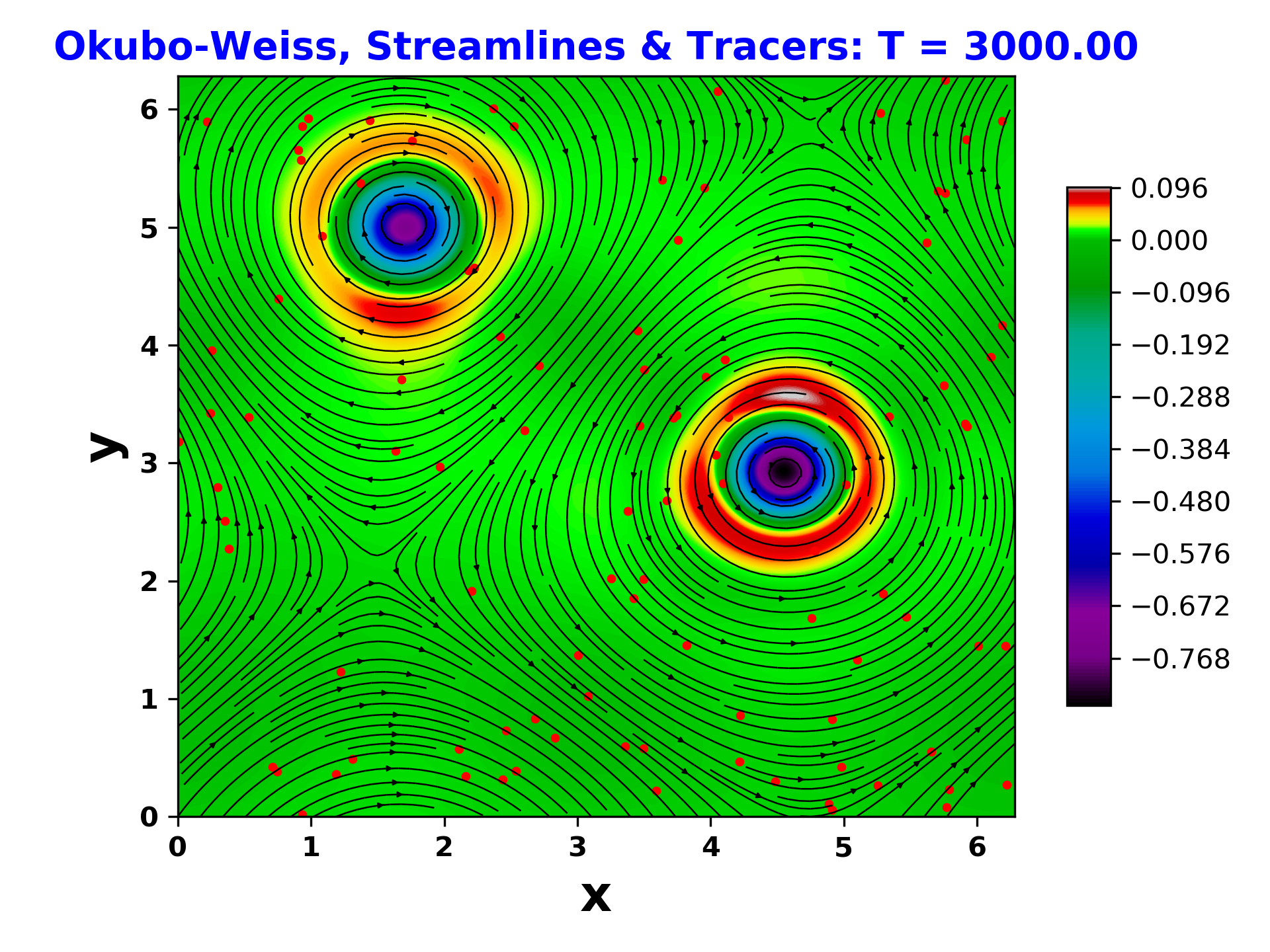}
        \caption{\textbf{T = 3000}}
       \label{fig:Okubui}
    \end{subfigure}
   \caption{Time evolution of the Okubo–Weiss field, stream function, and tracer particle distribution at selected time instants from $T = 0$ to 
$T = 3000$ for the 2-strip configuration of alternating-sign vorticity, corresponding to the lowest VPF (6.25\%).  Due to widely spaced vorticity strips, the KHI leads to the formation of only a small number of isolated vortex rolls, yielding a sparse and strongly anisotropic flow that evolves largely in isolation, with quick unidirectional merging and deformation primarily along the  $x$-direction and turbulence onset occurring relatively late ($T_{TO} = 45$). Successive vortex mergers then drive a rapid transition to turbulence characterized by the formation of larger-scale vortices and a short-lived, weak, anisotropic background, with the ensuing inverse cascade leading to the emergence of two dominant coherent vortices (a dipolar structure) by $T_D = 400$, embedded in small-scale fluctuations and executing large-scale orbital motion along nearly circular trajectories. An animation of the vorticity field, stream function field, and associated tracer particle distribution for the 2-strip configuration is provided as supplementary material with this article.}
    \label{fig:OW}
\end{figure*}

\begin{figure*}
    \centering
    \begin{subfigure}[b]{0.33\textwidth}
        \centering
         \includegraphics[width=\textwidth]{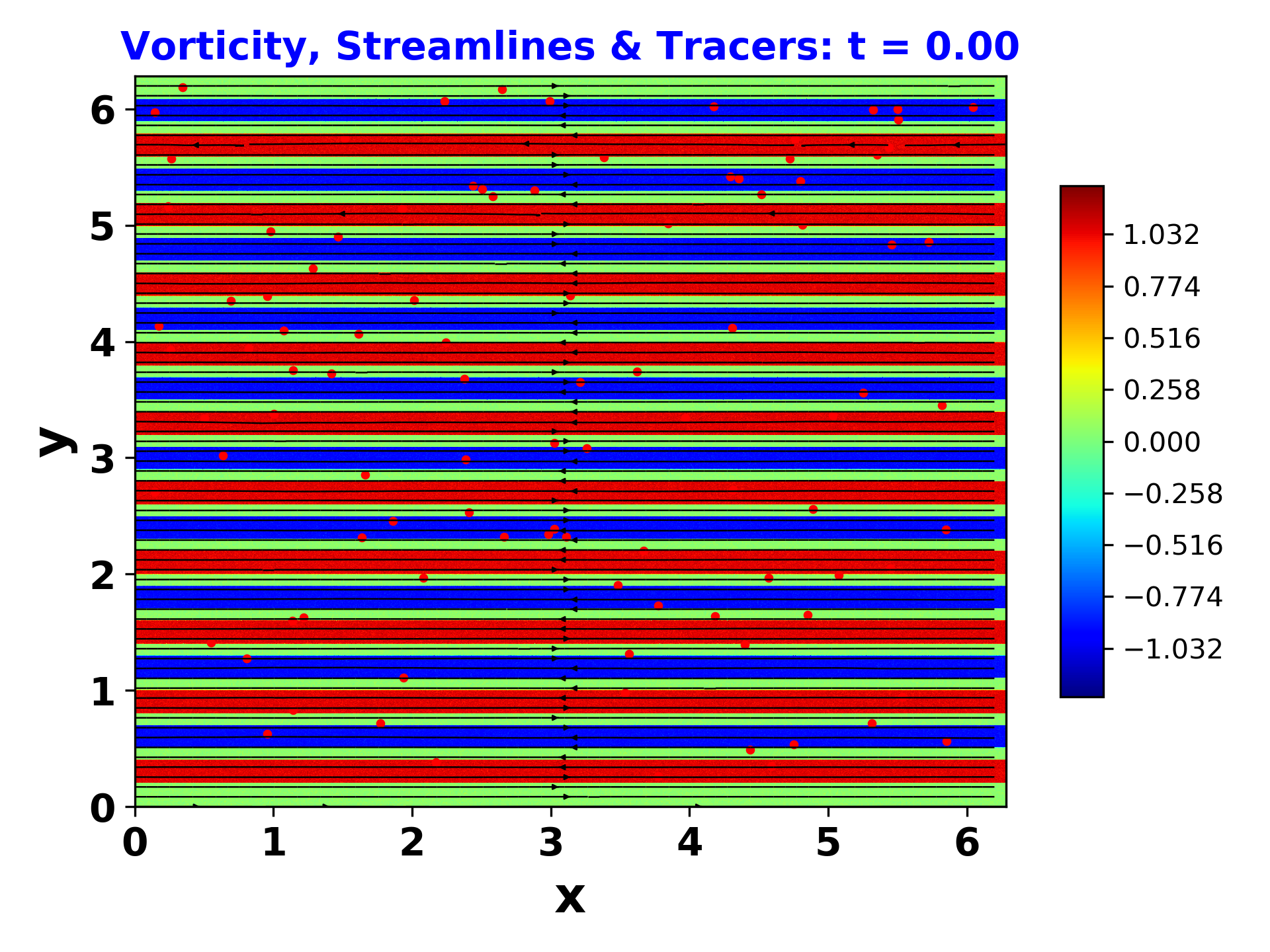}
        \caption{\textbf{T = 0}}
        \label{aa}
    \end{subfigure}
    \hfill
    \begin{subfigure}[b]{0.33\textwidth}
        \centering
         \includegraphics[width=\textwidth]{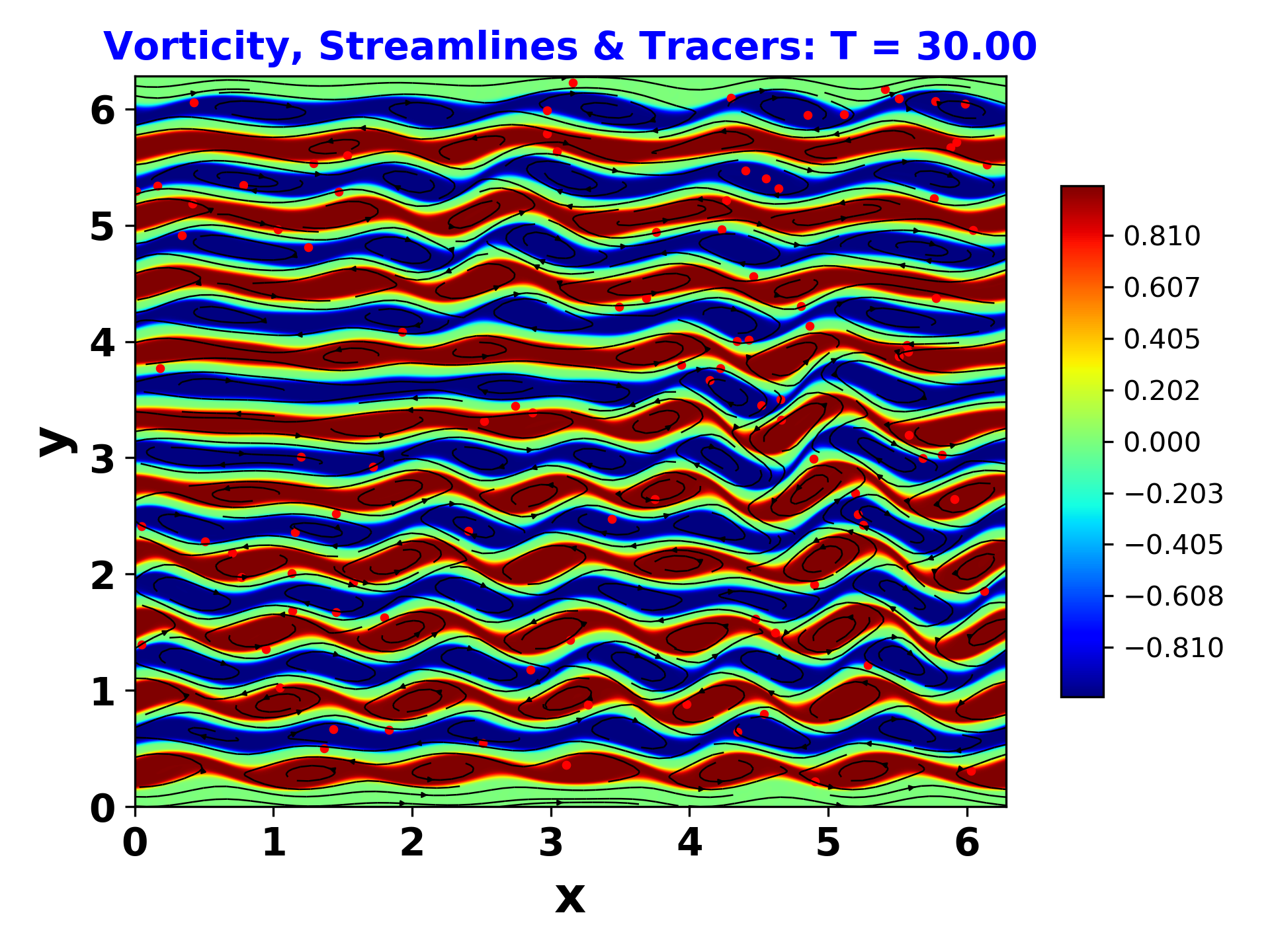}
        \caption{\textbf{T = 30}}
        \label{fig:Omegab}
    \end{subfigure}
    \hfill
    \begin{subfigure}[b]{0.33\textwidth}
        \centering
         \includegraphics[width=\textwidth]{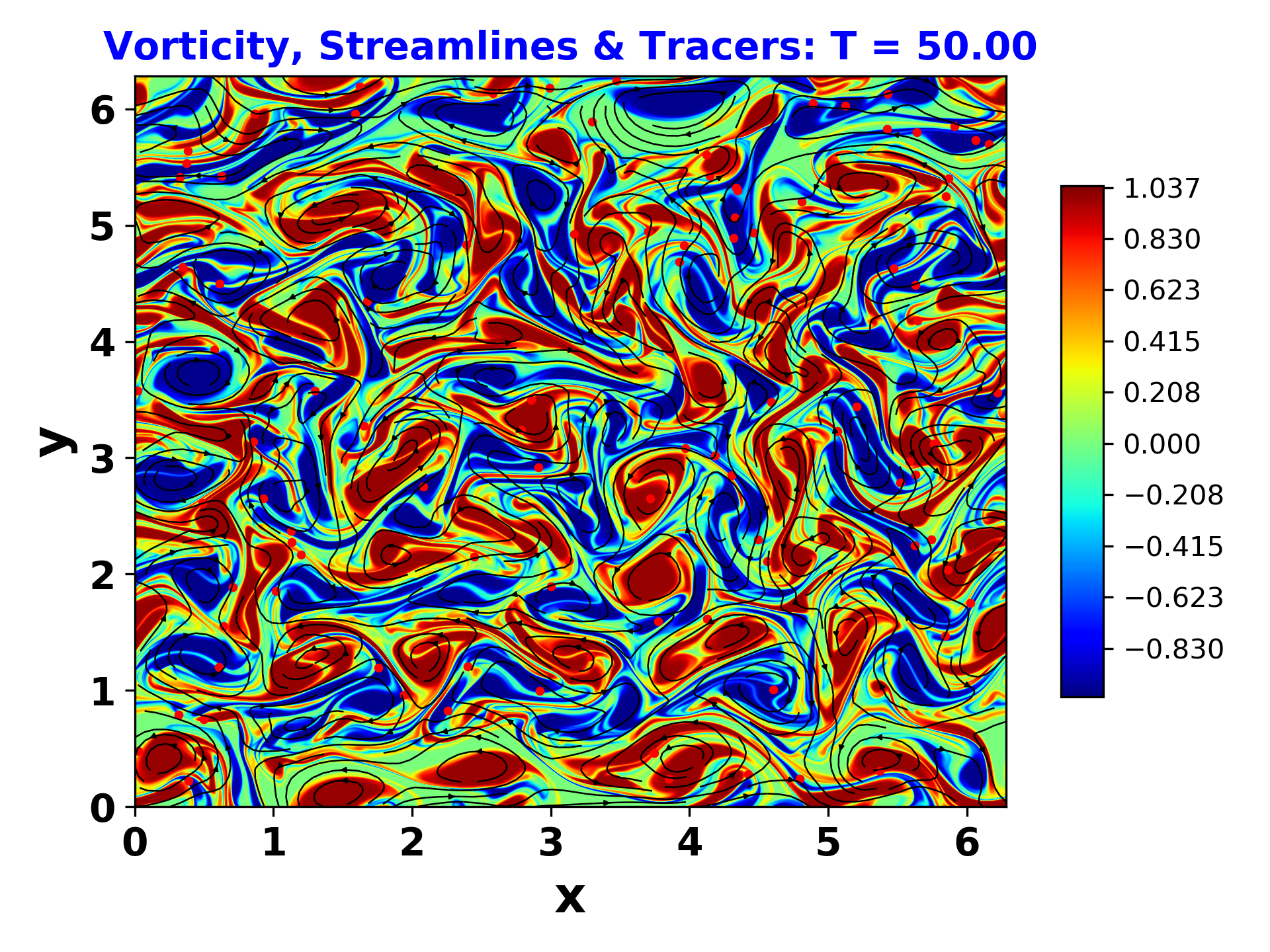}
        \caption{\textbf{T = 50}}
        \label{fig:Omegac}
    \end{subfigure}
    \hfill
    \begin{subfigure}[b]{0.33\textwidth}
        \centering
         \includegraphics[width=\textwidth]{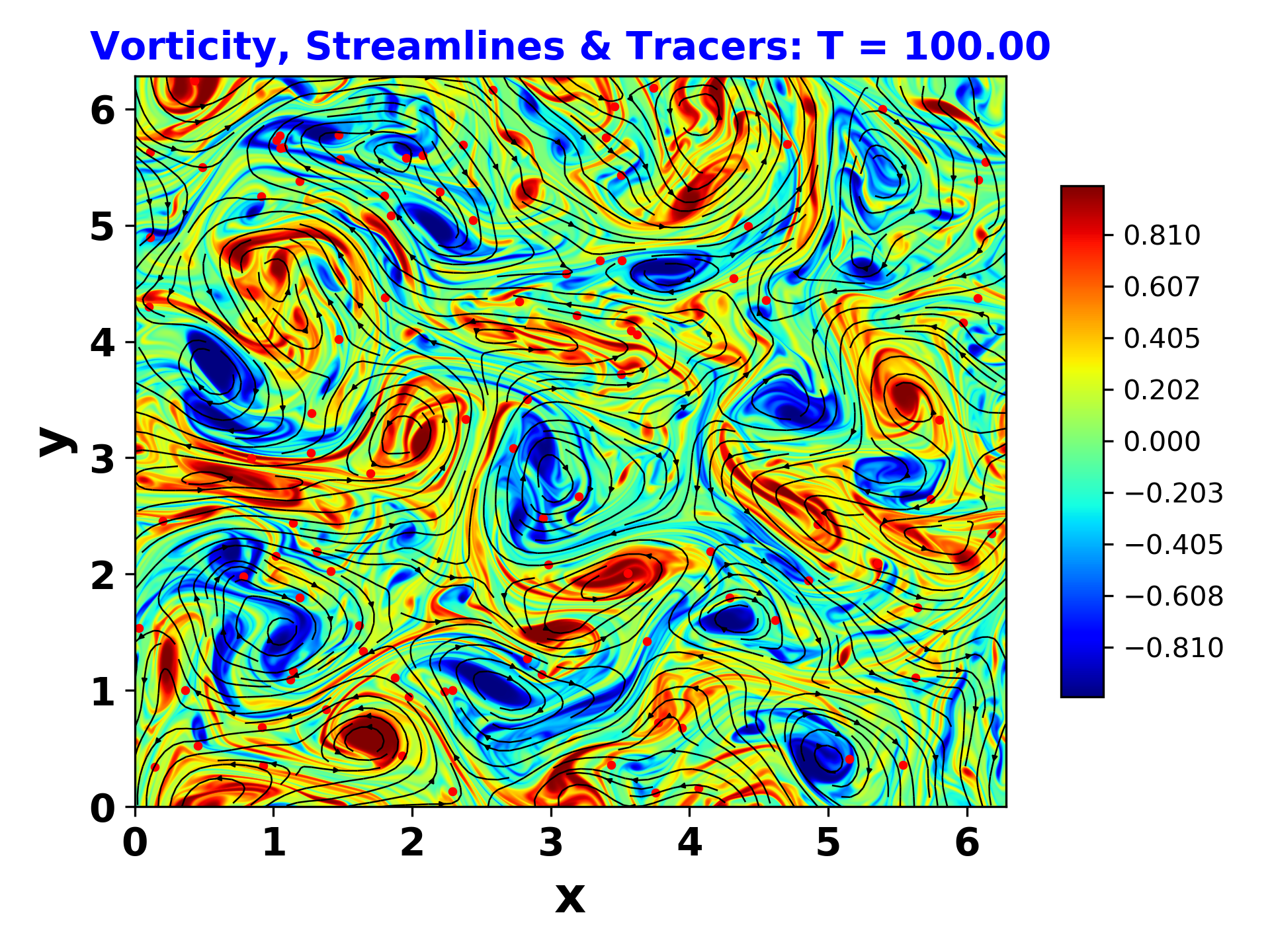}
        \caption{\textbf{T =100}}
        \label{fig:Omegad}
    \end{subfigure}
    \hfill
    \begin{subfigure}[b]{0.33\textwidth}
        \centering
         \includegraphics[width=\textwidth]{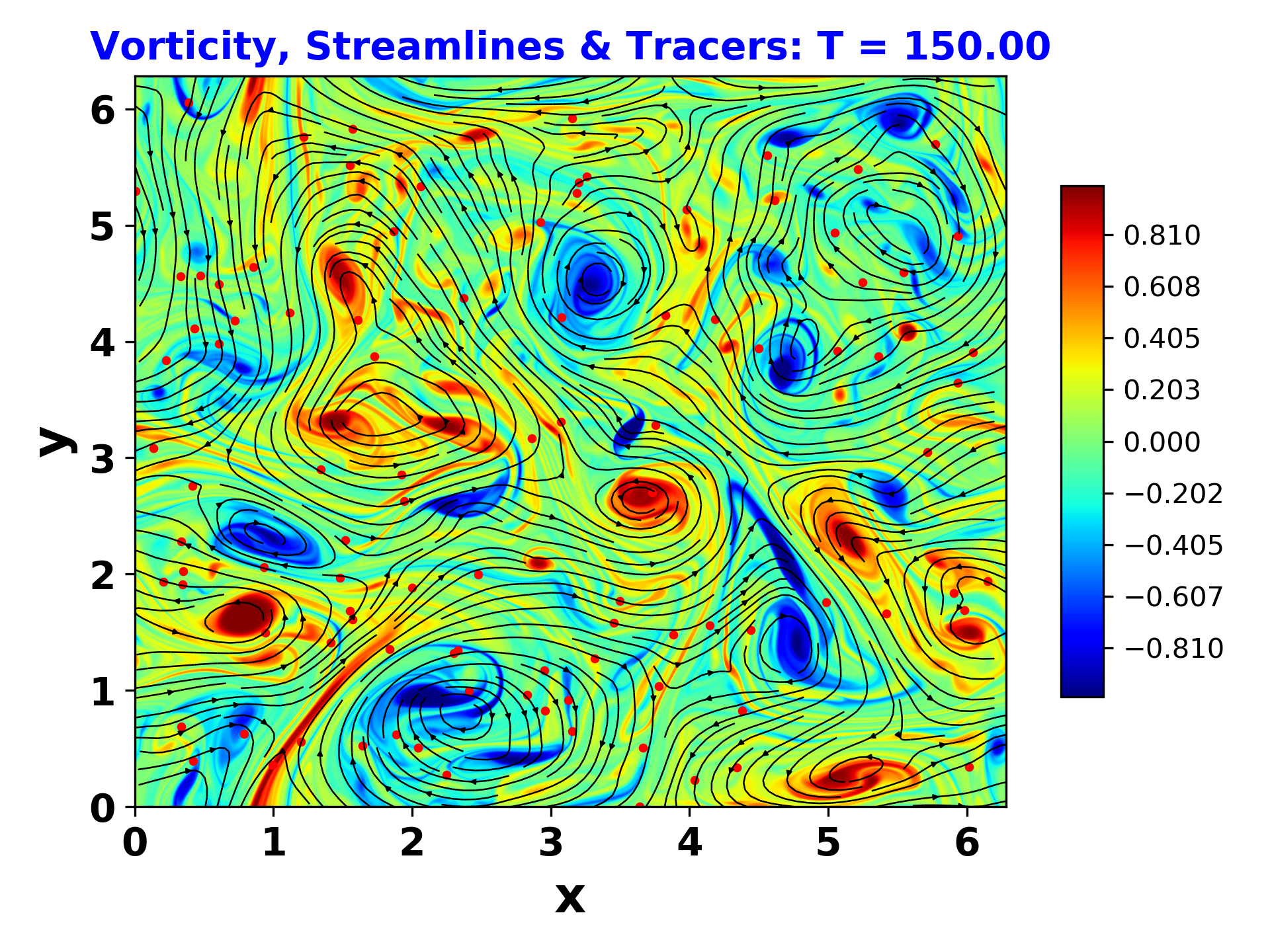}
        \caption{\textbf{T = 150}}
        \label{fig:Omegae}
    \end{subfigure}
    \hfill
    \begin{subfigure}[b]{0.33\textwidth}
        \centering
         \includegraphics[width=\textwidth]{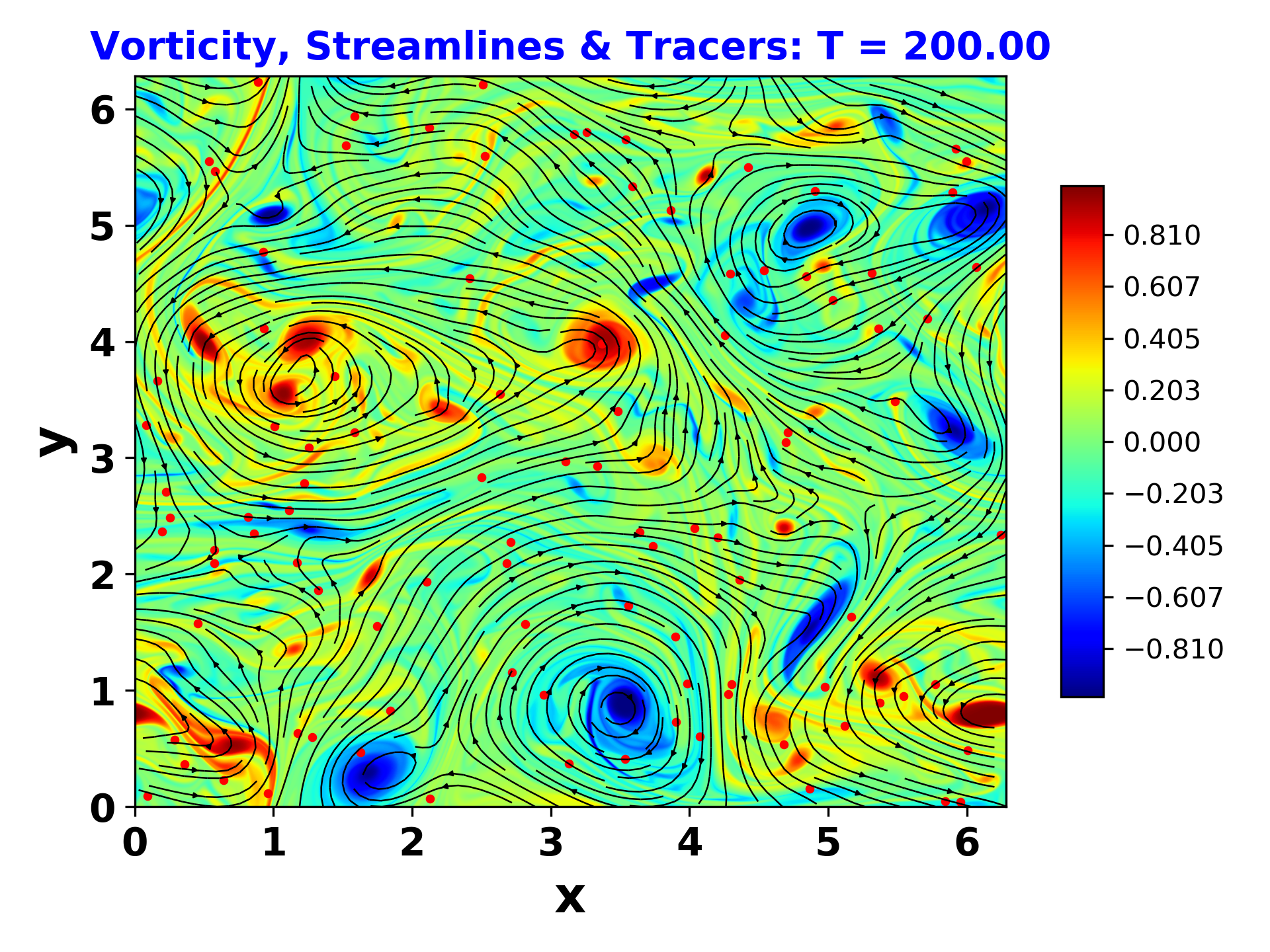}
        \caption{\textbf{T = 200}}
        \label{fig:Omegaf}
    \end{subfigure}
    \hfill
    \begin{subfigure}[b]{0.33\textwidth}
        \centering        \includegraphics[width=\textwidth]{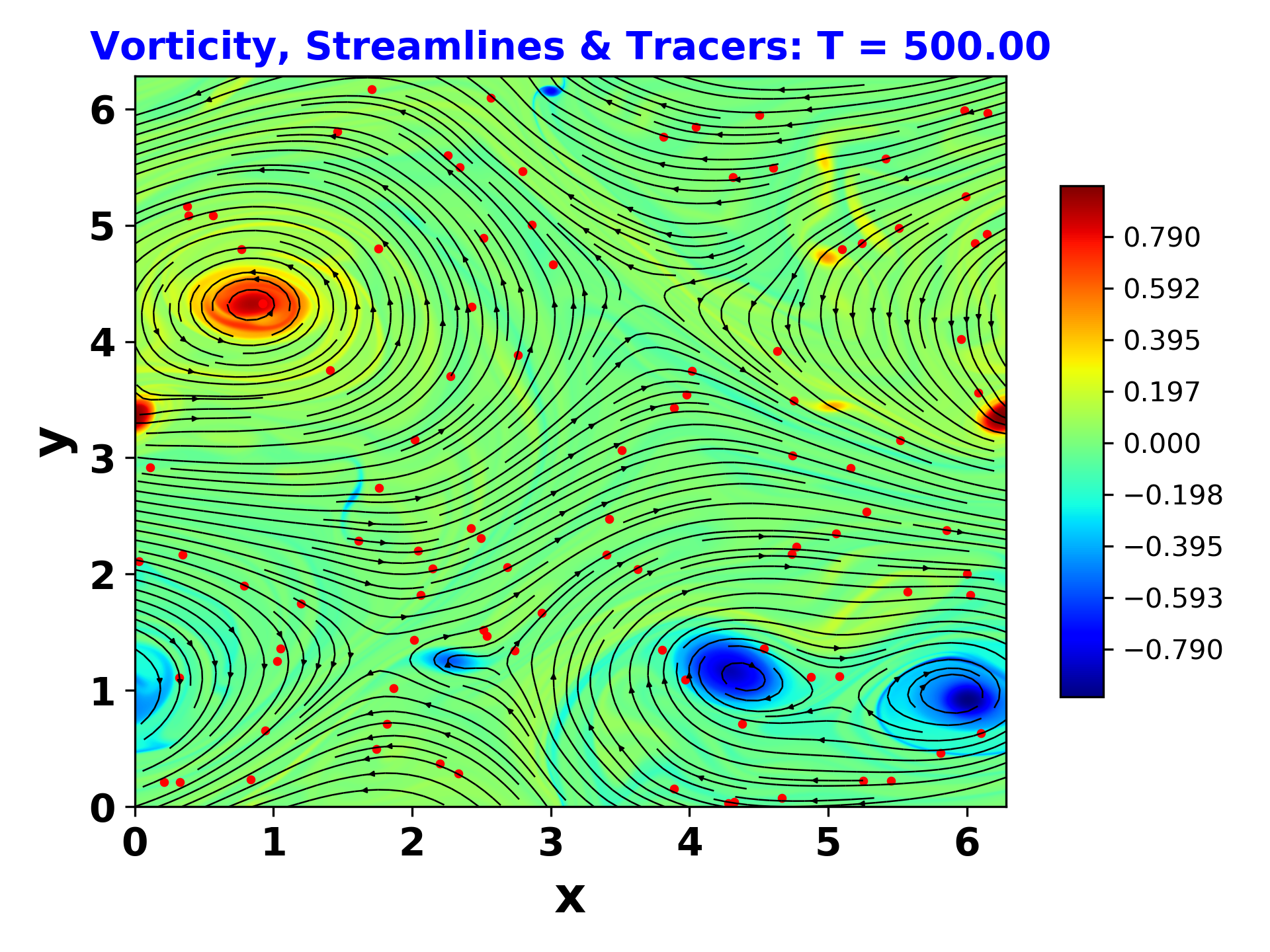}
        \caption{\textbf{T = 500}}
        \label{fig:Omegag}
    \end{subfigure}
    \hfill
    \begin{subfigure}[b]{0.33\textwidth}
        \centering
         \includegraphics[width=\textwidth]{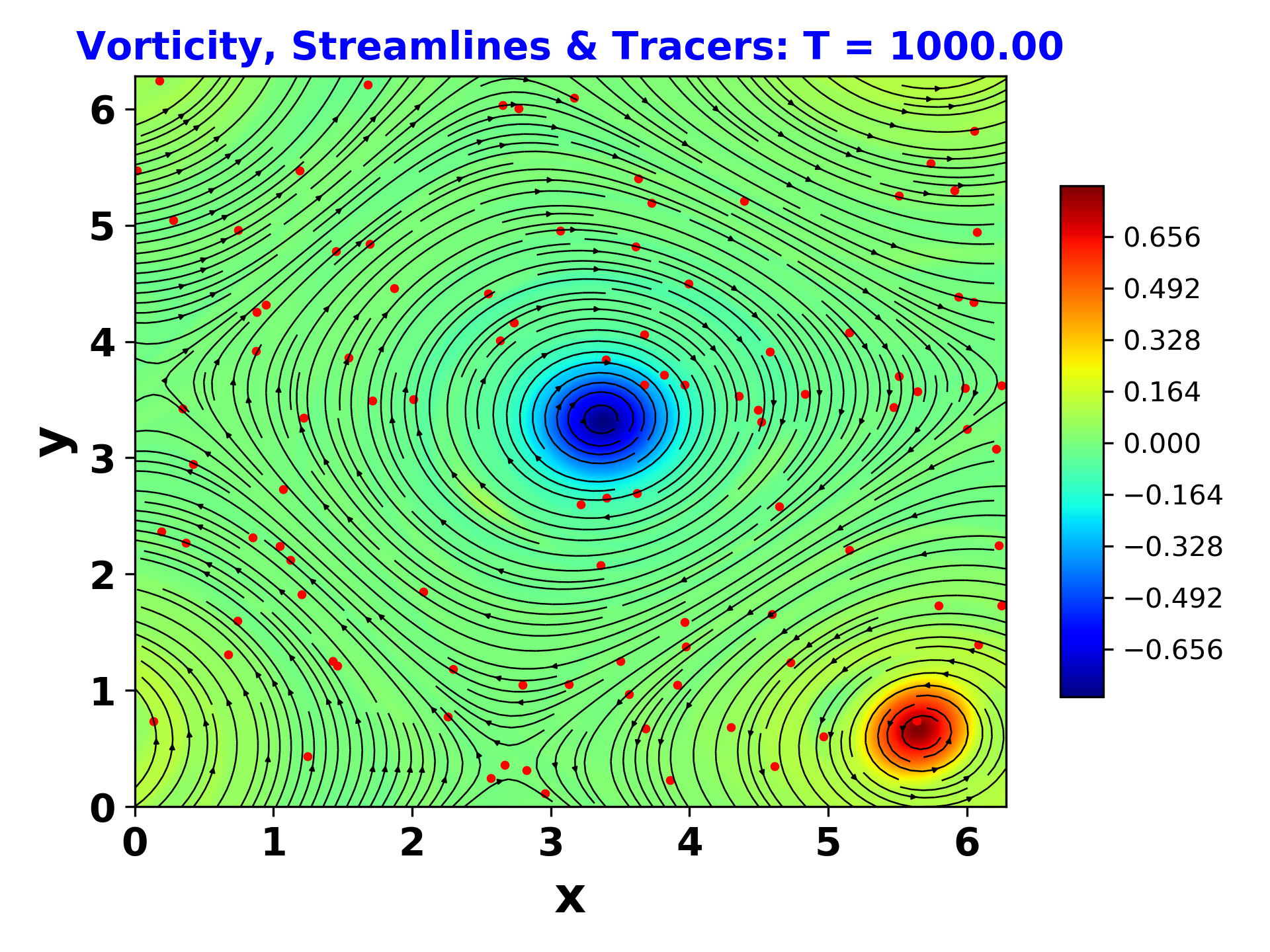}
        \caption{\textbf{T = 1000}}
        \label{fig:Omegah}
    \end{subfigure}
    \hfill
    \begin{subfigure}[b]{0.33\textwidth}
        \centering
         \includegraphics[width=\textwidth]{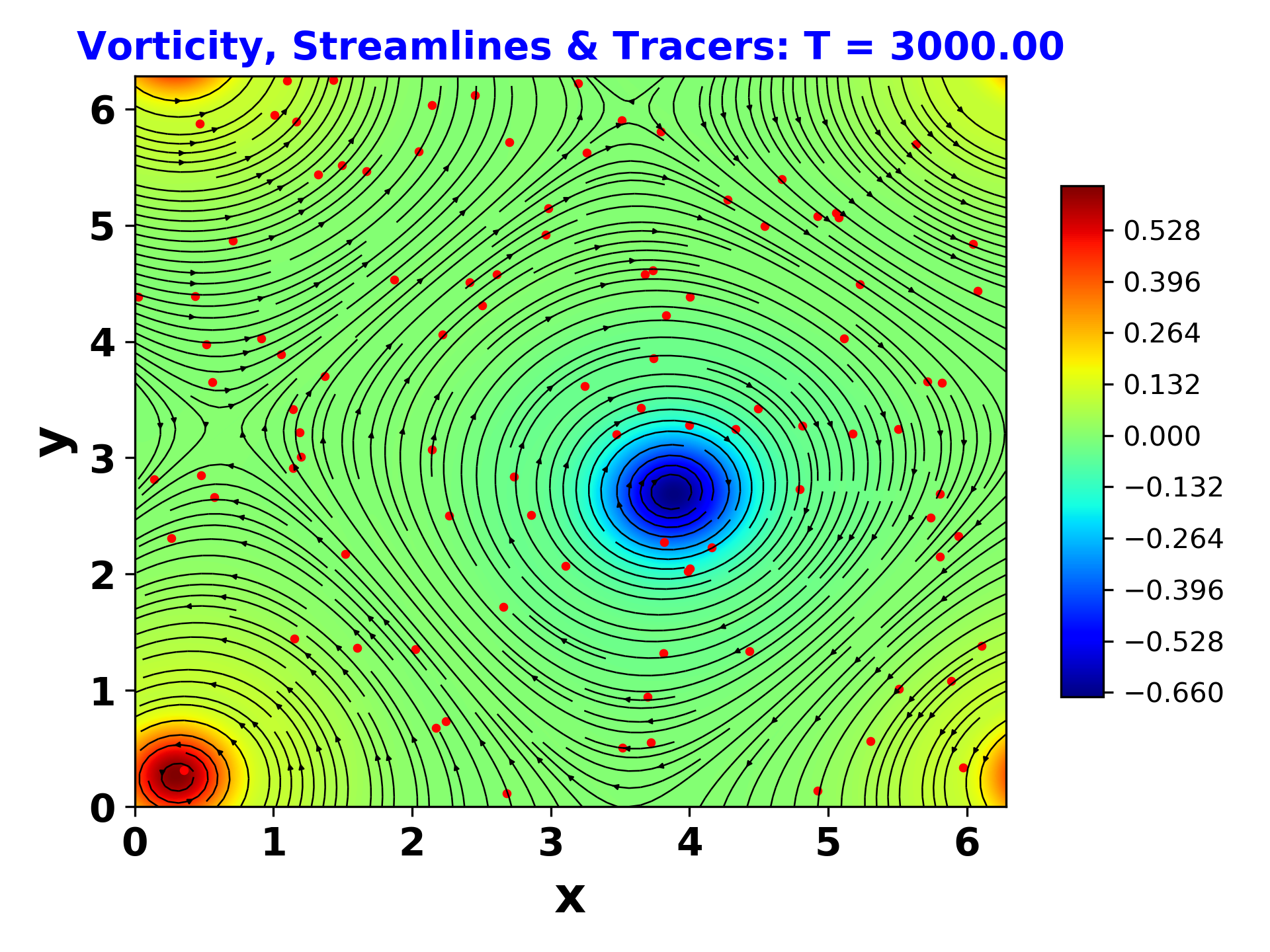}
        \caption{\textbf{T = 3000}}
       \label{fig:Omegai}
    \end{subfigure}
   \caption{
   Time evolution of the vorticity field, stream function field, and tracer particle distribution at selected time instants from $T = 0$ to $T = 3000$ for the 20-strip configuration of alternating-sign vorticity, corresponding to the highest vorticity packing fraction (VPF = 62.5\%). Due to the large number of closely packed strips, the KHI leads to the formation of numerous interacting vortex rolls, producing a dense and highly isotropic flow that interact quickly, with strong bidirectional merging and rapid fluid deformation along both $x$- and $y$- directions and turbulence onset occurring at the earliest times ($T_{TO} = 36$). Numerous large-scale vortices interact, merge and cascade over extended periods ($T = 36$ – $750$), generating a persistent, strong, and fully isotropic background, with the ensuing inverse cascade leading to the emergence of the largest-scale coherent dipolar vortices at late times ($T_D \approx 750$), embedded in small-scale fluctuations and executing small-scale orbital motion along circular trajectories. An animation of the vorticity field, stream function field, and associated tracer particle distribution for the 20-strip configuration is provided as supplementary material with this article.}
    \label{fig:VS}
\end{figure*}

{This ballistic regime persists until shear amplifies small perturbations at the vorticity strip interfaces via the Kelvin--Helmholtz instability (KHI). Driven by the growing transverse kinetic energy of the KHI (see Fig.~\ref{fig:KE_y}), the strip boundaries deform and roll up into large scale vortices (see $T = 30$, Fig.~\ref{fig:Okubub}, \ref{fig:Omegab}). As vortex rolls begin to emerge, they interact with their nearest neighbors, marking the transition to nonlinear turbulence at ($T \approx T_{TO}$) (see $T = 100$, Fig.~\ref{fig:Okubuc}, and $T=50$ Fig.~\ref{fig:Omegac})}.

The growth of the Kelvin–Helmholtz (KH) instability generates vortex rolls whose number, spacing, and interactions are determined by the initial number of vortex strips and their separation, making the onset and character of these interactions strongly dependent on the initial vorticity packing fraction. As packing increases, roll interactions occur more frequently and earlier, leading to a faster transition to turbulence.

For the lowest VPF (6.25\%), only a few widely spaced rolls form, resulting in a sparse, highly anisotropic configuration that evolves largely in isolation, with quick unidirectional merging and fluid deformation primarily along the $x$-direction, and turbulence onset occurring relatively late ($T_{TO} = 45$) (see Fig ~\ref{fig:OW} and Fig ~\ref{fig:KE_y}). With a moderate increase in VPF (12.5\%), more rolls form with reduced spacing, producing a less sparse, less anisotropic configuration that still evolves largely in isolation, with slow bidirectional merging and fluid deformation, and turbulence onset occurs slightly earlier ($T_O$ = 43) compared to the lowest VPF (6.25\%) case.
With a further increase in VPF (25\%), more closely spaced rolls form, producing a dense, nearly isotropic configuration; interactions and mergers occur earlier, driving faster bidirectional fluid deformation, with turbulence onset at $T_{TO} = 40$, earlier than in the lower VPF cases. At the highest VPFs (50\% and 62.5\%), the maximum number of closely packed strips produces a highly isotropic, dense configuration; rolls interact almost immediately, allowing minimal independent evolution, triggering strong bidirectional merging and rapid fluid deformation in both $x$- and $y$-directions, with turbulence onset at the earliest times ($T_{TO} = 36$ and $T_{TO} = 33$, respectively) (see Fig ~\ref{fig:VS} and Fig ~\ref{fig:KE_y}).

Vortex rolls with the same sense of circulation tend to merge, deforming the surrounding fluid manifold in the process. This merging leads to the formation of progressively larger vortex structures, consistent with the inverse energy cascade characteristic of two-dimensional turbulence. As a result, the flow evolves into a complex state composed of a turbulent background interspersed with relatively long-lived large-scale vortices that grow via the inverse cascade. 

For the lowest VPF (6.25\%), turbulence is characterized by the rapid formation of a few large-scale vortex rolls that undergo a quick inverse cascade, producing a short-lived, weak, and anisotropic background (see $T = 50$–200, Figs.~\ref{fig:Okubub}–\ref{fig:Okubuf}). With a moderate increase in VPF (12.5\%), turbulence involves the formation of a greater number of vortex rolls that inverse cascade more slowly, yielding a longer-lived, stronger, and less anisotropic background. For a further increase in VPF (25\%), still more large-scale vortex rolls form and undergo an even slower inverse cascade over extended times, resulting in a long-lived, strong, and nearly isotropic background. At the highest VPFs (50\% and 62.5\%), numerous large-scale vortex rolls develop and cascade over prolonged periods, generating a persistent, strong, and fully isotropic background (see $T = 30$–200, Figs.~\ref{fig:Omegad}–\ref{fig:Omegaf}). Overall, with increasing vortex packing fraction, turbulence involves the formation of an increasing number of large-scale vortices with longer inverse-cascade durations and becomes progressively longer-lived, stronger, and more isotropic.

The inverse cascade continues until the formation of the largest-scale coherent dipolar vortices at late times ($T \approx T_D$), after which the flow becomes increasingly dominated by the motion of these coherent dipoles (for VPF 6.25\%, see $T = 1000$ and $2000$ in Fig.~\ref{fig:Omegah} and Fig.~\ref{fig:Omegai}; for VPF 62.5\%, see $T = 500$, $1000$, and $3000$ in Fig.~\ref{fig:Okubug}, Fig.~\ref{fig:Okubuh}, and Fig.~\ref{fig:Okubui}). The largest-scale dipoles form at $T_D \approx 400$ for the lowest VPF (6.25\%), $T_D \approx 1000$ for VPF 12.5\%, $T_D \approx 900$–2300 for VPF 25\% (as two same-sign vortices remain separated for an extended period before merging), and at $T_D \approx 750$ and $600$ for the higher VPFs of 50\% and 62.5\%, respectively (see Table~\ref{tab:turbulence_table}).

During the transition from the linear to the nonlinear turbulent regime, the initially shear-driven motion in the $x$-direction is progressively suppressed by instability-driven nonlinear dynamics in the $y$-direction, leading to vortex roll formation, successive mergers, and enhanced mixing. As coherent shear flow breaks down through these interactions, $x$-direction transport slows and becomes sub-diffusive. The stage at which this transition from ballistic to sub-diffusive motion occurs in the $x$-direction depends strongly on the initial vorticity packing fraction (VPF)—relatively late for low VPF, earlier for moderate VPF, and earliest for high VPF. In contrast, $y$-direction transport grows in a super-diffusive regime as instability-driven fluctuations intensify. Both the growth rate and degree of $y$-direction super-diffusion increase with VPF, with the highest VPF cases exhibiting the strongest instability-enhanced mixing and flow deformation. Consequently, total transport after the onset of turbulence is governed by the coupled evolution of $x$- and $y$-displacements, whose relative contributions—specifically, the suppression of $x$-transport and the corresponding enhancement of $y$-transport—are jointly determined by the initial vorticity packing fraction.

{After the onset of turbulence, the flow evolves into a complex state composed of a turbulent background interspersed with long-lived coherent structures, which together drive the dispersion and mixing of tracer particles (see $T = 100, 150, 200$, Figs.~\ref{fig:Okubud}–\ref{fig:Okubuf} and Figs.~\ref{fig:Omegad}–\ref{fig:Omegaf}). Notably, the cores of long-lived coherent structures often act as trapping regions, limiting scalar advection across their enclosing elliptic manifolds. In contrast, nearby hyperbolic saddle points enhance dispersion by stretching and deforming fluid elements into elongated filaments along their unstable manifolds, thereby promoting mixing.}

During the vortex-merging and turbulence development phase, the evolution of $x$- and $y$-transports depends strongly on the initial vorticity packing fraction, the flow structure, and the stage of evolution, with total transport exhibiting a range of behaviors—sub-diffusive, diffusive, or super-diffusive—and transitions between anisotropic and isotropic regimes. 
For the lowest VPF (6.25\%), the $y$-directed Kelvin–Helmholtz instability (KHI) induces super-diffusive motion that gradually suppresses the ballistic $x$-motion at very late times; however, the $x$-transport continues to dominate due to sustained shear-driven flow, resulting in strongly anisotropic, sub-diffusive total transport (see 2-strip case in Fig ~\ref{fig:tr_all_xy}). With a moderate increase in VPF (12.5\%), the KHI drives stronger and faster super-diffusive $y$-motion than in the lowest VPF case, leading to a more pronounced suppression of $x$-motion; consequently, the system becomes less anisotropic, although the total transport remains sub-diffusive (see 4-strip case in Fig ~\ref{fig:tr_all_xy}). With a further increase in VPF (25\%), the KHI-driven $y$-transport grows even faster than in the lower-VPF cases, further reducing x-motion to sub-diffusive levels; consequently, $y$-transport slightly exceeds $x$-transport at later times, reflecting weaker anisotropy and a total transport approaching normal diffusion (see 8-strip case in Fig ~\ref{fig:tr_all_xy}). At the highest VPFs (50\% and 62.5\%), the $y$-directed KHI drives the fastest super-diffusive motion, producing the strongest suppression of $x$-motion and yielding overall super-diffusive total transport. The 50\% case shows mild super-diffusion, while the 62.5\% case exhibits the most intense and sustained super-diffusive behavior, with x- and $y$-transports closely matching at intermediate times, signifying a nearly isotropic turbulent regime (see 16 and 20-strip case in Fig ~\ref{fig:tr_all_xy}). Overall, increasing the initial vorticity packing progressively shifts the system from strongly anisotropic, sub-diffusive transport toward more isotropic, super-diffusive turbulent transport, enhancing transverse mixing and accelerating the breakdown of coherent shear motion.

{As small-scale random perturbations develop across multiple scales, vortices evolve and interact, rendering the flow chaotic and eventually turbulent. Tracer particles are advected in a seemingly stochastic manner, undergoing numerous small, random-like displacements driven by turbulent eddies. This motion resembles a random walk, causing the position probability distribution function (PDF), $n(x)$ and $n(y)$, to approach a normal or Gaussian distribution from the initial uniform distribution. The emergence of this normal distribution is explained by the central limit theorem, as the cumulative effect of many independent, small-scale fluctuations governs particle displacement.}

For the lowest VPF (6.25\%), $n(y)$ attains Gaussianity at a relatively long time ($T \approx 200$), while $n(x)$ remains bimodal under dominant shear until an even longer time ($T \approx 400$) before becoming near-Gaussian (see Fig.~\ref{fig:pdf_2}). With a moderate increase in VPF (12.5\%), both directions evolve faster—$n(y)$ approaches a Gaussian distribution earlier ($T \approx 100$), while $n(x)$ loses its shear-induced bimodality much sooner ($T \approx 200$) and becomes nearly Gaussian. Even though both $n(x)$ and $n(y)$ later approach Gaussian forms, they remain highly unequal with minimal overlap, reflecting strong anisotropy for the 6.25\% and 12.5\% cases, except for a brief isotropic phase around $T \approx 1200$–1600 for VPF 12.5\%, indicating reduced anisotropy compared to the 6.25\% case. The slower evolution of $n(x)$ and $n(y)$ toward normal distributions in both cases reflects the associated sub-diffusive particle motion. For a further increase in VPF (25\%), the trend continues, with $n(y)$ attaining Gaussianity even earlier ($T \approx 90$) and $n(x)$ breaking its bimodal structure within a shorter time ($T \approx 50$), becoming Gaussian around $T \approx 75$—significantly earlier than in the 12.5\% case. Both $n(x)$ and $n(y)$ achieve near-isotropy around $T \approx 500$, reflecting only very slight anisotropic transport. The comparatively faster evolution of $n(x)$ and $n(y)$ toward Gaussian distributions in this case corresponds to normal-diffusive particle motion. At the highest VPFs (50\% and 62.5\%), both $n(y)$ and $n(x)$ rapidly become Gaussian ($T \approx 40$ for $n(y)$, $T \approx 50$ for $n(x)$), achieving isotropy by $T \approx 250$, marking the fastest evolution toward Gaussianity and isotropic turbulence across all cases (see Fig.~\ref{fig:pdf_16} and ~\ref{fig:pdf_20}). This fastest evolution of $n(x)$ and $n(y)$ toward Gaussian distributions corresponds to super-diffusive particle motion. 

As turbulence develops, small-scale random velocity perturbations emerge and interact across multiple scales, causing the velocity PDFs $n(v_x)$ and $n(v_y)$ to gradually approach Gaussian distributions, with the timescale for this transition decreasing with increasing initial VPF. For the lowest VPF (6.25\%), $n(v_y)$ becomes approximately Gaussian around $T \approx 75$--$100$, while $n(v_x)$ remains shear-dominated and bimodal until $T \lesssim 100$, gradually broadening thereafter to reach Gaussianity (see intermediate times, Fig.~\ref{fig:vdf_2}). For moderate VPF (12.5\%), $n(v_y)$ Gaussianizes earlier ($T \approx 50$--$75$) and $n(v_x)$ transitions from bimodal to Gaussian by $T \lesssim 50$. For higher VPF (25\%), $n(v_y)$ and $n(v_x)$ become Gaussian at $T \approx 30$--$50$ and $T \lesssim 40$, respectively. At the highest VPFs (50\% and 62.5\%), $n(v_y)$ Gaussianizes by $T \approx 10$--$30$ and $n(v_x)$ catches up by $T \lesssim 30$ (see intermediate times, Fig.~\ref{fig:vdf_16}). In all cases, the PDFs eventually converge to Gaussian forms, with the $x$-direction catching up to the $y$-direction progressively faster as the initial VPF increases, reflecting the growing dominance of multiscale random fluctuations characteristic of fully developed turbulence. \\

%\newpage

\textbf{D. Largest scale coherent Dipole motion}\\

{At late times, transport is dominated by the slow drift of the largest scale coherent vortices or dipoles, each composed of two large, oppositely rotating vortices that remain spatially separated and coherent over long periods. Their long-term dynamics are influenced by the initial vorticity packing. For VPFs (6.25\%, 12.5\%, 25\%, 50\%), the vortices exhibit orbital rotation along nearly circular trajectories, producing oscillatory motion in both the $x$ and $y$ directions (see late times in Fig.~\ref{fig:tr_all_xy}). For the lower VPFs (6.25\%, 12.5\%, 25\%), the orbital rotation occurs on large scales (large-radius orbital motion), whereas for high VPF (50\%), the orbital rotation is confined to smaller scales (small-radius orbital motion). In these cases, the amplitude of oscillation and the anisotropy between $x$ and $y$ decrease with increasing packing fraction, becoming completely isotropic for the VPF (50\%) case (see late times in Fig.~\ref{fig:tr_all_xy}). The combined effect of tracer rotation around the dipoles and their orbital motion leads to sub-diffusive trapping, as seen in Fig.~\ref{fig:tr_all_xy} and ~\ref{fig:TotTransport}. In contrast, for the highest packing fraction (62.5\%), the vortices undergo linear translational dipole motion along diagonal paths, resulting in super-diffusive transport, as seen in Fig.~\ref{fig:tr_all_xy} and ~\ref{fig:TotTransport}.}

At long times, the particle position distributions $n(x)$ and $n(y)$ in both the $x$ and $y$ directions reflect coherent dipole motion transport and therefore vary with the initial VPF. For low VPF (6.25\%), the distributions are near- Gaussian but unequal, with minimal overlap, indicating anisotropic transport (see Fig.~\ref{fig:pdf_2}, from $T \gtrsim 750$). As the VPF increases (12.5\%, 25\%, 50\%), the nearly Gaussian distributions become more equal, overlapping, and symmetric, reflecting a transition towards isotropic transport (see Fig.~\ref{fig:pdf_16}, from $T \gtrsim 500$). For the highest VPF (62.5\%), the distributions develop heavy tails from $T \approx 2000$ onwards, signaling a departure from normality due to the onset of super-diffusive behavior, while still maintaining overlap consistent with isotropic transport (see Fig.~\ref{fig:pdf_20}, from $T \gtrsim 750$).

The counter-rotation of dipoles produces two distinct populations of velocity states, leading to bimodal velocity PDFs $n(v_x)$ and $n(v_y)$ with peaks at $\pm v_{dp}$, each corresponding to one of the counter-rotating vortex populations and representing their typical rotational velocity (see late times   in Fig.~\ref{fig:vdf_2} for the VPF 6.25\% case and Fig.~\ref{fig:vdf_16} for the VPF 50\% case).

\begin{table*}
\centering
\renewcommand{\arraystretch}{1.2} % Improves vertical spacing
\begin{tabular}{
|>{\centering\arraybackslash}p{1.9 cm}
|>{\centering\arraybackslash}p{1.8 cm}
|>{\centering\arraybackslash}p{2.4 cm}
|>{\centering\arraybackslash}p{2.7 cm}
|>{\centering\arraybackslash}p{2 cm}
|>{\centering\arraybackslash}p{2.7 cm}|}
\hline
\textbf{Initial Condition} & \textbf{ Turbulence Onset} & \textbf{Turbulence Transport Evolution} & \textbf{MSD Power-law Exponent $\boldsymbol{\alpha}$} & \textbf{Large Scale Structure Formation} & \textbf{MSD Power-law Exponent $\boldsymbol{\alpha}$} \\
\hline
\makecell{2 strips; \\ VPF: 6.25\%} & $T_{TO}$ = 45 & \makecell{\textbf{Sub-diffusion}; \\High anisotropy \\ } & $\alpha < 1$ & \makecell {$T_D$ = 400; \\ \textbf{Sub-diffusion}} & $\alpha < 1$ \\
\hline
\makecell{4 strips; \\ VPF: 12.5\%}& $T_{TO}$ = 43 &\makecell{\textbf{Sub-diffusion}; \\Less anisotropy \\ } & $\alpha < 1$ & \makecell {$T_D$ = 1000; \\ \textbf{Sub-diffusion}} & $\alpha < 1$ \\
\hline
\makecell{8 strips; \\ VPF: 25\%}& $T_{TO}$ = 40 &\makecell{\textbf{Normal diffusion};\\Slight anisotropy \\ } & $\alpha \approx 1$ & \makecell {$T_D$ = 2300; \\ \textbf{Sub-diffusion}} & $\alpha < 1$ \\
\hline
\makecell{16 strips; \\ VPF: 50\%} & $T_{TO}$ = 36 & \makecell{ \textbf{Mild} \\ \textbf{Super-diffusion}; \\ Isotropy \\} & $\alpha > 1$ &   \makecell{$T_D$ = 750; \\ \textbf{Sub-diffusion}} & $\alpha < 1$ \\
\hline
\makecell{20 strips;\\ VPF: 62.5\%} & $T_{TO}$ = 33 &  \makecell{ \textbf{Strong} \\ \textbf{Super-diffusion}; \\ Isotropy \\} & $\alpha > 1$ & $T_D$ = 600; \textbf{Super-diffusion} & $\alpha > 1$ \\
\hline
\end{tabular}
\caption{Summary of turbulence evolution across different initial strip configurations. The first column lists the initial number of vorticity strips and their corresponding VPFs within the simulation domain. The second column presents the turbulence onset time, $T_{TO}$, marking the transition of the flow from the linear to the nonlinear regime, which occurs progressively earlier as the VPF increases. The third column describes the turbulence transport evolution, indicating the dominant transport regime for each case along with the degree of anisotropy between $x$ and $y$ transport. The fourth column lists the corresponding MSD power-law exponent $\alpha$ obtained from the time scaling of the mean-square displacement, where $\alpha < 1$ denotes subdiffusion, $\alpha \approx 1$ normal diffusion, and $\alpha > 1$ superdiffusion. The fifth and sixth columns summarize the large-scale dipole formation time $T_D$ and the corresponding MSD power-law exponent $\alpha$ for dipole motion, which similarly transitions from subdiffusive to superdiffusive with increasing VPF.}
\label{tab:turbulence_table}
\end{table*}

The general features of turbulent transport and its dependence on the initial vorticity packing fraction (VPF), as obtained from this study and the above discussions, are summarized in the table (see Table~\ref{tab:turbulence_table}). The table shows that with increasing VPF, the onset of turbulence—i.e., the transition from the linear to the nonlinear regime—is accelerated. During the intermediate stage of turbulence development, particle transport evolves from a sub-diffusive and anisotropic regime to a super-diffusive and isotropic regime. At later stages, the coherent dipole motion also transitions from sub-diffusive to super-diffusive behavior with increasing VPF.

\begin{figure*}

     \begin{subfigure}{0.49\textwidth}
       \centering\includegraphics[width=\textwidth]{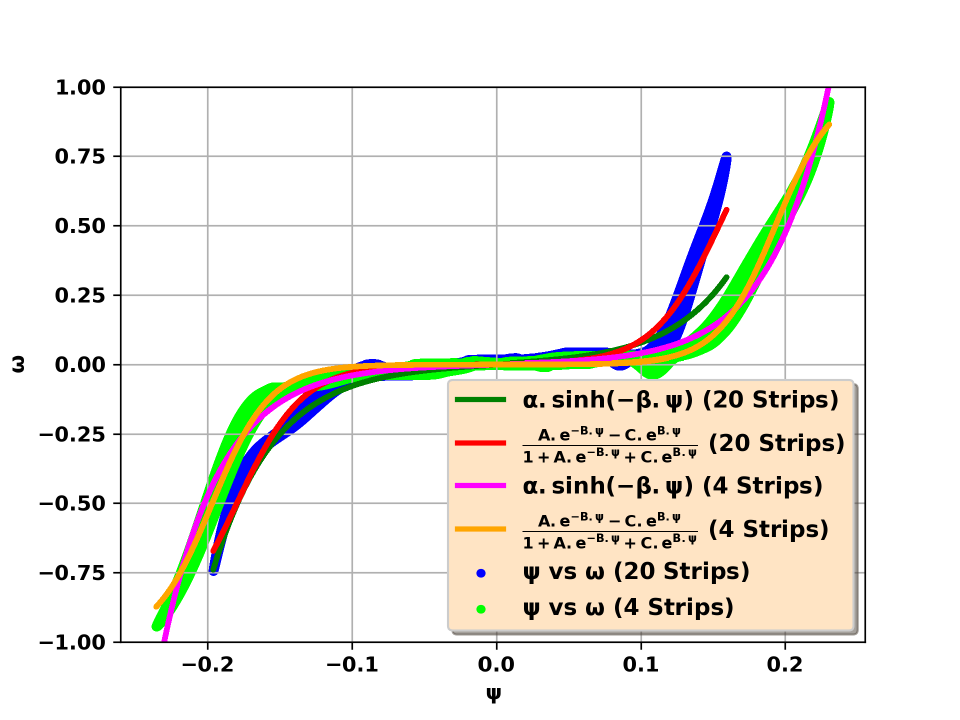}
        \caption{}
        \label{fig:scatter}
    \end{subfigure}
    \hfill
    \begin{subfigure}{0.49\textwidth}
         \centering \includegraphics[width=\textwidth]{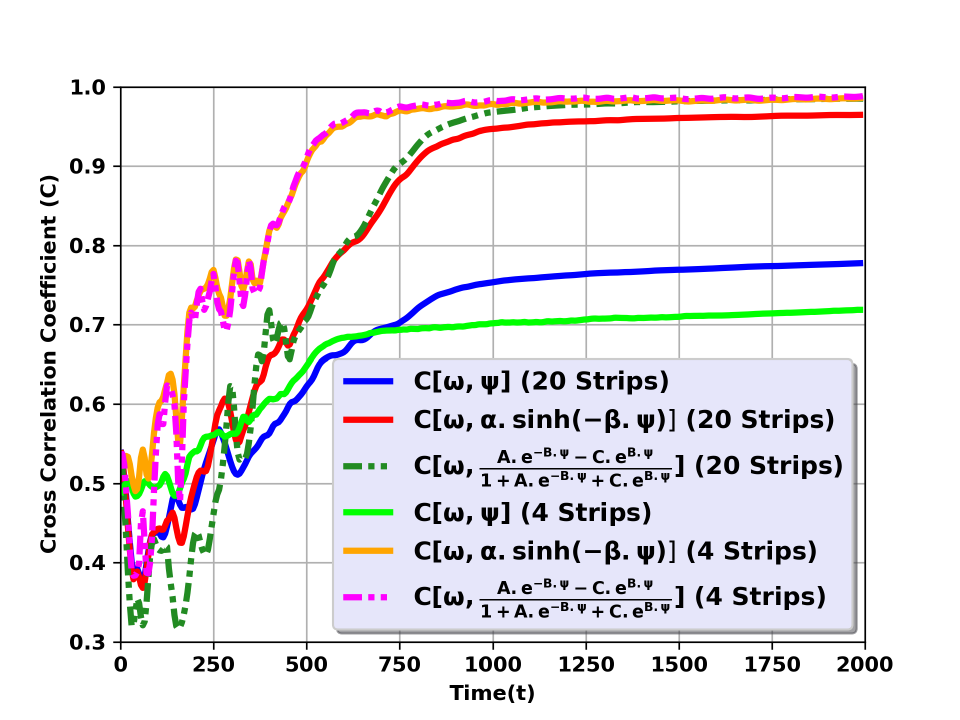}
        \caption{}
         \label{fig:crosscorrelations}
     \end{subfigure}
     \hfill
    \begin{subfigure}{0.49\textwidth}
         \centering \includegraphics[width=\textwidth]{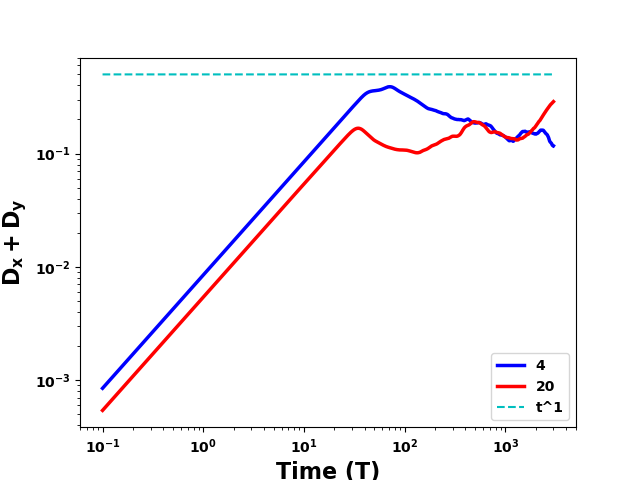}
        \caption{}
         \label{fig:latetimetr}
     \end{subfigure}
        \caption{ (a) The $\omega$–$\psi$ scatter plots (adapted from Figs.~10(b) and 16(c) of Biswas \textit{et al.}\cite{BG2022} with the corresponding author's permission) show good agreement with the point-vortex theory for low VPF (4-strip, 12.5\%) and with the KMRS predictions for high VPF (20-strip, 62.5\%), indicating a transition from point-vortex–dominated to finite-size–vortex–dominated dynamics with increasing vortex packing . 
        (b) Time evolution of spatially averaged cross-correlations between  $C[\omega(x,y,t), \psi(x,y,t)]$, $C[\omega(x,y,t), \omega_{\mathrm{PV}}(x,y,t)]$, and $C[\omega(x,y,t), \omega_{\mathrm{KMRS}}(x,y,t)]$ showing that, for low VPF (4 strips, 12.5\%), both point- and patch-vortex models correlate well with DNS, whereas for high VPF (20 strips, 62.5\%), the KMRS (finite-size vortex) model shows a stronger correlation, highlighting the dominance of finite-size effects with increasing vortex packing. (This plot has been adapted from Figs.~11 and 17(c) of Biswas \textit{et al.}\cite{BG2022} with the corresponding author's permission)). (c) The transport analysis exhibits a consistent trend, transitioning from sub-diffusive behavior in the low-VPF (4-strip) case to super-diffusive behavior in the high-VPF (20-strip) case at late times  establishing a strong correlation between the late-time Eulerian statistical equilibrium structures and the underlying Lagrangian transport in decaying two-dimensional incompressible Navier–Stokes turbulence.}
           \label{fig:L}
\end{figure*}

From the above discussions, we summarize our findings on the correlations at late times between the Eulerian equilibrium statistics examined by Biswas et al.\cite{BG2022} and the associated Lagrangian transport investigated in the present study, along with the role of the initial total circulation quantified by the vortex packing fraction (VPF). Biswas et al.\cite{BG2022} analyzed the fluid fields using $\omega$–$\psi$ scatter plots to assess correspondence with the point-vortex and patch-vortex theories, and through cross-correlations between $C[\omega(x,y,t), \psi(x,y,t)]$, $C[\omega(x,y,t), \omega_{\mathrm{PV}}(x,y,t)]$, and $C[\omega(x,y,t), \omega_{\mathrm{KMRS}}(x,y,t)]$. The $\omega$–$\psi$ scatter plots in Fig.~\ref{fig:scatter} (adapted from Biswas et al.\cite{BG2022}, reproduced with permission from the corresponding author) show good agreement with the point-vortex theory for low VPF (4-strip, 12.5\%) and with the KMRS predictions for high VPF (20-strip, 62.5\%), indicating a transition from point-vortex–dominated to finite-size–vortex–dominated dynamics with increasing vortex packing. Similarly, the dynamics of the spatially averaged cross-correlations between $C[\omega(x,y,t), \psi(x,y,t)]$, $C[\omega(x,y,t), \omega_{\mathrm{PV}}(x,y,t)]$, and $C[\omega(x,y,t), \omega_{\mathrm{KMRS}}(x,y,t)]$ reveal that Fig.~\ref{fig:crosscorrelations} (adapted from Biswas et al.\cite{BG2022}, reproduced with permission from the corresponding author), for the loosely packed case (4 strips, 12.5\%), both $C[\omega(x,y,t), \omega_{\mathrm{PV}}(x,y,t)]$ and $C[\omega(x,y,t), \omega_{\mathrm{KMRS}}(x,y,t)]$ remain high (close to unity) throughout the evolution, indicating that the flow dynamics are well captured by the point-vortex approximation. In contrast, for the tightly packed case (20 strips, 62.5\%), the correlation between $\omega(x,y,t)$ and the finite-size (KMRS) model, $C[\omega(x,y,t), \omega_{\mathrm{KMRS}}(x,y,t)]$, remains markedly stronger than that with the point-vortex model, underscoring the increasing importance of finite-size vortex effects at higher VPF. Our Lagrangian transport analysis exhibits a consistent trend, transitioning from sub-diffusive behavior in the low-VPF (4-strip) case to super-diffusive behavior in the high-VPF (20-strip) case at late times Fig.~\ref{fig:latetimetr}. This establishes a strong correlation between the late-time Eulerian statistical equilibrium structures and the underlying Lagrangian transport in decaying two-dimensional incompressible Navier–Stokes turbulence.

\section{Summary and Conclusions}

In this paper, we presented results from numerical simulations of decaying two-dimensional incompressible Navier-Stokes turbulence driven by Kelvin–Helmholtz instability with varying initial vorticity packing fractions, examining how they influence turbulence evolution across the linear, nonlinear, and dipolar vortex regimes. 
Using a tracer particle solver developed and thoroughly benchmarked for this study, and integrated with our existing fluid solver GHD2D, we analyzed, over long times, particle trajectories, dispersion, and position–velocity PDFs to identify cross-diagnostic correlations. Our findings show that the initial vorticity packing fraction critically governs the onset, nonlinear development, and steady-state dynamics of turbulence. The key results are summarized below: \\

1) The initial vorticity packing fraction and circulation
yeild distinct fluid turbulence characteristics. The KH shear instability generates vortex rolls that merge through an inverse energy cascade to form large scale dipoles. At low vorticity packing, turbulence onset is delayed and remains weak, anisotropic, and short-lived with dipoles forming quickly and undergoing orbital rotation along large radius circular paths. In contrast, higher packing produces a rapid onset of strong, long-lived, and isotropic turbulence, with dipole formation slightly delayed and their motion characterized by orbital rotation along small-radius circular paths. In the extreme case of the highest packing fraction (62.5\%), the dipoles instead undergo long, linear translational motion along diagonal trajectories. \\

2) The initial vorticity packing fraction and circulation lead to qualitatively distinct particle transport behaviors across early (short), intermediate, and late (long) timescales. 
In the early stage, the onset of turbulence—marking the transition from the linear to the nonlinear regime—and the corresponding change in particle transport from linear ballistic to nonlinear motion are primarily governed by the instability growth rate, which increases with the vorticity packing fraction.   
During the intermediate turbulent regime, transport exhibits a range of behaviors—sub-diffusive, diffusive, or super-diffusive—and undergoes transitions between anisotropic and isotropic states, depending on the initial packing, flow structure, and stage of evolution.  At late times, particle transport is found to be dominated by the slow drift of the largest coherent vortices, whose dynamics are also influenced by the initial vorticity packing, exhibiting either orbital rotation along nearly circular trajectories-leading to sub-diffusive trapping-or linear translational dipole motion along diagonal trajectories, associated with super-diffusive events. \\

3) While the overall particle transport across the entire transport regime gradually decreases with increasing vortex packing fraction, in the extreme case of the highest VPF (62.5\%), the total transport in the late stages becomes the largest among all cases due to anomalous superdiffusive behavior. \\

4) The initial vorticity packing fraction shapes the directional position PDFs, $n(x)$ and $n(y)$, governing both their evolution and long-time behavior and reflecting the associated particle transport. Low vorticity packing delays Gaussianization from the initial uniform distribution, leaving $n(x)$ and $n(y)$ unequal and indicative of sub-diffusive, anisotropic transport over long times. Increasing the vorticity packing fraction accelerates Gaussianization and overlap ($n(x) \approx n(y)$), marking a transition from sub-diffusive to super-diffusive and from anisotropic to isotropic transport during the intermediate stages, while remaining sub-diffusive at long times. At the highest packing fraction (62.5\%), the distributions develop heavy tails at very long times, reflecting fully super-diffusive particle transport.
 \\
  
5) The initial vorticity packing fraction shapes the directional velocity PDFs, $n(v_x)$ and $n(v_y)$, governing both their evolution and long-time behavior and reflecting the associated particle velocity transport.
Initially shear-dominated, with $n(v_x)$ exhibiting a bimodal structure and $n(v_y)$ a sharp central spike at $T = 0$, the PDFs evolve toward Gaussian distributions as turbulence develops and later regain bimodality due to the motion of counter-rotating dipoles. At low packing, Gaussianization is delayed, leaving $n(v_x)$ and $n(v_y)$ unequal and reflecting anisotropic, sub-diffusive transport during the intermediate stages, while at long times the bimodal distributions remain non-overlapping, indicating persistent anisotropic transport. At high packing, both $n(v_x)$ and $n(v_y)$ rapidly Gaussianize, overlap, and reflect isotropic, super-diffusive transport during the intermediate stages, and at long times the bimodal distributions overlap and remain equal, indicating sustained isotropic transport. \\
   
6) These distinct long-time transport characteristics, which vary consistently with increasing vorticity packing fraction from sub-diffusive to super-diffusive late time behavior, suggest a strong correlation between late-time Eulerian statistical equilibrium structures and the underlying Lagrangian transport in 2D decaying turbulence.

\appendix

\begin{acknowledgments}
The simulations and visualizations presented here are
performed on GPU nodes and visualization nodes of the ANTYA cluster at the Institute for Plasma Research (IPR), India. The authors are grateful to
the HPC support team of IPR for extending their help related to the ANTYA cluster. The authors are also grateful to Prof. Guglielmo Lacorata of the Institute of Marine Sciences (CNR–ISMAR), Rome, for his valuable insights in benchmarking the particle solver used in this study against results from the kinematic 2D chaotic flow investigated by his group.
\end{acknowledgments}

\nocite{*}
\bibliography{References}% Produces the bibliography via BibTeX.

\end{document}